\documentclass{aa}

\usepackage{natbib}
\usepackage[varg]{txfonts}
\usepackage{booktabs}
\usepackage{bm}
\usepackage{graphicx}  
\usepackage{subfig}
\usepackage{upgreek}
\bibpunct{(}{)}{;}{a}{}{,} 

\usepackage{colortbl}

\begin{document}

\title{Probing the hidden atomic gas in Class I jets with SOFIA}
 
\author{T.\,Sperling\inst{1}
\and J.\,Eislöffel\inst{1}
\and C.\,Fischer \inst{3}
\and B.\,Nisini\inst{2} 
\and T.\,Giannini\inst{2}
\and A.\,Krabbe \inst{3}}
\institute{Thüringer Landessternwarte, Sternwarte 5, D-07778, Tautenburg, Germany
\and INAF—Osservatorio Astronomico di Roma, via Frascati 33, I-00040 Monte Porzio, Italy
\and Deutsches SOFIA Institut University of Stuttgart, D-70569 Stuttgart, Germany}
\date{Received: 03 December 2019  / Accepted: 28 August 2020}

\abstract {We present SOFIA/FIFI-LS observations of five prototypical, low-mass Class I outflows (HH111, SVS13, HH26, HH34, HH30) in the far-infrared [O\,I]$_{63\upmu\text{m}}$ and [O\,I]$_{145\upmu\text{m}}$ transitions.} {Spectroscopic [O\,I]$_{63\upmu\text{m}, 145\upmu\text{m}}$ maps enable us to study the spatial extent of warm, low-excitation atomic gas within outflows driven by Class I protostars.    These [O\,I] maps  may potentially allow us to measure the mass-loss rates ($\dot{M}_\text{jet}$) of this warm component of the atomic jet.    } {A fundamental tracer of warm (i.e. $T$ $\sim$ $500$--$1500$\,K), low-excitation atomic gas is the
[O\,I]$_{63\upmu\text{m}}$ emission line, which is predicted to be the main coolant of dense dissociative J-type shocks   caused by decelerated wind or jet shocks associated with protostellar outflows. Under these conditions, the [O\,I]$_{63\upmu\text{m}}$ line can be  directly connected to the instantaneous mass ejection rate. Thus, by utilising spectroscopic [O\,I]$_{63\upmu\text{m}}$ maps, we wish to determine the atomic mass flux rate $\dot{M}_\text{jet}$ ejected from our target outflows.} {Strong [O\,I]$_{63\upmu\text{m}}$ emission is detected at the driving sources HH111IRS, HH34IRS, SVS13, as well as at the bow shock region, HH7. The detection of the [O\,I]$_{63\upmu\text{m}}$ line at HH26A and HH8/HH10 can be attributed to jet deflection regions. The far-infrared counterpart of the optical jet is detected in [O\,I]$_{63\upmu\text{m}}$ only for HH111, but not for HH34. We interpret the [O\,I]$_{63\upmu\text{m}}$ emission at HH111IRS, HH34IRS, and SVS13 to be coming primarily from a decelerated wind shock, whereas multiple internal shocks within the HH111 jet may cause most of the [O\,I]$_{63\upmu\text{m}}$ emission seen there. At HH30, no [O\,I]$_{63\upmu\text{m}, 145\upmu\text{m}}$ was detected. The [O\,I]$_{145\upmu\text{m}}$ line detection is at noise level  almost everywhere in our obtained maps. The observed outflow rates of our Class I sample are to the order of $\dot{M}_\text{jet}\sim 10^{-6} M_\odot\text{yr}^{-1}$, if proper shock conditions prevail.  Independent calculations connecting the [O\,I]$_{63\upmu\text{m}}$ line luminosity and observable jet parameters with the mass -loss rate are consistent with the applied shock model and lead to similar mass-loss rates. We discuss  applicability and caveats of both methods.   } {High-quality spectroscopic [O\,I]$_{63\upmu\text{m}}$ maps of protostellar outflows at the jet driving source potentially allow a clear determination of the mass ejection rate.  }     
\maketitle

{\renewcommand{\arraystretch}{1.2}
\begin{table*}[!htb]
\caption{\small{Target information.}}\label{table:objects}
\centering
\begin{tabular}{c c c c   c c c }
\hline\hline
Source & Cloud & R.A.\,(J2000)\tablefootmark{a} & Decl.\, (J2000)\tablefootmark{a} & $D$\tablefootmark{b} & $L_\text{bol}$\tablefootmark{c} & P.A.   \\[1.5pt]
 & & (h m s) & ($\degr\,\arcmin\,\arcsec$) &   (pc)  & ($L_\odot$) & ($^\text{o}$) \\ 
\hline  
HH111 IRS & \small{Orion B, L1617}   & \small{$05\,51\,46.1$} & \small{$+02\,48\,30$} & 420 & $20.8$\tablefootmark{d} & 275 \\ [1.5pt]
\hline  
SVS13     & \small{Perseus, L1450} & \small{$03\,29\,03.7$} & \small{$+31\,16\,04$} & 235 & $29.8-36.1$\tablefootmark{e} & 125  \\ [1.5pt]
\hline  
HH26 IRS  & \small{Orion B, L1630}   & \small{$05\,46\,03.6$} & \small{$-00\,14\,50$} & 420 & $4.0-8.0$\tablefootmark{f} & 45 \\ [1.5pt]
 \hline  
HH34 IRS  & \small{Orion A, L1641}   & \small{$05\,35\,29.8$} & \small{$-06\,26\,58$} & 430 & $10.8-17.4$\tablefootmark{g} & 165 \\ [1.5pt]
 \hline  
HH30 IRS  & \small{Taurus, L1551}  & \small{$04\,31\,37.5$} & \small{$+18\,12\,24$} & 140 & $0.2$\tablefootmark{h} & 30 \\ [1.5pt]
\hline 
\end{tabular}
\tablefoot{ 
\tablefoottext{a}{\small{taken from the Two Micron All Sky Survey (2MASS);}}
\tablefoottext{b}{\small{rounded from \citet{zucker_2019} based on GAIA DR2, except SVS13, taken from \citet{hirota_2008, hirota_2011} and based on VLBI observations of the associated maser;}}
\tablefoottext{c}{\small{we correct the luminosities taken from the cited papers for our assumed distances: $L_\text{bol} = (D_{\text{adopted}}/D_\text{paper})^2 L_\text{bol}^\text{paper}$.}}
\tablefoottext{d}{\small{\citet{reipurth_1989a}, $L_\text{bol}/L_\odot=25 $ at 460\,pc}}
\tablefoottext{e}{\small{\citet{cohen_1985} measures $L_\text{bol}/L_\odot= 66 $, whereas \citet{reipurth_1993} and \citet{harvey_1998} give $L_\text{bol}/L_\odot= 80 $ at 350\,pc in each case. More recently, \citet{tobin_2016} measured $L_\text{bol}/L_\odot= 32.5$ at 230\,pc}}
\tablefoottext{f}{\small{\citet{antoniucci_2008}, $L_\text{bol}/L_\odot= 4.6-9.2 $ at 450\,pc}}
\tablefoottext{g}{\small{\citet{antoniucci_2008}, $L_\text{bol}/L_\odot= 12.4-19.9 $ at 460\,pc}}
\tablefoottext{h}{\small{\citet{wood_2002}, \citet{cotera_2001}, \citet{molinari_1993}.}}
}
\end{table*}

\section{Introduction}

Jets powered by young stellar objects (YSOs) are an integral part of star formation and can  extend up to parsec distances from the driving source \citep[e.g.][]{eisloeffel_ppp_2000, ray_2007, frank_2014, bally_2016}. These outflows play an important role in transporting the angular momentum accumulated in the accretion disc away from the forming star, and therefore offer a unique opportunity to investigate the accretion properties of  protostars.

Based on their infrared spectral energy distribution (SED), the evolutionary sequence of protostars is broadly divided into the three Classes 0, I, and II  \citep[e.g.][]{lada_1987, andre_1993, greene_1994}.
 In the earliest evolutionary phase, the newly formed protostar (Class 0, lifetime: $\tau_\text{life}$\,$\sim$\,$10^4\,\text{yr}$) lies deeply embedded in its natal cloud, accreting the main part of its final mass and showing the strongest outflow activity \citep{bally_2016}. Due to the high extinction, Class 0 objects and their associated molecular outflows (e.g. detected in CO, SiO) are studied at submillimetre and far-infrared (FIR) wavelengths \citep[e.g.][]{gueth_1999, codella_2007}. In the consecutive Class I phase ($\tau_\text{life}$\,$\sim$\,$10^5\,\text{yr}$), the central object, though still embedded in a dusty envelope, becomes visible in the near-infrared (NIR). Outflows from Class I sources are detected in various collisionally excited optical and NIR atomic lines indicating an increasing atomic jet component. The more evolved Class II objects (referred to as classical T\,Tauri stars (CTTSs), $\tau_\text{life}$\,$\sim$\,$10^6\,\text{yr}$) have accreted and blown away so much of the surrounding material that they are becoming visible in the optical, but still are far from reaching the main sequence (the following stages, i.e. Class III and Post T\,Tauri stars, have lifetimes to the order of $\tau_\text{life}$\,$\sim$\,$10^7\,\text{yr}$).
 
The efficiency of the accretion-ejection  process during star formation can be estimated from observations that allow reliable conclusions on the  the mass-accretion rate $\dot{M}_\text{acc}$ and the mass-ejection rate $\dot{M}_\text{loss}$. Such studies of YSOs demonstrate a correlation between both quantities indicating a physical mechanism  behind it \citep[e.g.][]{hartigan_1995, white_2004}.
 
Theoretical models are consistent with that finding \citep[e.g.][]{shu_1994, konigl_2000}, and it is expected that as YSOs go through their stages of evolution, that is, they evolve  from a Class 0  to Class II object, their  mass-loss rate and mass-accretion rates decrease \citep{botemps_1996, saraceno_1996, caratti_2012}. Furthermore, the ratio of both quantities ($f=\dot{M}_\text{loss}/\dot{M}_\text{acc}$) provides decisive information on  proposed jet acceleration mechanisms. Ratios to the order of $f$\,$\sim$\,$0.3$  would give reason to justify an X-wind scenario  \citep{shu_1988, shu_1994}, whereas $f$\,$\sim$\,$0.01$--$0.5$ suggest that magnetohydrodynamical (MHD) disc wind models \citep{casse_2000, ferreira_1997} are more suitable describing the jet launching process.
 In this context, measurements of mass-loss rates can provide useful insights into the energy budget of young stars, the jet launching mechanism, and the evolution of protostellar outflows. 
 
However, large extinction in the earliest  stages of star formation prevents an accurate determination of essential physical properties   of YSOs such as the mass-flux rate  or the particle densities close to the star. Consequently, detailed studies of extended jets from these objects are usually performed only far from the central source (i.e. $\theta > 10\arcsec$). In these regions, however,
the jet has already interacted with the ambient medium through multiple shocks, loosing
the pristine information about its acceleration mechanism and connection with the accretion
events. 

 The far-infrared [O\,I]$_{63\upmu\text{m}}$ emission line (here [O\,I]$_{63}$) offers a unique opportunity to study the above-mentioned dynamical properties of the atomic jet, since this line is a) directly connected to the occuring shock, b) less affected by extinction, c) expected to be comparably bright amongst other shock tracers. With the traditional approach by using  CO rotational lines,  one can only indirectly estimate the time-averaged mass-loss rate from young embedded protostars, whereas the [O\,I]$_{63}$ line potentially allows a direct determination of the instantaneous mass loss rate \citep{hollenbach_1989}.

In this paper, we present extensive SOFIA FIFI-LS observations (Sect.\,\ref{observations}) of five prototypical low-mass Class I outflows  (HH111, SVS13 (also referred to as HH7-11 or SSV13), HH26, HH34, HH30, see Table\,\ref{table:objects})  starting close to their respective driving source. A brief description of the observed Herbig-Haro (HH) objects can be found in Appendix A. All targets have been mapped along their outflows in the atomic fine structure [O\,I]$_{63}$ (${}^{3}$P$_1$$-$${}^{3}$P$_2$) and [O\,I]$_{145}$ (${}^{3}$P$_0$$-$${}^{3}$P$_1$) transitions (abbreviation for the [O\,I]$_{145\upmu\text{m}}$ emission line). Since the excitation energies of the involved states ${}^{3}$P$_1$ and  ${}^{3}$P$_0$ are $\Delta E({}^{3}\text{P}_1-{}^{3}\text{P}_2)/k_\text{B} = 228\,\text{K}$ and  $\Delta E({}^{3}\text{P}_0-{}^{3}\text{P}_1)/k_\text{B} = 99\,\text{K}$, they can easily be excited via collisions with atomic or molecular hydrogen tracing the presence of warm (i.e. $T\sim$\,$500-1500\,\text{K}$), dense, low-excitation atomic gas. In comparison, optical lines such as [S\,II]$\uplambda\uplambda$6731,6716, H$\alpha$, [N\,II]$\uplambda\uplambda$6548,6583, or [O\,I]$\uplambda$6300 trace the hot atomic gas ($T$\,$\sim$\,$10^4\,\text{K}$) and are commonly used to investigate extended protostellar outflows \citep[e.g.][]{hartigan_1994, bacciotti_1999, hartigan_2011}. On the other hand NIR emission lines (e.g.\,[Fe\,II], H$_2$) have widely been used to derive the physical conditions of warm ionised gas in Class 0/I sources \citep[e.g.][]{eisloeffel_2000, davis_2003, giannini_2004, nisini_2005, takami_2006, garcia_2013, giannini_2013}. However, NIR lines fail to probe the warm low excitation atomic gas, that can indeed play a central role in the energetics of embedded jets. NIR  lines from singly ionised iron (e.g.\,[Fe\,II] 1.644\,$\upmu$m) trace shock excited, partially ionised gas at $T_\text{ex}$\,$\sim$\,$2000-15000\,\text{K}$ in dense regions ($n_\text{cr} > 10^{4}\,\text{cm}^{-3}$); for example,\,\citet{nisini_2002, pesenti_2003}, whereas ro-vibrational H$_2$ lines, such as at 2.122\,$\upmu$m, are connected to the warm ($T$\,$\sim$\,$2000-3000\,\text{K}$), dense ($n_\text{H}$\,$\geq$\,$10^3\,\text{cm}^{-3})$, molecular component of the outflow, \citep[e.g.][]{garcia_lopez_2010, davis_2011}. 
To complete the picture, observations of low-$J$  pure rotational CO-lines  in the millimetre range (e.g. $J=1-0$ at 2.6\,mm, $J=2-1$ at 1.3\,mm, $J=3-2$ at 0.86\,mm) have been used to investigate large scale morphologies of outflows tracing the cold swept-up gas ($T$\,$\sim$\,$10-100\,\text{K}$), for example, \citet{fukui_1993} and \citet{raga_1993}.  In the end, all those observations at different wavelengths probe different physical conditions and  thus complement each other providing a robust picture of outflow dynamics in protostellar systems. 

The FIR [O\,I]$_{63}$ line is predicted to be the main coolant of dense dissociative J-shocks in a wide range of shock velocities and gas densities \citep{hollenbach_1989}. As such, this line is the best tracer of the interactions between the high velocity primary jet and the dense ambient medium. 
In addition to dense shocks, [O\,I]$_{63}$ emission is recognised to be strong in photo-dissociation regions (PDR) due to  possible UV illumination by a present UV field \citep{hollenbach_1985, ceccarelli_1997}. Although a contribution  from  UV excitation cannot be ruled out, recent mid-infrared observations of protostellar outflows in [Fe\,II], [Fe\,III], and [Si\,II]  emission lines indicate that they are negligible \citep{watson_2016}. 
 
Unfortunately, the water vapour in the Earth's atmosphere prevents ground-based observations of FIR emission lines such as [O\,I]$_{63,145}$. Whereas the  [O\,I]$_{63}$ line is thought to be strong in outflows, the [O\,I]$_{145}$ line is fainter usually by a factor of 10--20, making a detection technically challenging. Past observations of [O\,I]$_{63}$ emission associated with protostellar outflows have been obtained by the Kuiper Airborne Observatory \citep{ceccarelli_1997}, ISO \citep{giannini_2001, nisini_2002, liseau_2006}, and Herschel/PACS \citep{green_2013, karska_2013, podio_2012, benedettini_2012, santangelo_2013, nisini_2015}, among others. Surprisingly, jets from Class I objects were not mapped in any of these studies, leaving a gap in our understanding of the protostellar evolution from Class 0 to Class II objects. Fortunately, the FIFI-LS instrument aboard the  flying observatory  SOFIA allows deep observations of both [O\,I]$_{63,145}$ emission lines simultaneously enabling measurements of the mass-loss rates, if proper conditions prevail  (Sect.\,\ref{sec:mass_flux_rates}).  

\section{Observations}\label{observations}

All five prototypical Class I objects (HH111, SVS13, HH26, HH34, HH30: see Table\,\ref{table:objects} \& \ref{table:atranparameters}) were mapped along their outflow axes at the [O\,I]\,\,63.1852\,$\upmu$m  (4.7448\,THz)  and the [O\,I]\,\,145.535\,$\upmu$m  (2.060\,THz)  far-infrared fine structure lines with the FIFI-LS instrument operating aboard the Stratospheric Observatory For Infrared Astronomy \citep[SOFIA,][]{young_2012}.
SOFIA is a modified Boeing 747SP aircraft with a 2.5\,m telescope (effective aperture) and has a nominal pointing accuracy of $0\farcs 5$.
 
As a key feature of FIFI-LS is that both [O\,I]$_{63,145}$ emission lines were observed simultaneously with two independent grating spectrometers with wavelength ranges of 51--120\,$\upmu$m (blue channel) and 115--200\,$\upmu$m (red channel) \citep{fifi_looney_2000, fischer_2018, fifi_colditz_2018}. FIFI-LS provides an array of $5 \times 5$ spatial pixels (spaxels) in each channel covering a field of
view of $30\arcsec \times 30\arcsec$  in the blue (pixel size: $6\arcsec\times 6\arcsec$)
and $1\arcmin \times 1\arcmin$ (pixel size:$12\arcsec\times 12\arcsec$) in the red.  

The diffraction-limited FWHM beam size at $63\,\upmu$m  ($145\,\upmu$m) is $\sim 5\farcs 4$ ($12\farcs 4$).
The spectral resolutions $R=\lambda/\Delta\lambda$ are  $ 1300$ at  $63\,\upmu$m  and  $1000$ at $145\,\upmu$m, which correspond to a medium velocity resolution of 231\,km\,$\text{s}^{-1}$ and 300\,km\,$\text{s}^{-1}$ , respectively. The data cubes of the blue (red) channel  feature a sampling of 34\,km\,$\text{s}^{-1}$ (42\,km\,$\text{s}^{-1}$) per spectral element. The spatial sampling is specified by 1$\arcsec$/spaxel in the blue channel and 2$\arcsec$/spaxel  in the red channel. 
The data were  acquired via  three SOFIA flights in Cycles 3 and 5 (program IDs: 03\verb|_|0073, 05\verb|_|0200) in two point symmetric chop   FIFI-LS  mode.

\begin{table}[!htb]
\caption{\small{SOFIA flight information and chosen ATRAN parameters for the five observed targets (zenith angle $\theta$, flight height $H$, water vapour overburden $wvp$, see Fig.\,\ref{fig:sofia}).}}\label{table:atranparameters}
\centering
\begin{tabular}{c c c c c c c }
\hline\hline
Object              & Obs. date         & Flight            &  Obs. time\tablefootmark{a}     & $\theta$  & $H$   & $wvp$     \\
                    &                   &                   & (min)                           & ($\degr$) & (kft) & ($\upmu$m) \\
\hline
\small{HH111}       & \tiny{2017-03-07} & \tiny{383}        & \tiny{73}                       & $40$      & $40$  & $8$        \\
\small{SVS13}       & \tiny{2017-03-07} & \tiny{383}        & \tiny{61}                       & $55$      & $39$  & $6$        \\
\small{HH26}        & \tiny{2015-03-12} & \tiny{199}        & \tiny{39}                       & $50$      & $41$  & $6$         \\
\small{HH34}        & \tiny{2015-10-22} &\tiny{249}         & \tiny{71}                       & $55$      & $42$  & $7$         \\
\small{HH30}        & \tiny{2017-02-25} & \tiny{378}        & \tiny{35}                       & $55$      & $42$  & $6$         \\
\hline
\end{tabular}
\tablefoot{ 
\tablefoottext{a}{\small{Total effective on source integration time.}}
}
\end{table}

\section{Data reduction}\label{data_reduction}
 
Although SOFIA operates 12--14 kilometres above the ground the interfering impact of the Earth's atmosphere has to be mitigated (see Fig.\,\ref{fig:sofia}).
We applied the SOFIA/FIFI-LS data reduction pipeline (REDUX) to our data cubes excluding the telluric correction step, since the atmospheric transmission at about $\lambda =63\,\upmu$m is below 0.6 (see Fig.\,\ref{fig:minispectrum_hh111A}) causing difficulties in the telluric correction \citep{vacca_2016}. 
Instead, we used our own Python script JENA.py,  which firstly  cuts off irregularities on the edges of the FIFI-LS data cubes that arise from the mosaic mapping. Furthermore, JENA.py applies  an optimal spectrum extraction procedure on each data cube spaxel (portrayed in Fig.\,\ref{beam}) to increase their signal-to-noise ratio \citep[SNR][]{horne_1986}. Finally, JENA.py mitigates the impact of the atmosphere as described in the following. 

\begin{figure} 
\resizebox{\hsize}{!}{\includegraphics[trim=110 420 90 50, clip, width=0.9\textwidth]{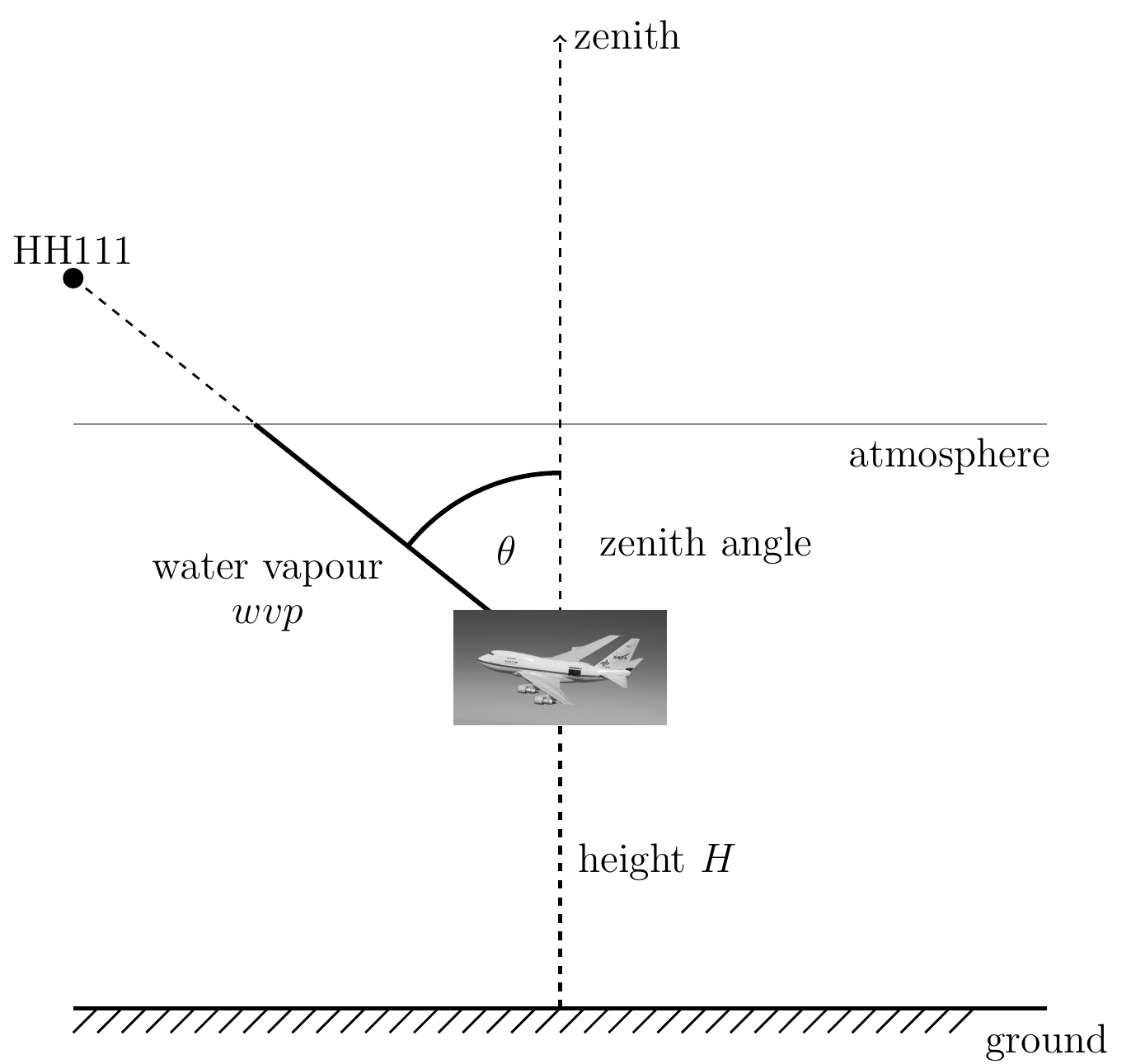}}
\caption{\small{SOFIA flight and relevant atmospheric parameters.}}\label{fig:sofia} 
\end{figure}

\begin{figure*}[!htb] 
\centering
\subfloat{\includegraphics[width=0.5\textwidth]{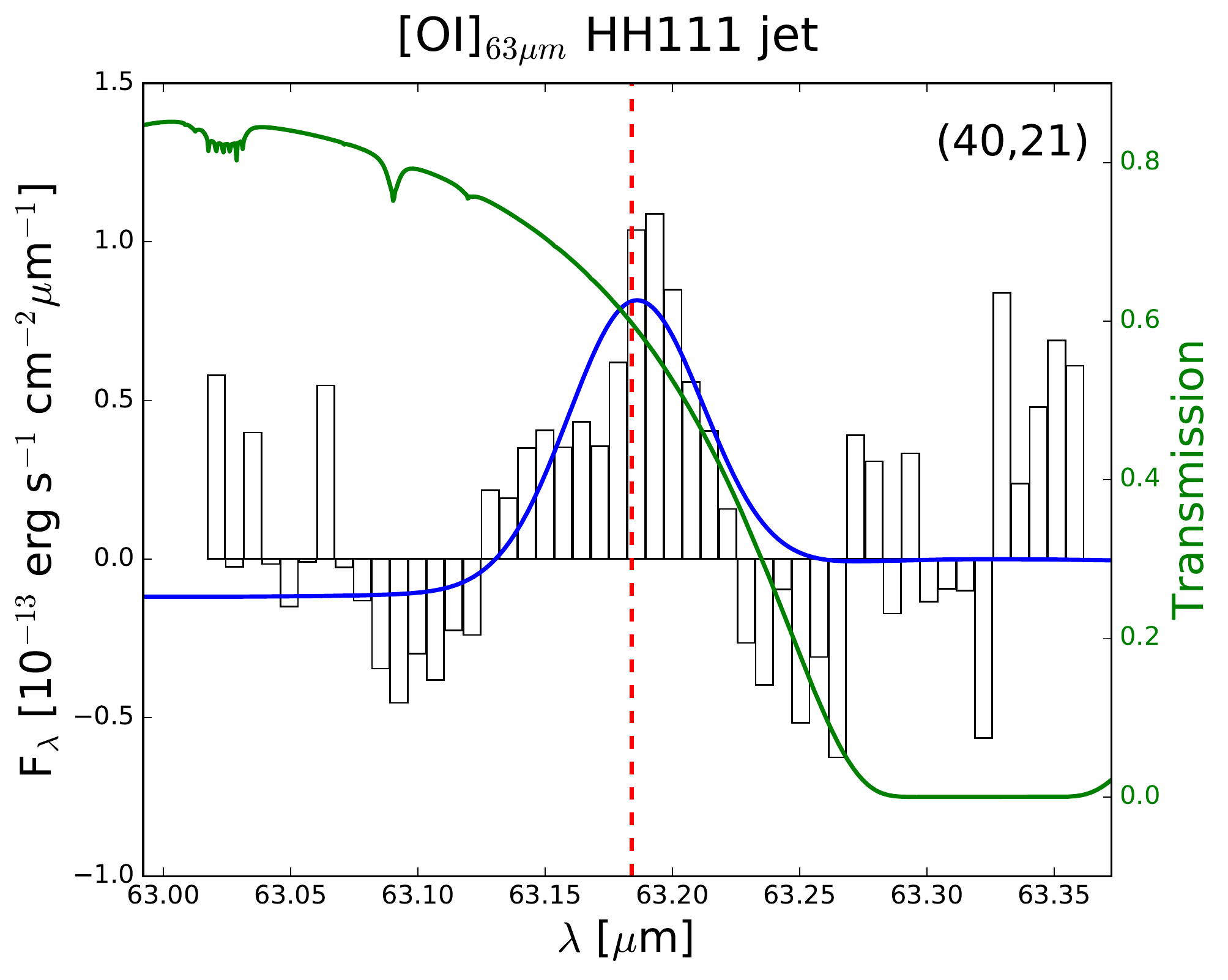}\label{fig:minispectrum_hh111A}}
\hfill
\subfloat{\includegraphics[width=0.5\textwidth]{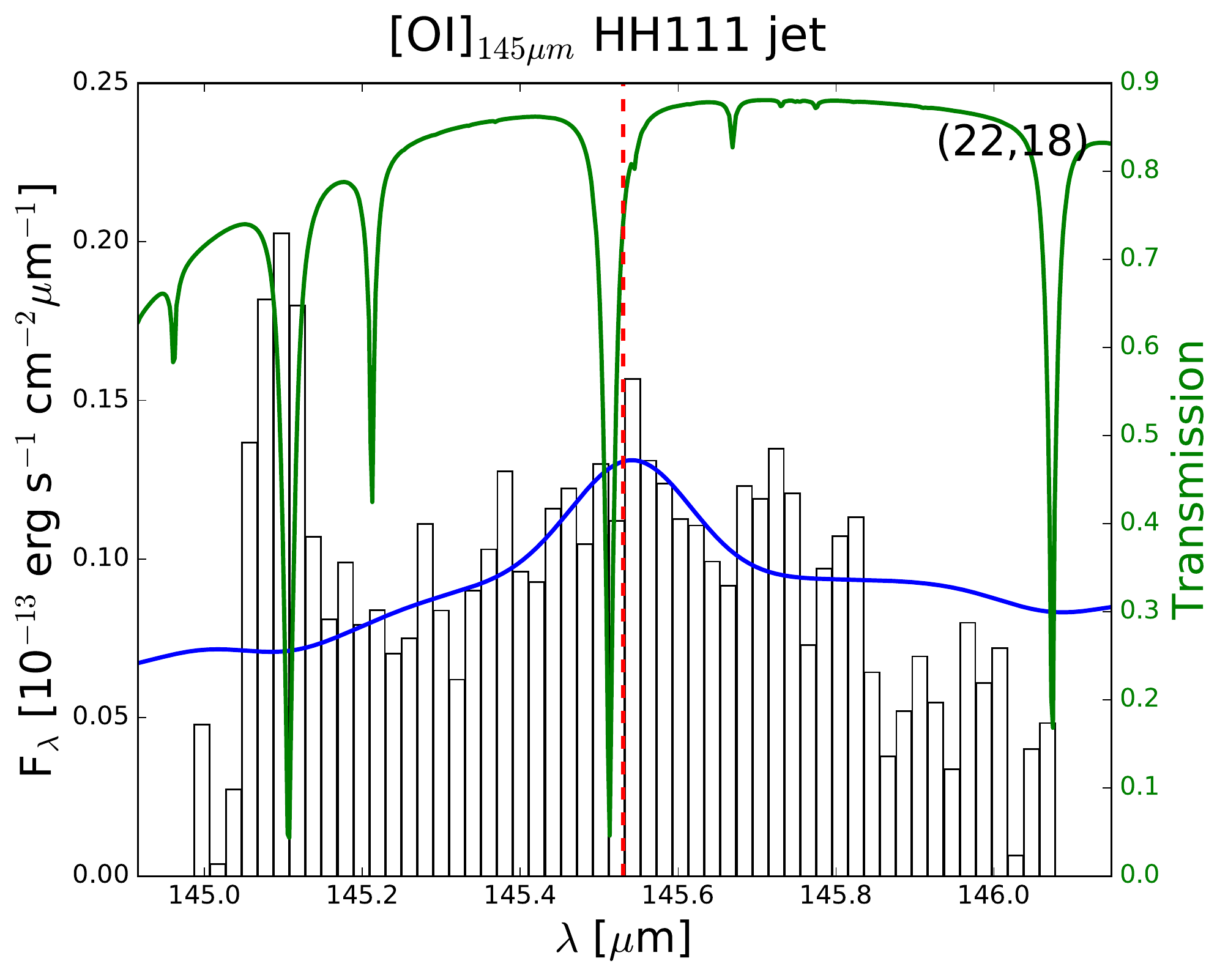}\label{fig:minispectrum_hh111_red}}
\hfill 
\caption{\small{Atmospheric transmission at both [O\,I]$_{63,145}$ lines together with two sample spectra from HH111.}}\label{fig:transmission}
\end{figure*} 

Considering that various synthetic spectra of atmospheric transmission  (ATRAN-models \citep{atran_lord_1992}) are accessible, we chose one specific ATRAN-model for each observed object out of all relevant three parametric ATRAN-models $\tau(\lambda; \bm{a})$ according to the flight parameters $\bm{a} :=  \left \{H, \theta, wvp \right \}$ during their observations  (Table \ref{table:atranparameters}). Here we denote  $H$, $\theta,$ and $wvp$ as flight height, zenith angle, and water vapour overburden, respectively (Fig.\,\ref{fig:sofia} and Table\,\ref{table:atranparameters}).

\begin{figure}{}
\resizebox{\hsize}{!}{\includegraphics[trim=55 100 100 60, clip, width=\textwidth]{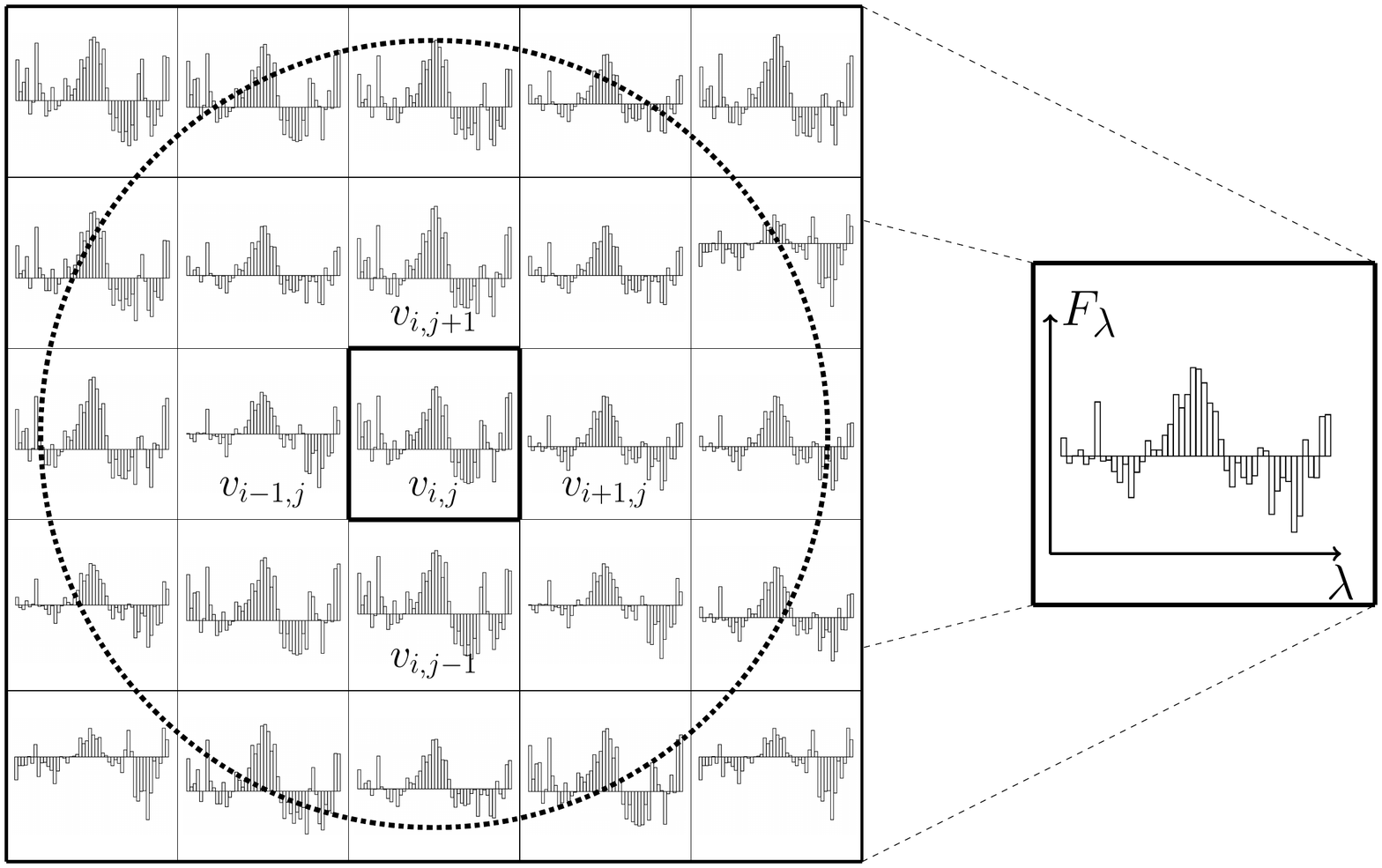}}
\caption{\small{An illustration of the optimal spectrum extraction procedure implemented in JENA.py. The dotted circle around spaxel $(i,j)$ in the middle of the $5\times 5$ spaxel field shows exemplary the FWHM of the spatial beam.  For each surrounding  spaxel, that is (partly) covered by the beam, a factor (e.g. $v_{i+1,j}$ for spaxel $(i+1,j)$) is calculated representing the enclosed volume under the normalised 2D-Gaussian beam. Spaxel $(i,j)$  is replaced by the sum of all beam covered spaxels weighted non-uniformly  by their specific volume factors (to see on the right side of the picture) while simultaneously preserving its photometric accuracy.}}
\label{beam}
\end{figure}

Assuming now that the observed object radiates an emission line in the form of a 1D-Gaussian function,
\begin{equation}
\Phi(\lambda; \bm{b}) = \frac{A}{\sqrt{2\pi}\sigma}\,\text{exp}\Bigg[-\frac{1}{2} \left(\frac{\lambda - \mu}{\sigma}\right)^2 \Bigg] + B
,\end{equation}
\noindent
with the four parameters $\bm{b} := \left \{ A, \sigma, \mu, B \right \}$, and the FIFI-LS spectral instrument  function $\text{SIF}\left(\lambda; R\right)$ is given by a 1D-Gaussian depending on the spectral resolution  $R$, we ideally expect to detect the discrete signal  
\begin{equation}\label{equ:S}
y(\lambda_k; \bm{a}, \bm{b}, R) =  S\left( \bigg[ \Phi(\lambda; \bm{b})\cdot  \tau(\lambda; \bm{a})   \bigg]  *\text{SIF}\left(\lambda; R  \right)\right)
\end{equation}
in each spaxel of a given data cube. The function $S$ in Eq.\,\ref{equ:S} just samples the modelled signal to equidistant wavelength grid points $\lambda_k$ predetermined by the individual data cubes.

To extract the emission line parameters  $\bm{b}$ in each spaxel of our data cubes, we used the Levenberg–Marquardt algorithm  \citep{lm_fit_python}. 
 Given the fact that the atmospheric transmission causes difficulties in the telluric correction, we chose to weight our  $\chi^2$ in the non-linear least-squares fit in each spaxel with the  atmospheric transmission  
\begin{equation}
\chi^2  =  \sum_k  \tau(\lambda_k; \bm{a}) \cdot \left( \frac{\text{data}(\lambda_k) - y(\lambda_k; \bm{a}, \bm{b}, R) }{ \epsilon (\lambda_k)}\right)^2 ,
\end{equation}
whereby data$(\lambda_k)$ are the flux measurements at $\lambda_k,$ and $\epsilon(\lambda_k)$ are the corresponding error values, which are given by the standard flux errors $\sigma(\lambda_k)$ multiplied by the number of spaxels in the spatial beam.

The continuum subtracted flux $f$ in one spaxel is then  determined  only by the parameter $A$, since  
 \begin{equation}
 f = \int_{-\infty}^\infty \left( \Phi(\lambda,\bm{b})-B\right) \,\text{d}\lambda = A.
  \end{equation} 
 Accordingly, the atmospherically adjusted continuum in one specific spaxel is given by the parameter $B$.

Errors in $A$ and $B$ were estimated from the covariance matrix. However, this method led to unreliable error values $\Delta A$ in a few spaxels with low signal-to-noise values. Therefore we applied the method of \citet{avni} to estimate the 1$\sigma$ confidence interval for $\Delta A$ only in these cases. 
 
On account of the medium spectral resolution of FIFI-LS we did not extract any velocity information from $\bm{b}$. 
The observed [O\,I]$_{63}$ linewidths $\Delta V_\text{obs}$ are to the order of 180--220\,$\text{km}\,\text{s}^{-1}$ , indicating that this line is spectrally  unresolved in all our targets ($\Delta V_\text{obs} = \sqrt{\Delta V_\text{line}^2 + \Delta V_\text{FIFI-LS}^2} $). We therefore constrained the intrinsic linewidth in the fitting procedure to be in the range of $\Delta V_\text{line} = 30-150\,\text{km}\,\text{s}^{-1}$. 
The total uncertainty in the absolute flux calibration for the integrated line fluxes amounts to approximately $20\,\%$.  

To evaluate the significance of the [O\,I]$_{63,145}$  line detection, we estimated the signal-to-noise ratio in each spaxel using the rms-method. Since atmospheric features at both [O\,I]$_{63,145}$ lines heavily corrupt our spectra (see Fig.\,\ref{fig:transmission}), we  determined the rms on the continuum around the line where the atmospheric transmission is above 0.6. 

Figure\,\ref{fig:all_minispectra} shows several sample spaxels at $63\,\upmu\text{m}$ indicating a clear [O\,I]$_{63}$ detection (SNR $\sim$ $4$--$15$) at the associated regions of HH111, SVS13, HH26, and HH34. In contrast, however, the [O\,I]$_{145}$ line detection is at noise level in most spaxels of our maps. This is not surprising since the [O\,I]$_{145}$ line lies deep in an ozone feature and is by its physical nature much fainter than the [O\,I]$_{63}$ line. Therefore, we do not present $145\,\upmu\text{m}$ maps here .

\begin{figure*} 
\centering
\subfloat{\includegraphics[trim=0 0 0 0, clip, width=0.33 \textwidth]{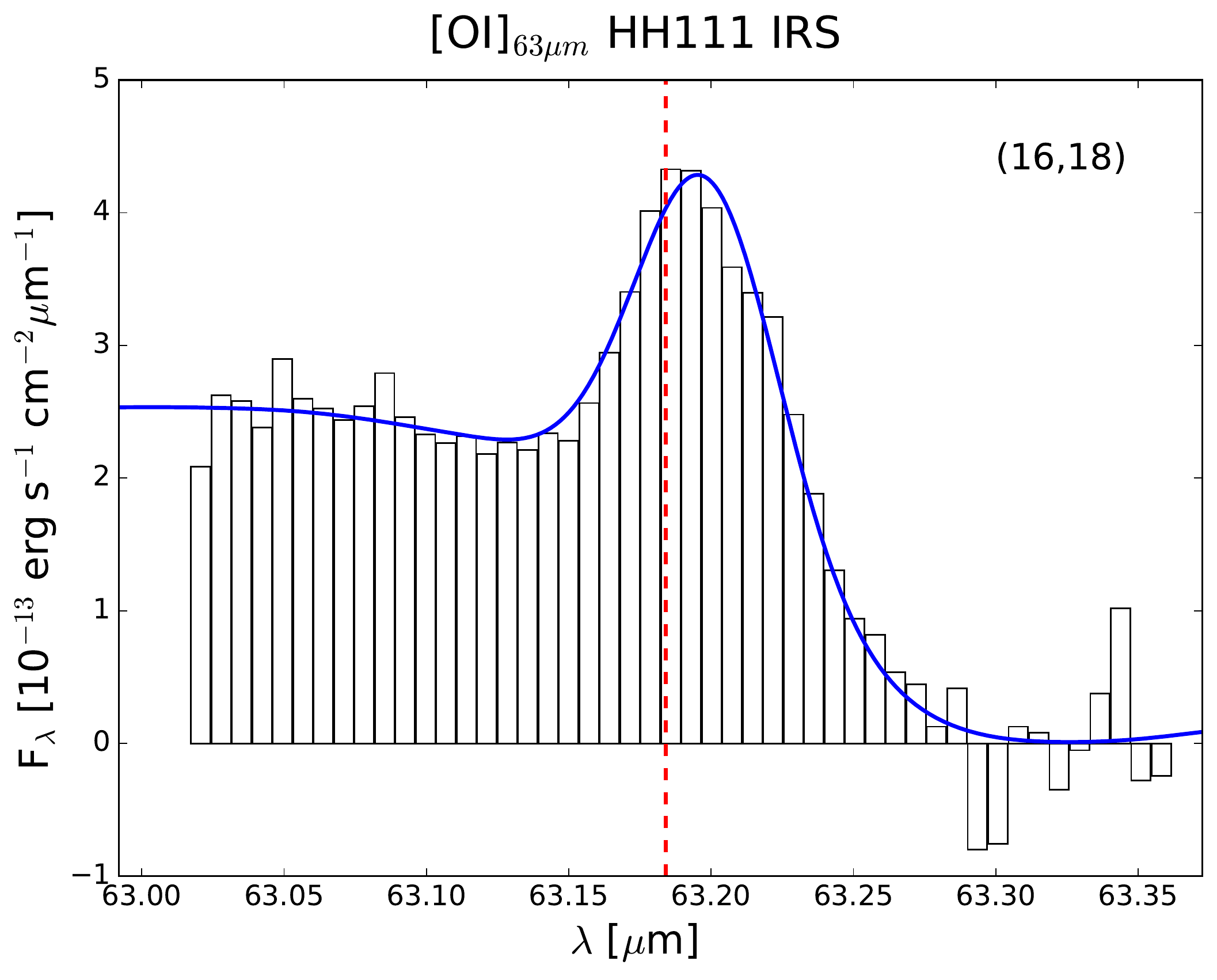}\label{fig:minispectrum_hh111B}}
\hfill
\subfloat{\includegraphics[trim=0 0 0 0, clip, width=0.33 \textwidth]{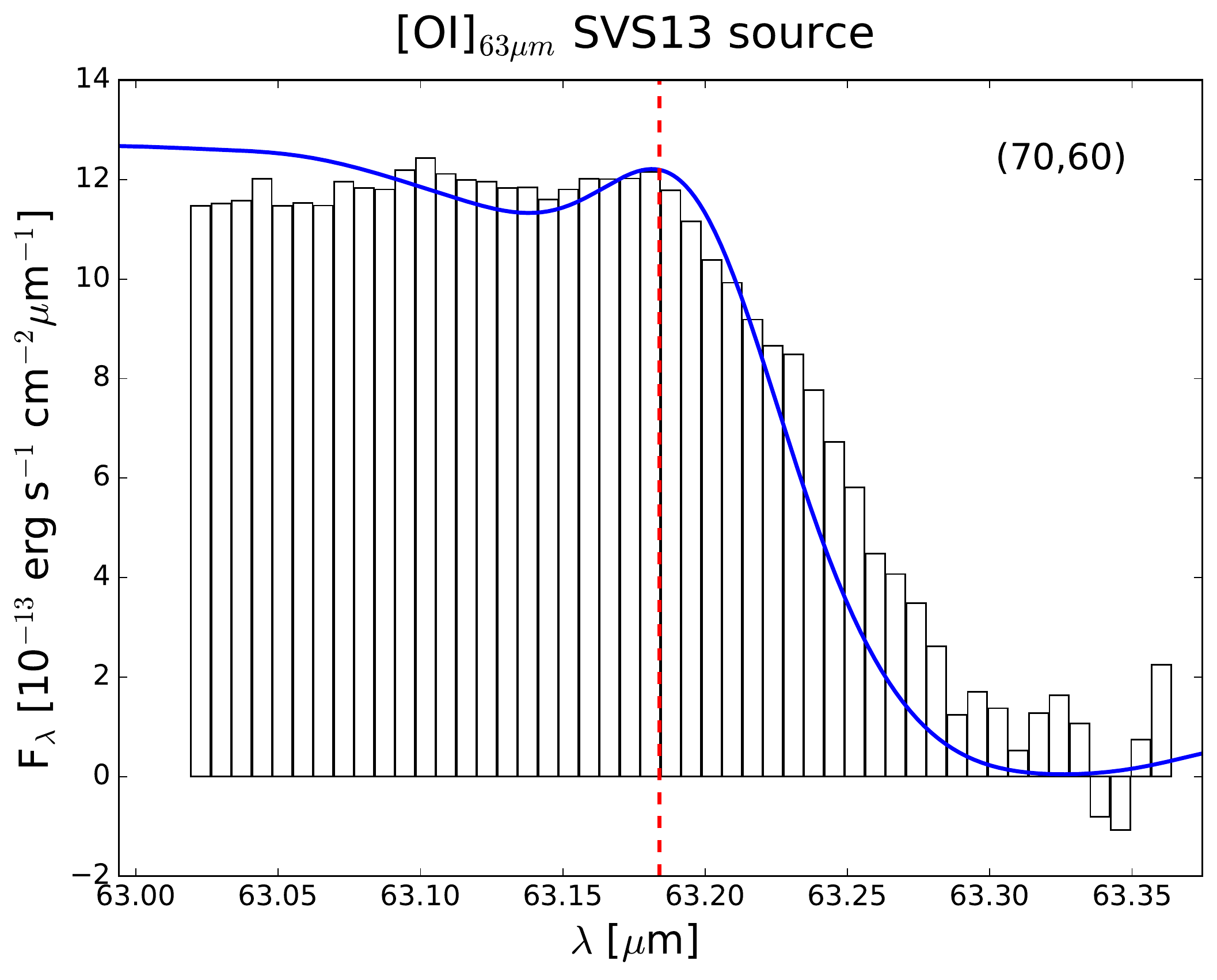}\label{fig:minispectrum_svs13A}}
\hfill
\subfloat{\includegraphics[trim=0 0 0 0, clip, width=0.33 \textwidth]{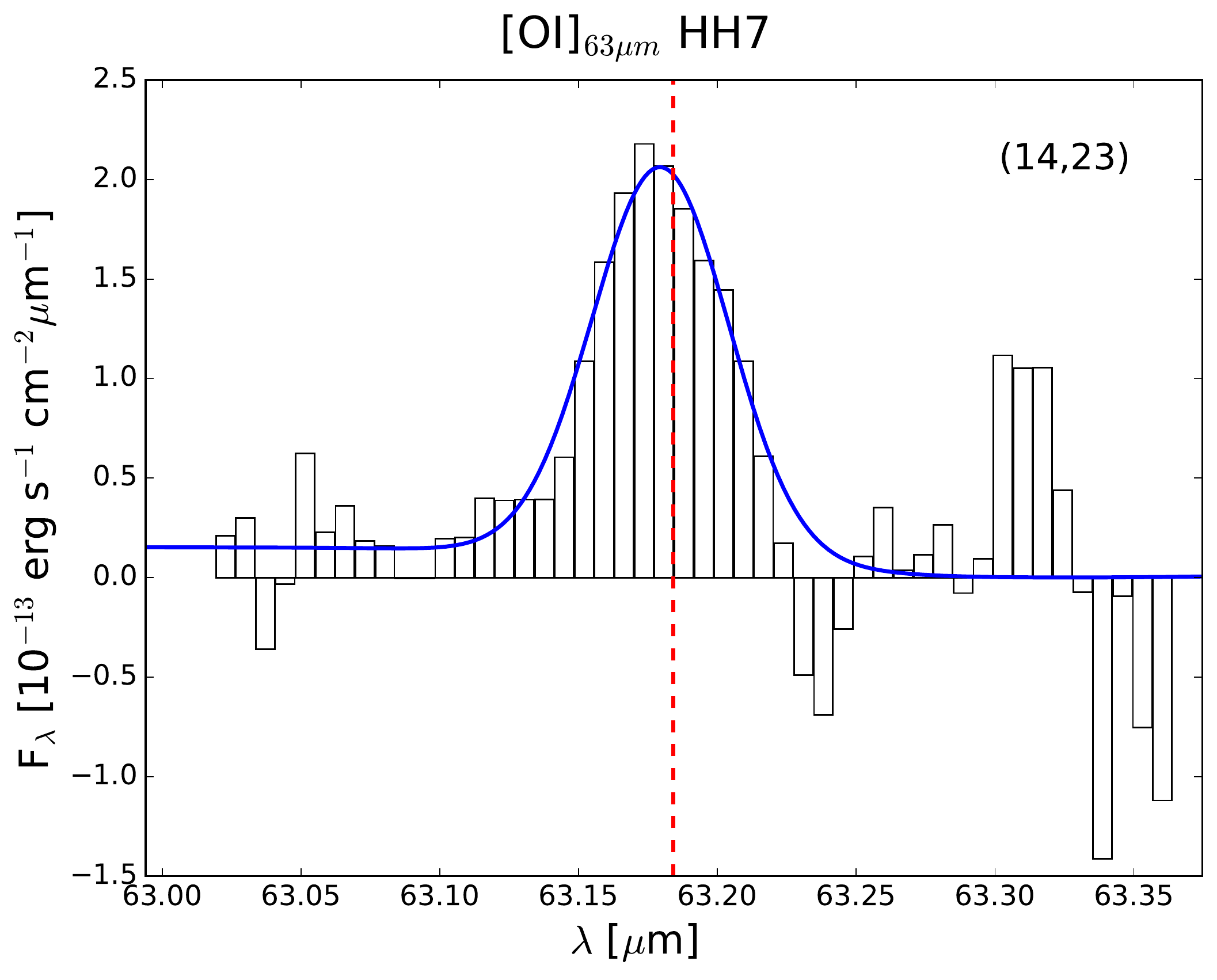}\label{fig:minispectrum_svs13B}}
\hfill
\subfloat{\includegraphics[trim=0 0 0 0, clip, width=0.33 \textwidth]{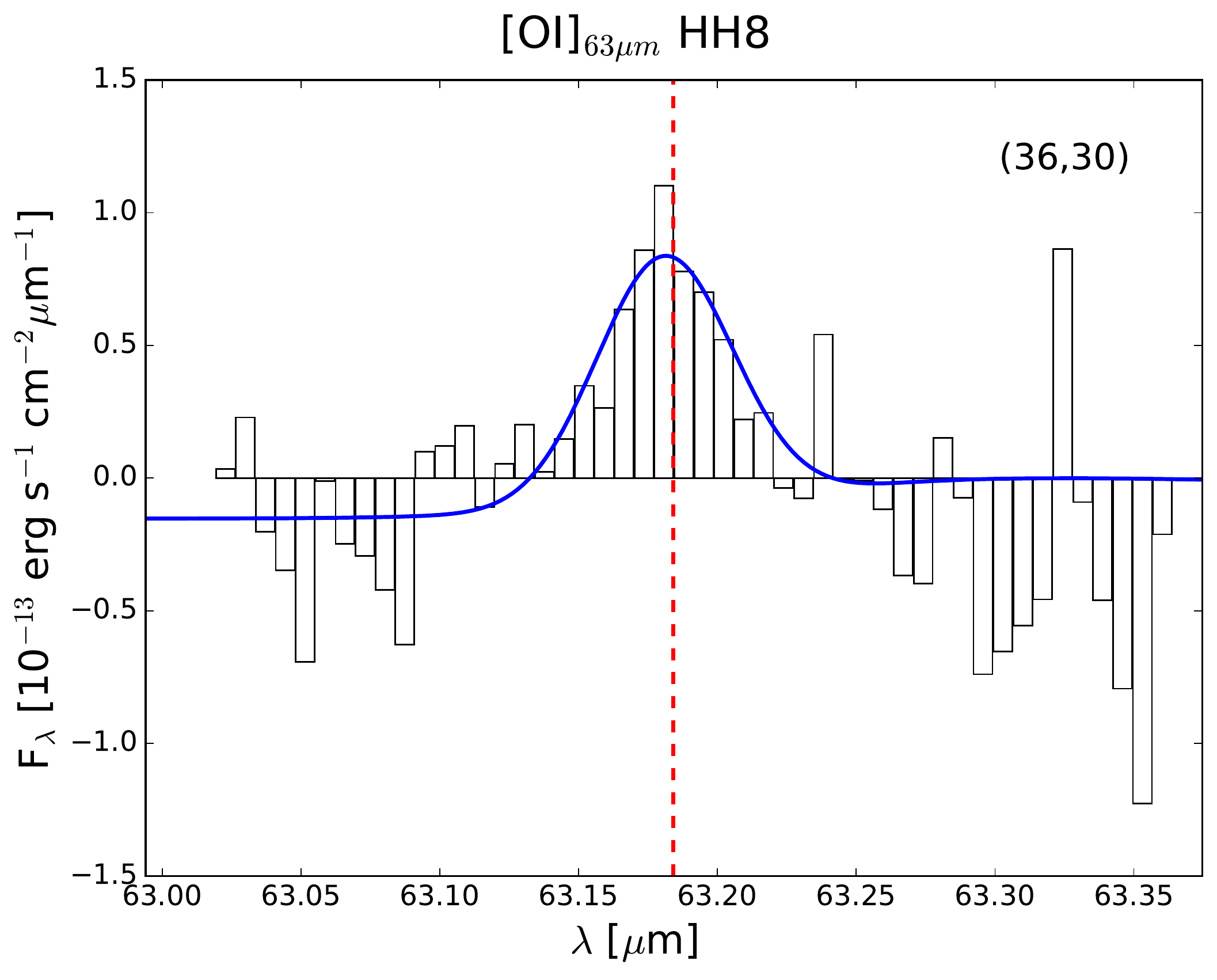}\label{fig:minispectrum_svs13C}}
\hfill
\subfloat{\includegraphics[trim=0 0 0 0, clip, width=0.33 \textwidth]{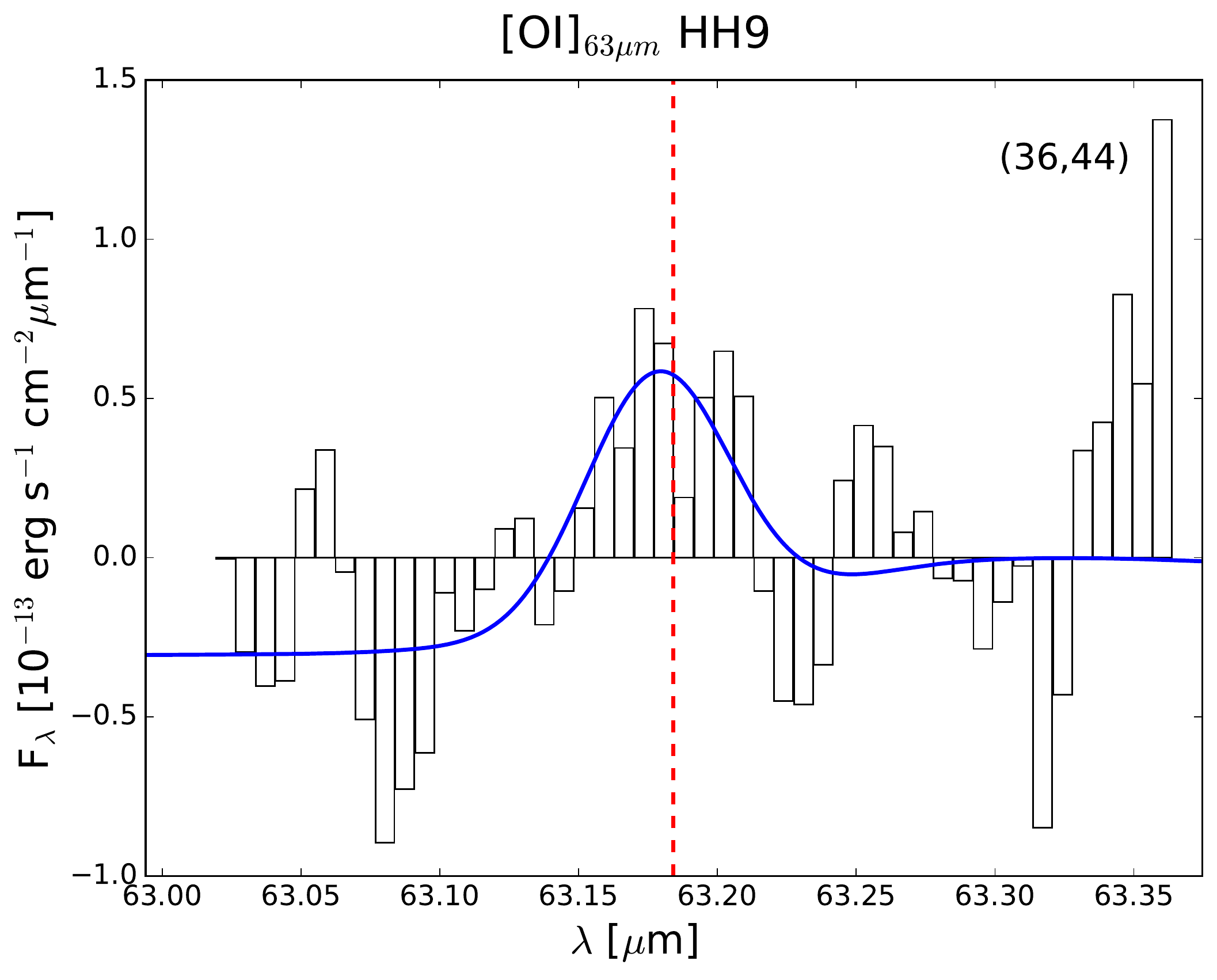}\label{fig:minispectrum_svs13D}}
\hfill
\subfloat{\includegraphics[trim=0 0 0 0, clip, width=0.33 \textwidth]{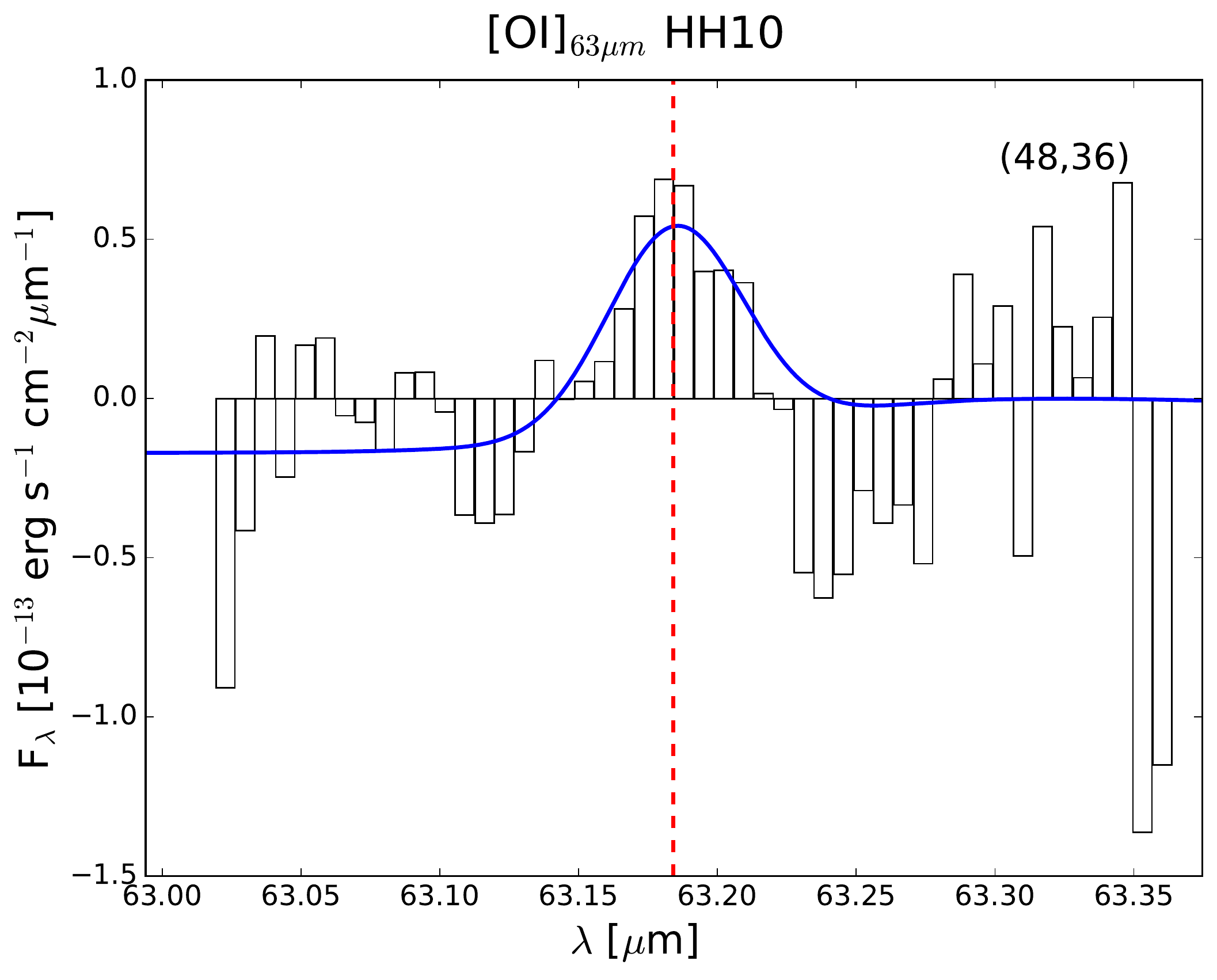}\label{fig:minispectrum_svs13E}}
\hfill
\subfloat{\includegraphics[trim=0 0 0 0, clip, width=0.33 \textwidth]{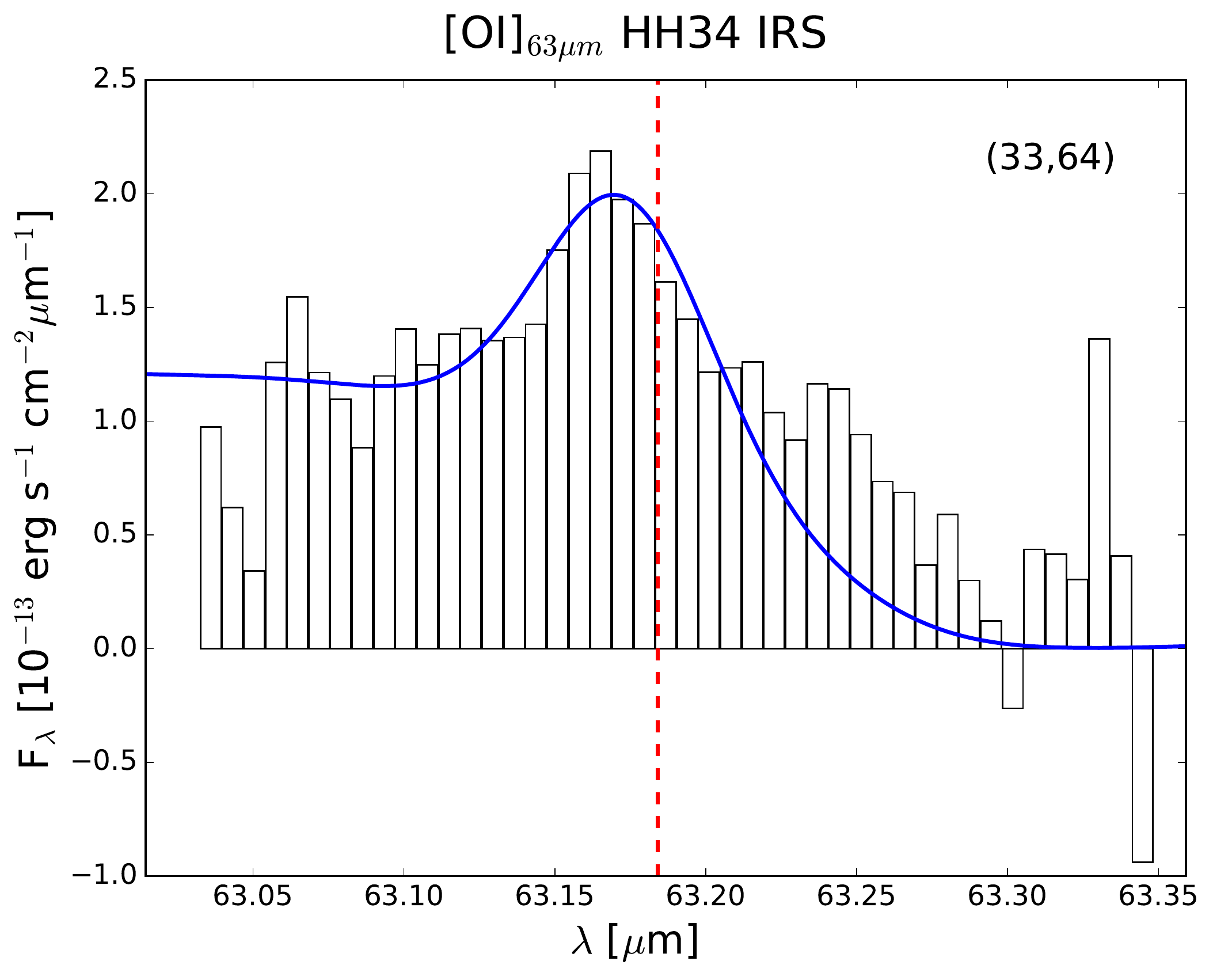}\label{fig:minispectrum_hh34A}}
\hfill
\subfloat{\includegraphics[trim=0 0 0 0, clip, width=0.33 \textwidth]{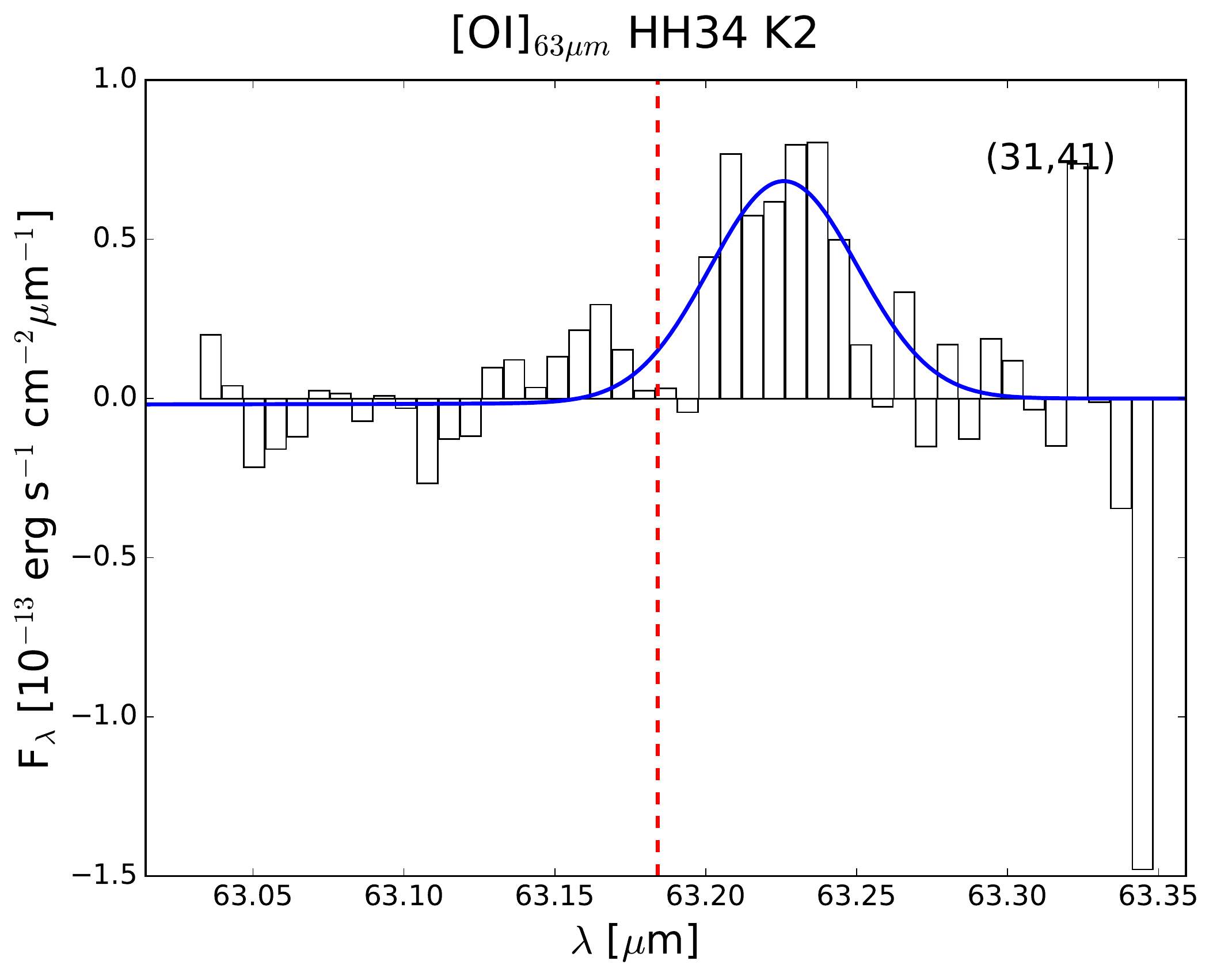}\label{fig:minispectrum_hh34B}}
\hfill
\subfloat{\includegraphics[trim=0 0 0 0, clip, width=0.33 \textwidth]{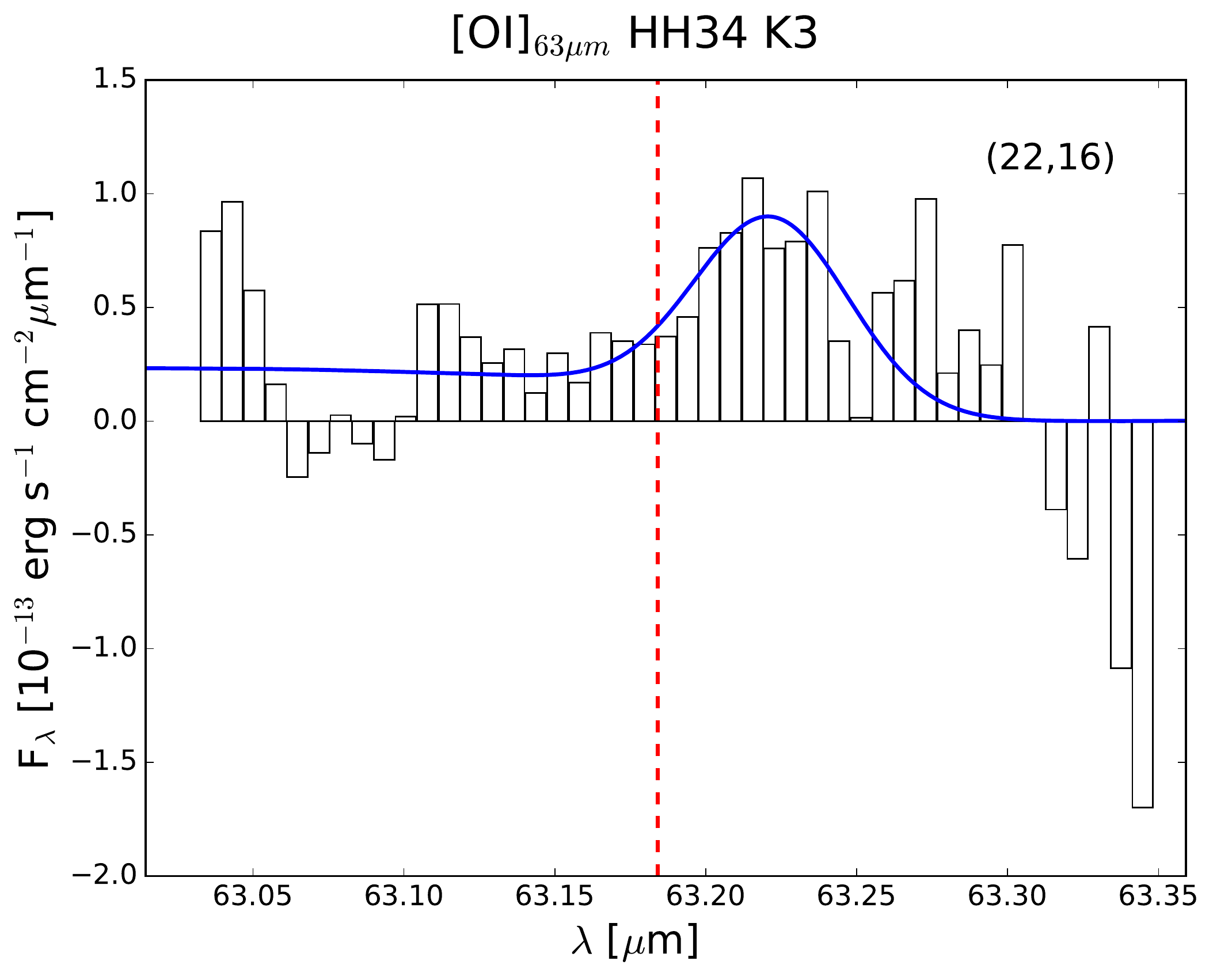}\label{fig:minispectrum_hh34C}}
\hfill
\subfloat{\includegraphics[trim=0 0 0 0, clip, width=0.33 \textwidth]{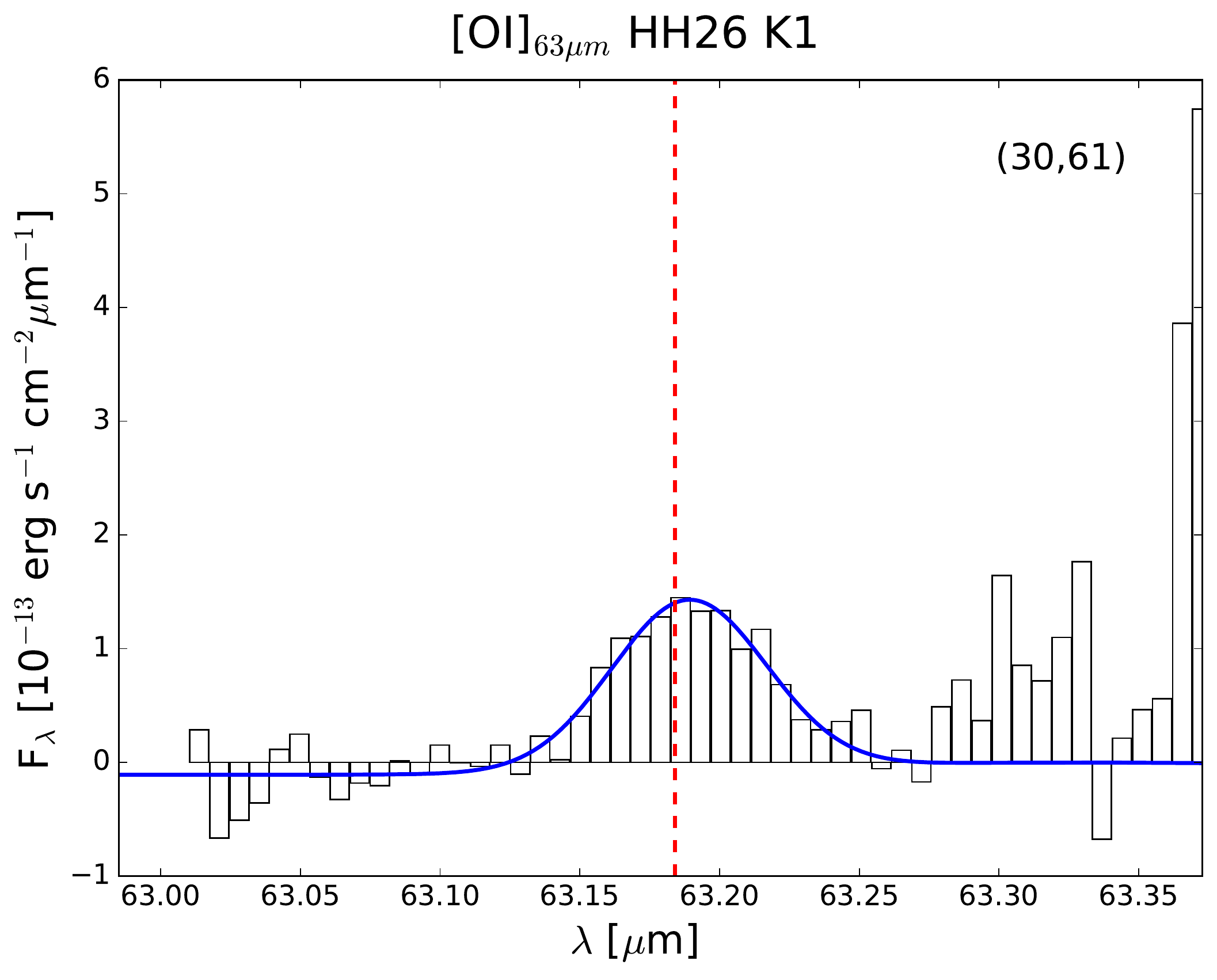}\label{fig:minispectrum_hh26A}}
\hfill
\subfloat{\includegraphics[trim=0 0 0 0, clip, width=0.33 \textwidth]{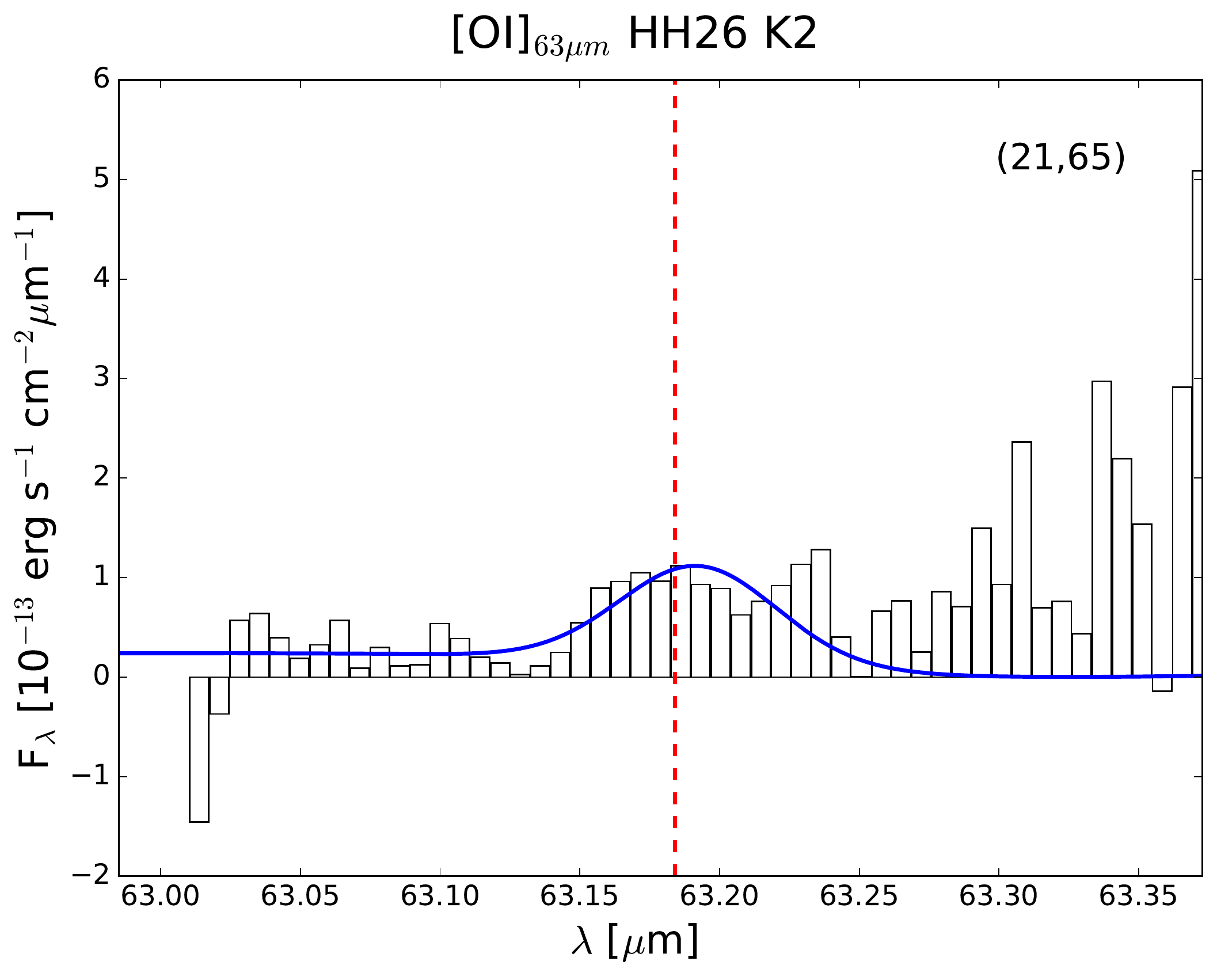}\label{fig:minispectrum_hh26B}}
\hfill
\subfloat{\includegraphics[trim=0 0 0 0, clip, width=0.33 \textwidth]{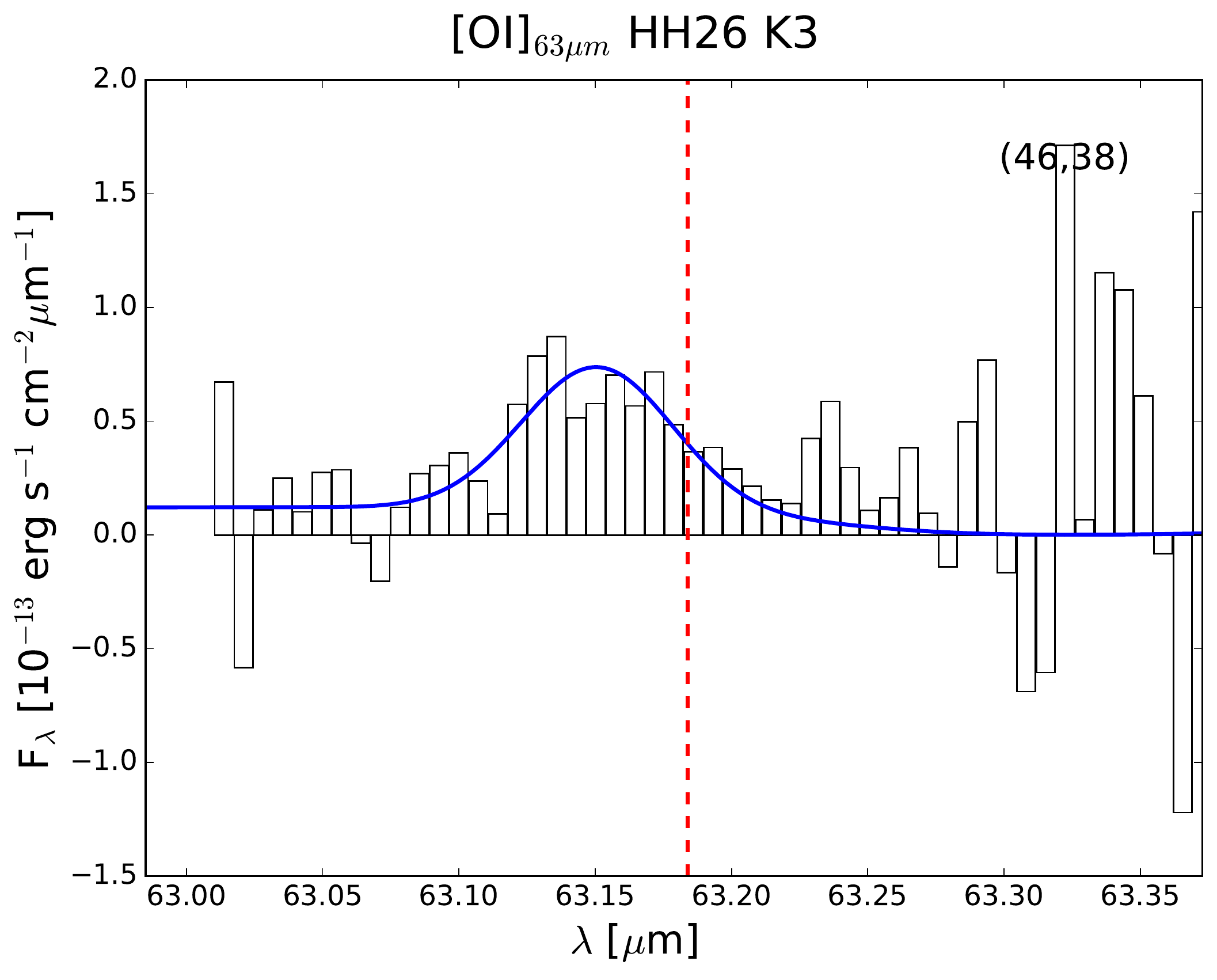}\label{fig:minispectrum_hh26C}}
\hfill
\subfloat{\includegraphics[trim=0 0 0 0, clip, width=0.33 \textwidth]{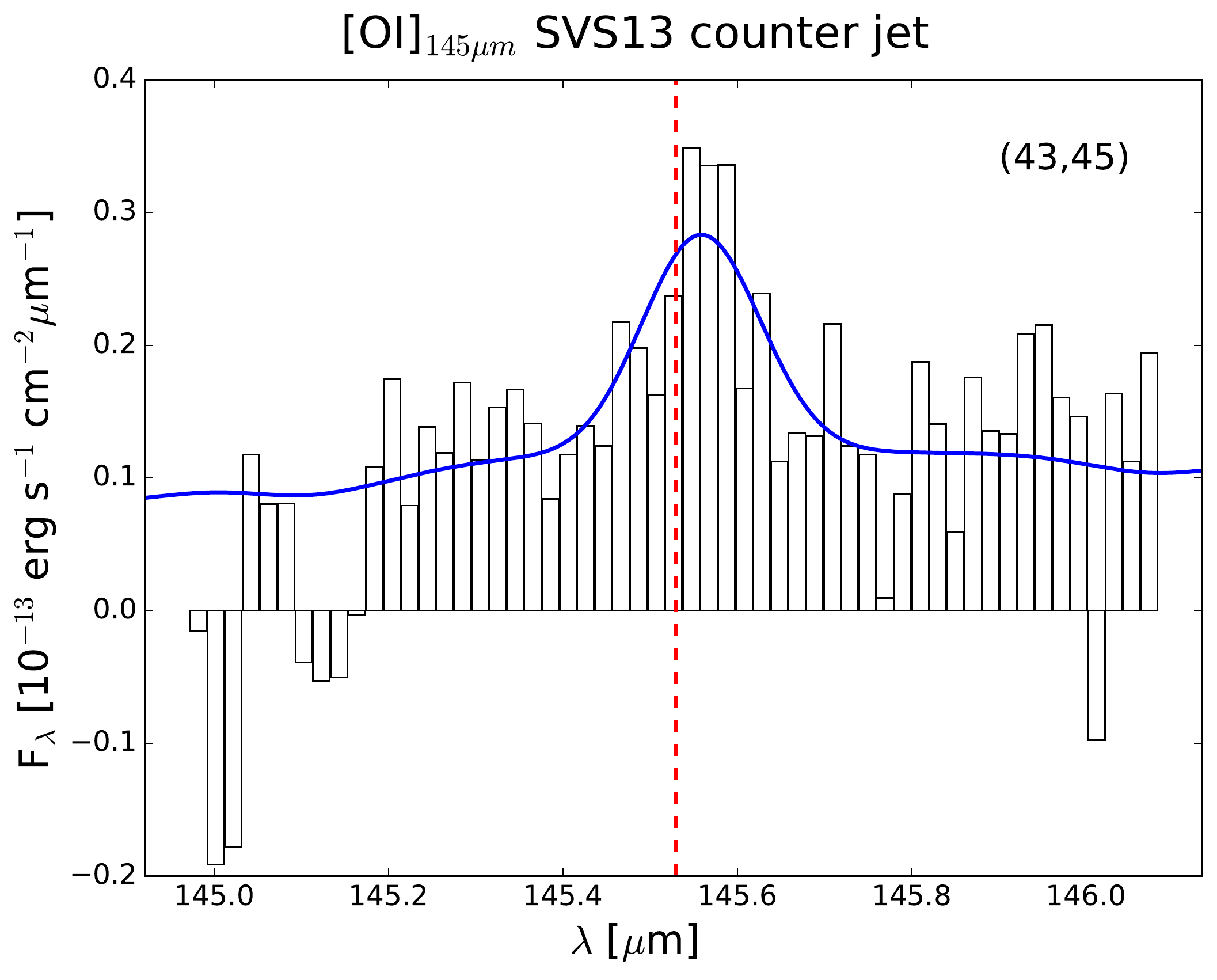}\label{fig:minispectrum_svs13_red_counter_jez}}
\hfill
\subfloat{\includegraphics[trim=0 0 0 0, clip, width=0.33 \textwidth]{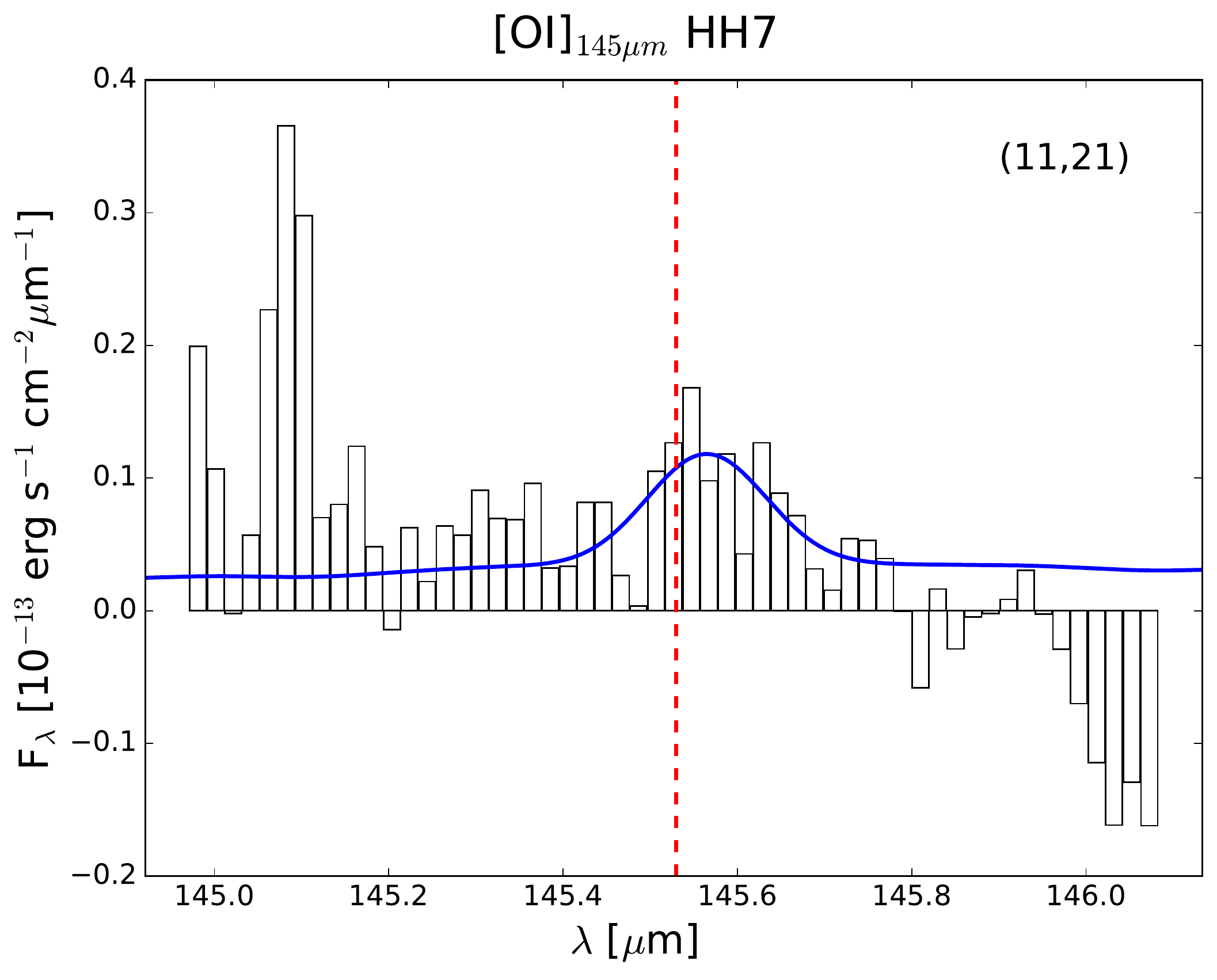}\label{fig:minispectrum_hh7_red}}
\hfill
\subfloat{\includegraphics[trim=0 0 0 0, clip, width=0.33 \textwidth]{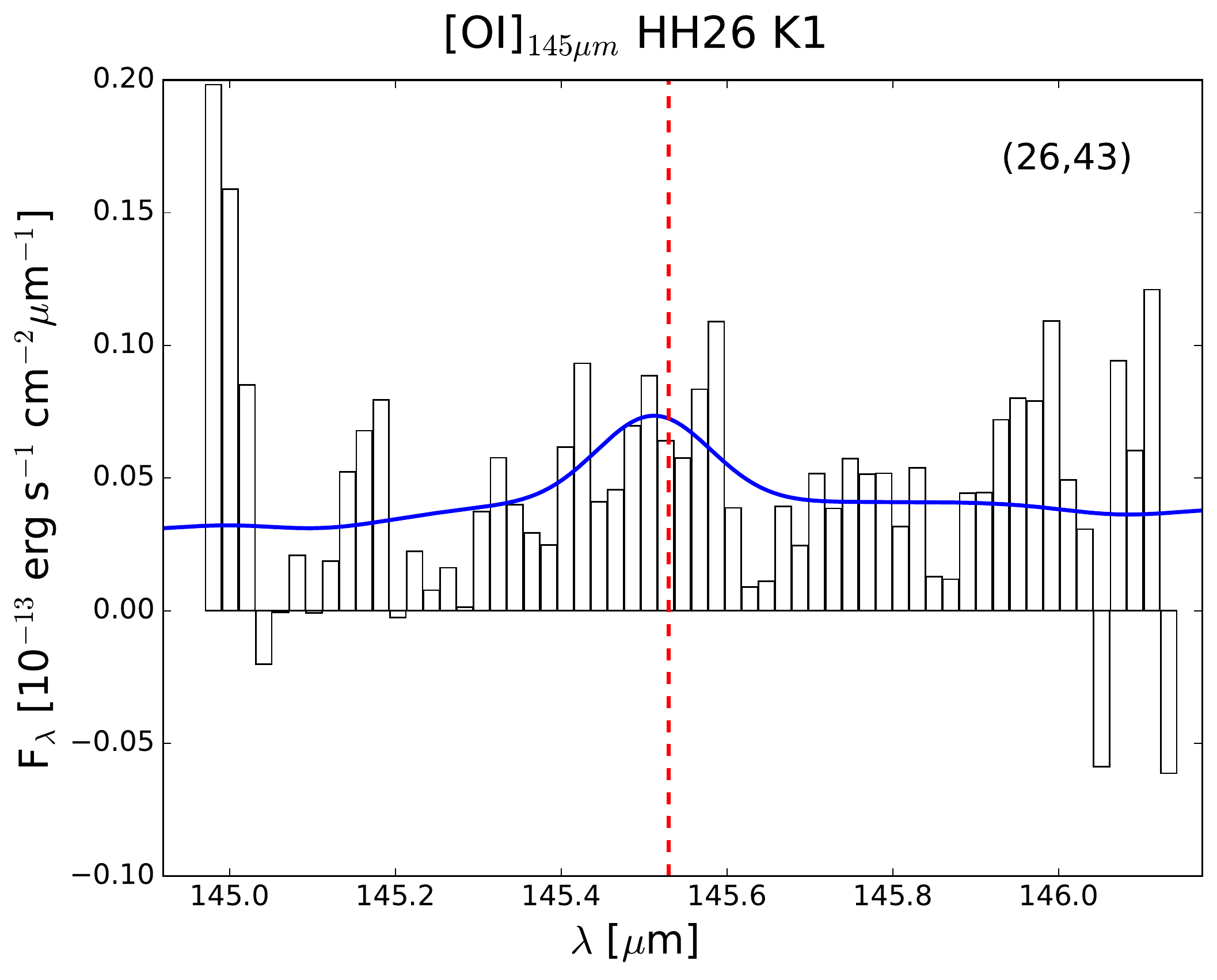}\label{fig:minispectrum_hh26_red}}
\caption{\small{Sample spectra of the detected [O\,I]$_{63,145}$ lines at different regions for of the observed targets.}}\label{fig:all_minispectra}
\end{figure*} 
 
\section{Results}

\subsection{Continuum sources at  $63\,\upmu$m and $145\,\upmu$m}\label{continuum_maps}

Here, we briefly describe the obtained continuum maps (presented in the Appendix C) of our five objects. 

For three objects in our sample (HH111 IRS, SVS13, HH34 IRS), a bright continuum source was detected in both channels. The location of these three continuum sources match within a $\sim$2\arcsec offset tolerance with the coordinates taken from 2MASS. For HH26, a bright continuum source was detected only at 145\,$\upmu$m, coinciding with HH26 IRS (Table\,\ref{table:objects}). Another bright region is located at the position of HH26A in the 145\,$\upmu$m continuum map. Towards HH26B, a possible faint continuum source is detected at $(\alpha, \delta)_{\text{J}2000} =$(5$^\text{h}$46$^\text{m}$01$\fs$9, -0$\degr$15$\arcmin$ $10\farcs 3$) at 63\,$\upmu$m. No continuum point sources were detected in both channels for HH30. However, for HH30 the continuum at 63\,$\upmu$m is very faintly elongated along the jet axis at P.A. $30^\text{o}$.

We fitted a 2D-Gaussian function,  
\begin{equation}
\Phi_\text{s}(r; A_\text{s}, \sigma_\text{s}, B_\text{s}) = \frac{A_\text{s}}{ 2\pi\sigma_\text{s}^2 }\,\text{exp}\Bigg[- \left(\frac{r^2 }{2\sigma_\text{s}^2}\right)  \Bigg] + B_\text{s}
,\end{equation}
to the detected continuum sources ($r$ as radial distance from the source  peak) to extract the continuum flux  $F_{\lambda}$, here  defined as background-corrected continuum flux within an aperture of radius 1.5$\sigma_\text{s}$  of the fitted 2D-Gaussian \citep{mighell_1999}.  
The quantified continuum fluxes of our sample are listed in Table\,\ref{table:continuum_fluxes}. These values are consistent with expected values from SED curves in the literature \citep[see e.g.][for HH34 IRS and HH26 IRS]{antoniucci_2008, benedettini_2000} or in the SIMBAD catalogue. \\
In the red channel, the measured FWHM are to the order of $20$--$22\arcsec$ for all the detected sources, whereas in the blue channel it is $\sim$\,$10\arcsec$ for HH111 and SVS13, and $\sim$\,$15\arcsec$ for HH34.   Since all measured FWHM are significantly greater than the corresponding beam sizes, we conclude that all detected continuum sources are extended.  

After subtracting the  specific 2D-Gaussian from the continuum maps, yet another potential continuum source  (also extended) in the red channel  of HH34 became apparent at $(\alpha, \delta)_{\text{J}2000} =$(5$^\text{h}$35$^\text{m}$30$\fs$3, -6$\degr$27$\arcmin$ $35\farcs 1$).

{\renewcommand{\arraystretch}{1.2}
\begin{table}
\caption{\small{Measured continuum fluxes of our objects at the observed [O\,I] transitions (flux in units:  $10^{-13}\,\text{erg}\,\text{s}^{-1}\,\text{cm}^{-2} \upmu\text{m}^{-1}$). The listed values correspond to the background-corrected flux within an aperture of radius 1.5$\sigma_\text{s}$.  }}\label{table:continuum_fluxes}
\centering
\begin{tabular}{c c c c}
\hline\hline
Obj.     &           & $F_{63\upmu\text{m}} \pm\Delta F_{63\upmu\text{m}}$   & $F_{145\upmu\text{m}} \pm\Delta F_{145\upmu\text{m}} $   \\[1.5pt]
\hline  
\small{HH111IRS} &           & $ 255.2 \pm   6.9$      & $ 61.9\pm  1.4 $       \\ [1.5pt]
\small{SVS13}    &           & $ 1335.6 \pm 14.5 $      & $ 245.1 \pm 1.9 $      \\[1.5pt]
\small{HH34IRS}  &           & $ 312.0 \pm 19.5 $      & $ 36.6 \pm 2.0 $        \\[1.5pt]
\small{HH26IRS}  &           & $  - $\tablefootmark{a}  & $ 13.7 \pm 0.9 $        \\[1.5pt]
\small{HH30}     &           & $  - $\tablefootmark{a}  & $  -  $\tablefootmark{a}  \\[1.5pt]
\hline\hline
\end{tabular}
 \tablefoot{
\tablefoottext{a}{\small{No source was detected in the continuum maps.}}
}
\end{table}

\begin{figure*}   
\centering
\subfloat{\includegraphics[width=0.99\textwidth]{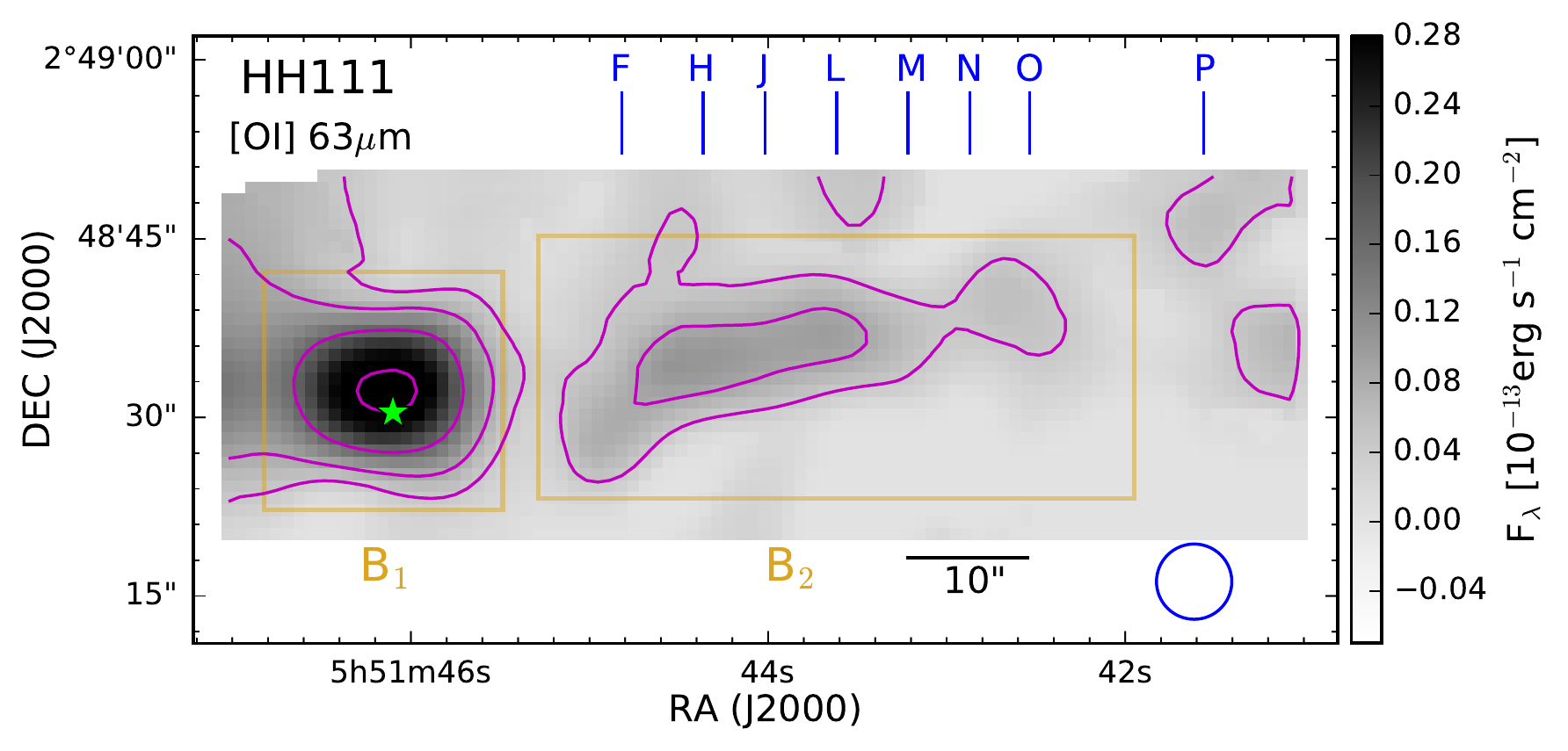}}\label{fig:e_map_hh111_blue}
\hfill
\subfloat{\includegraphics[width=0.99\textwidth]{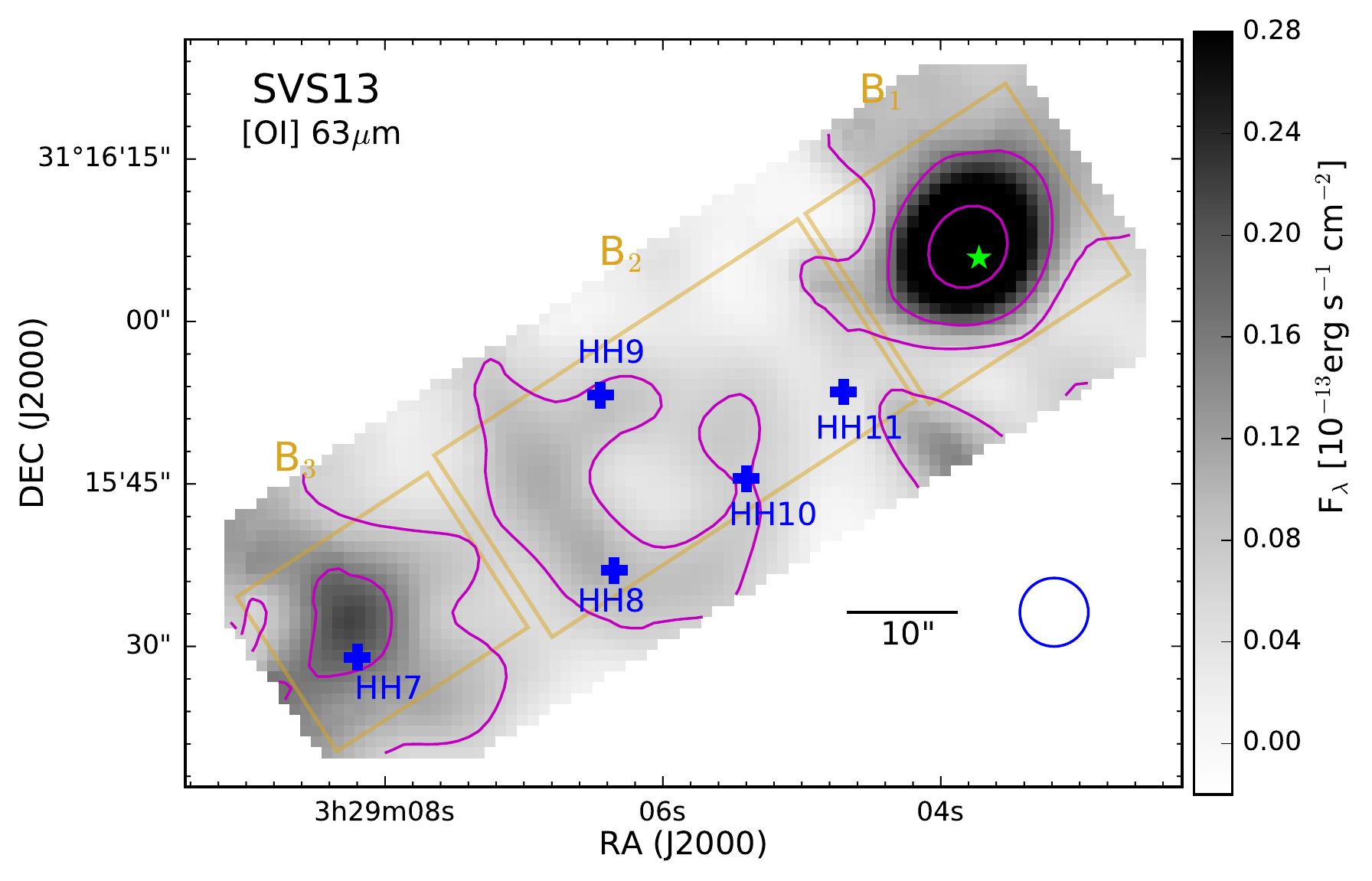}}\label{fig:e_map_svs13_blue}
\caption{\small{Continuum-subtracted [O\,I]$_{63}$ maps of HH111 and SVS13. Light green stars indicate the position of the jet driving sources HH111IRS and SVS13 (coordinates from Table\,\ref{table:objects}).   The  blue circle shows the FWHM spatial beam size in the blue channel of the FIFI-LS instrument. The golden boxes are the boundaries of the rectangular apertures, in which the fluxes $F_{63\upmu\text{m}}$ and $F_{145\upmu\text{m}}$ are measured (Table\,\ref{table:main_results}). \textit{Top panel:} blue lines at the top label the right ascension of the knots F--P associated with the optical jet \citep{reiputh_1989}. Contour lines are drawn in magenta in logarithmic scale at four levels between (0.0420--0.3200)$\times 10^{-13}\,\text{erg}\,\text{s}^{-1}\,\text{cm}^{-2}$.  \textit{Bottom panel:} blue crosses label the positions of HH 7-11 associated with the jet \citep[coordinates taken from][]{bally_1996}.  Contour lines are drawn in magenta in logarithmic scale at three levels between (0.068--0.400)$\times 10^{-13}\,\text{erg}\,\text{s}^{-1}\,\text{cm}^{-2}$.}}\label{fig:hh111_svs13}
\end{figure*} 

\begin{figure*}[htb!] 
\centering
\subfloat{\includegraphics[width=0.65\textwidth]{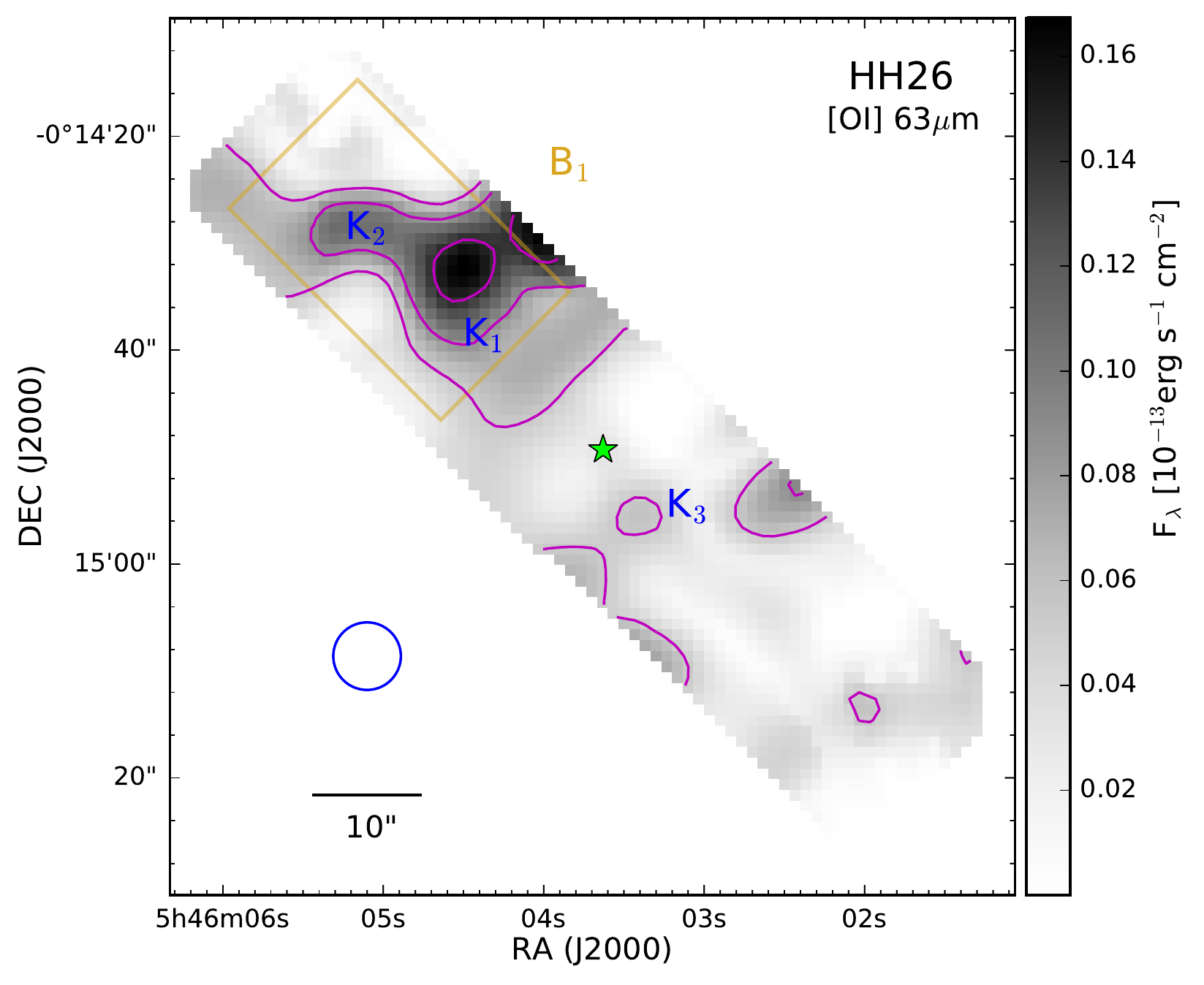}}\label{fig:e_map_hh26_blue}
\hfill
\subfloat{\includegraphics[width=0.60\textwidth]{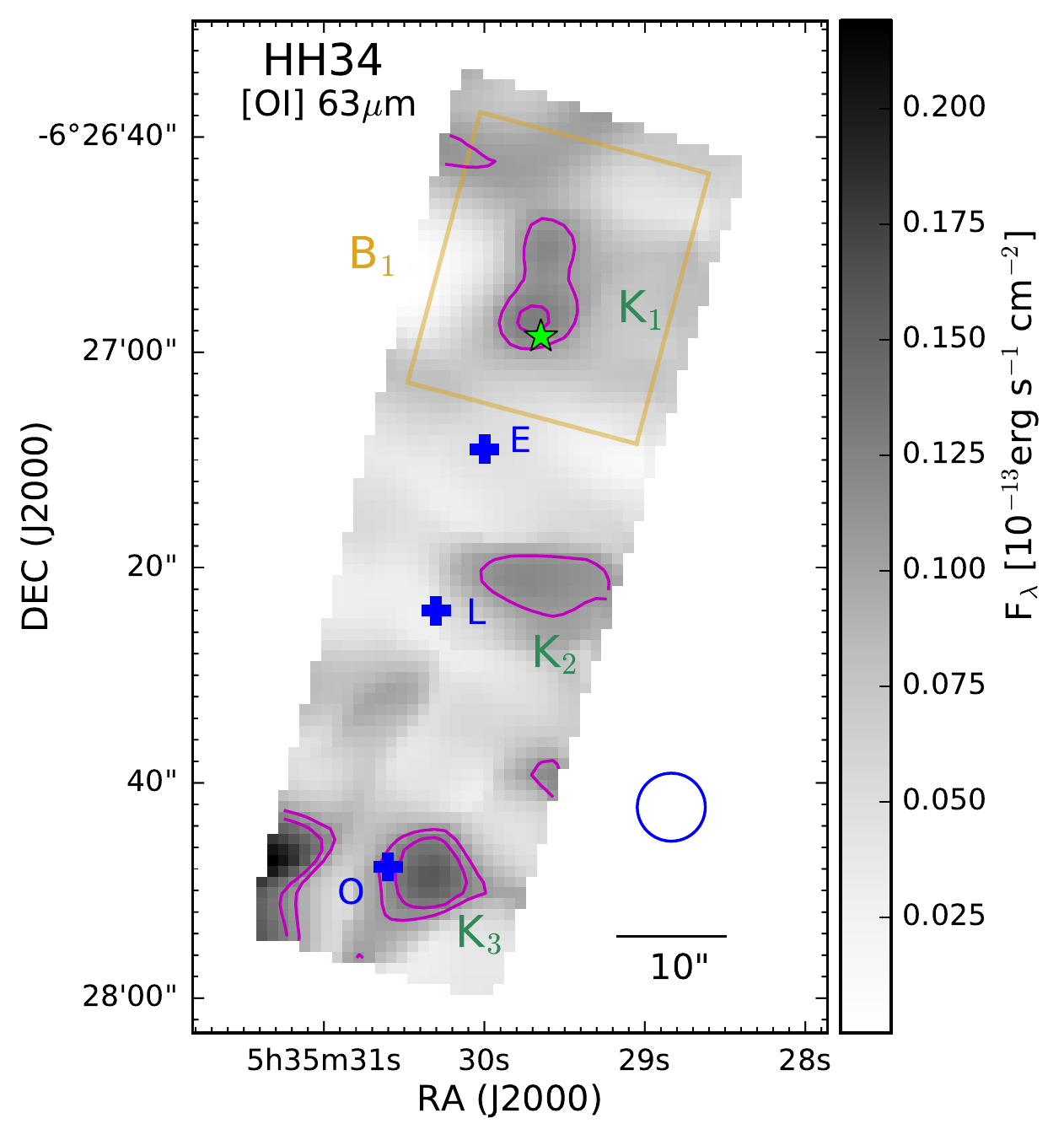}}\label{fig:e_map_hh34_blue}
\hfill
\caption{\small{Continuum-subtracted [O\,I]$_{63}$ maps of HH26 and HH34. The light green stars indicate  the position of the jet driving sources HH26IRS, HH34IRS, and HH30IRS (coordinates from Table\,\ref{table:objects}). The blue circle shows the FWHM spatial beam size in the blue channel of the FIFI-LS instrument. The golden boxes are the boundaries of the rectangular apertures, in which the fluxes $F_{63\upmu\text{m}}$ and $F_{145\upmu\text{m}}$ are measured (Table\,\ref{table:main_results}). \textit{Top panel:} contour lines are drawn in magenta in logarithmic scale at three levels between (0.052--0.142)$\times 10^{-13}\,\text{erg}\,\text{s}^{-1}\,\text{cm}^{-2}$. \textit{Bottom  panel:} blue crosses label the position of the two knots E and L (between them lies the optical jet)  and knot O \citep{heathcote_1992}. Contour lines  are drawn in magenta at two levels (0.105--0.125)$\times 10^{-13}\,\text{erg}\,\text{s}^{-1}\,\text{cm}^{-2}$.}}\label{fig:hh26_hh34_hh30}
\end{figure*} 

\begin{figure}[htb!] 
\resizebox{\hsize}{!}{\includegraphics[trim=0 0 0 0, clip, width=1.0\textwidth]{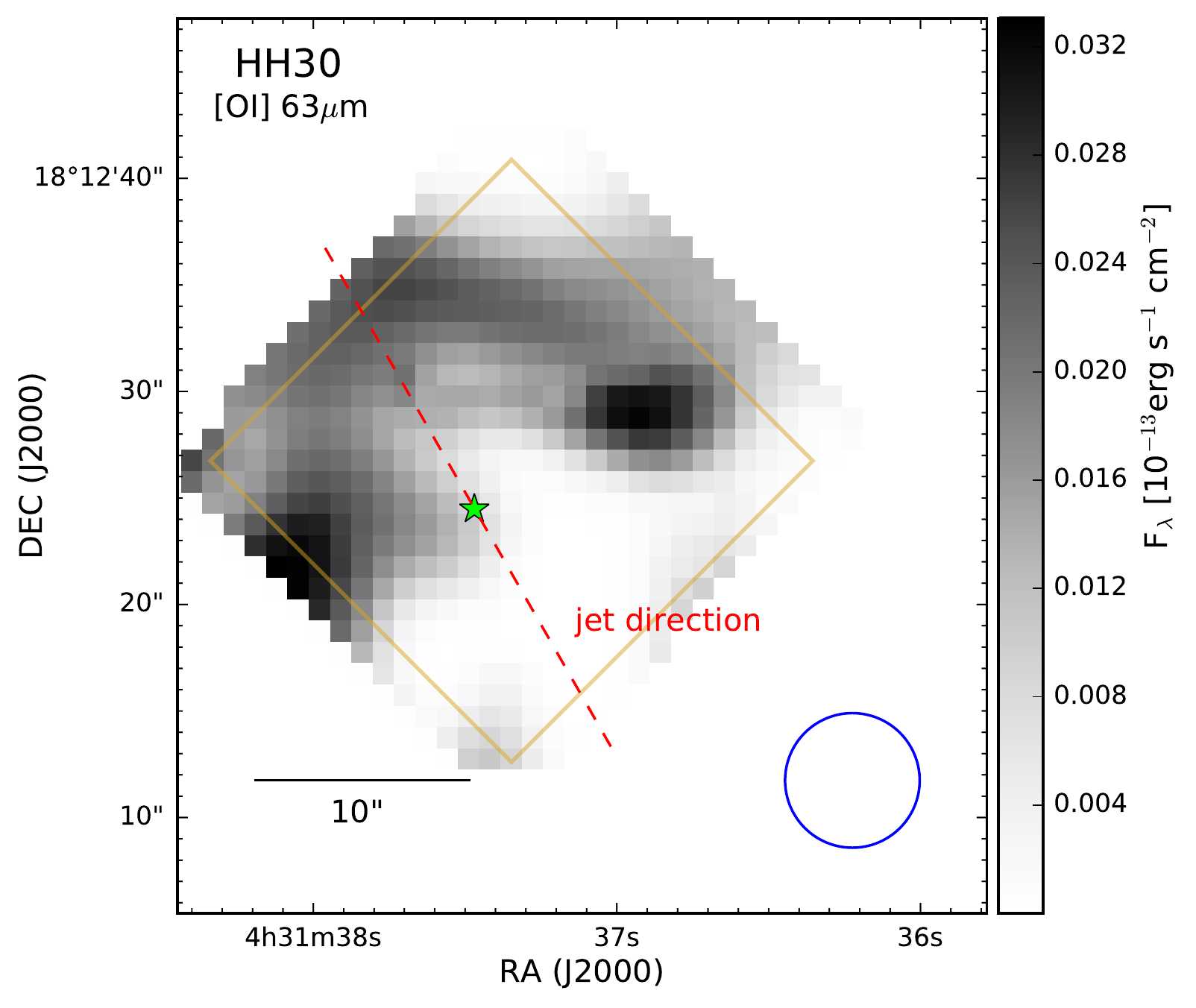}}
\caption{Continuum-subtracted [O\,I]$_{63}$ map of HH30. The dashed red line indicates the jet direction at P.A. 30$^\text{o}$ \citep{pety_2006}. The [O\,I]$_{63}$ line was not detected in any spaxel, meaning the features in this map are noise artefacts. }\label{fig:e_map_hh30_blue} 
\end{figure}

{\renewcommand{\arraystretch}{1.2}
\begin{table*} 
\caption{\small{Relevant information on the aperture boxes, which  are sketched in the continuum subtracted [O\,I]$_{63}$ emission maps (Fig.\,\ref{fig:hh111_svs13}, \ref{fig:hh26_hh34_hh30}). Fluxes measured within these boxes are listed in units of $10^{-13}\,\text{erg}\,\text{s}^{-1}\,\text{cm}^{-2}$.}}\label{table:main_results}
\centering
\begin{tabular}{c c c c c c c }
\hline\hline
&  &  &  & [O\,I]$_{63\,\upmu\text{m}}$ &     &    [O\,I]$_{145\,\upmu\text{m}}$    \\ \cmidrule{4-7}
Obj. & Region & Box & Size & $F_{63\upmu\text{m}}$  & $L\left(\text{[O\,I]}_{63}\right)/L_\odot$\tablefootmark{b}   &   $F_{145\upmu\text{m}}$    \\[1.5pt]
\hline  
\small{HH111} & \tiny{HH111IRS} & $B_1$ & \tiny{$20\arcsec\times 20\arcsec$} & $44.76\pm 3.17$ & $ 2.47\pm 0.18\times 10^{-2}$ & $<10.1$\tablefootmark{a}    \\ [1.5pt]
              & \tiny{jet (knots F-O)} & $B_2$ & \tiny{$50\arcsec\times 22\arcsec$} & $23.49\pm 5.40$ & $ 1.29\pm 0.30\times 10^{-2}$ & $<8.4$\tablefootmark{a}    \\[3pt]
\hline
\small{SVS13} & \tiny{SVS13}& $B_1$ & \tiny{$22\arcsec\times 21\arcsec$} & $84.38\pm 8.47$ & $ 1.46\pm 0.15\times 10^{-2}$ & $<3.7$\tablefootmark{a}    \\[1.5pt]
              & \tiny{HH8-11}& $B_2$ & \tiny{$40\arcsec\times 20\arcsec$} & $41.08\pm 5.47$ & $ 7.09\pm 0.94\times 10^{-3}$ & $<5.1$   \tablefootmark{a} \\[1.5pt]
              & \tiny{HH7}& $B_3$ & \tiny{$21\arcsec\times 17\arcsec$} & $39.69\pm 2.09$ & $ 6.85\pm 0.36\times 10^{-3}$ & $<3.4$\tablefootmark{a}    \\[1.5pt]
\hline
\small{HH26}  &   \tiny{HH26A}& $B_1$ & \tiny{$28\arcsec\times 17\arcsec$} & $29.18\pm 4.44$ & $ 1.61\pm 0.24\times 10^{-2}$ & $<3.7$\tablefootmark{a}   \\[1.5pt]   
\hline
\small{HH34}  &\tiny{counter jet, HH34IRS} & $B_1$ & \tiny{$26\arcsec\times 22\arcsec$} & $41.66\pm 5.87$ & $ 2.41\pm 0.34\times 10^{-2}$ & $<3.0$\tablefootmark{a}    \\[1.5pt]
\hline
\small{HH30}  & \tiny{HH30IRS} & $B_1$ & \tiny{$20\arcsec\times 20\arcsec$} & $<7.2$\tablefootmark{a}  & $<4.5 \times 10^{-4}$\tablefootmark{a} & $<3.1$\tablefootmark{a}  \\[1.5pt]
\hline\hline
\end{tabular}
\tablefoot{
\tablefoottext{a}{\small{The listed value corresponds to the $3\sigma$ upper limit.}}
\tablefoottext{b}{\small{calculated via $L([\text{OI]}_{63})=4\pi D^2 F_{63\upmu\text{m}}$ (distances see Table\,\ref{table:objects}).}}
}
\end{table*}

\subsection{[O\,I]$_{63}$ morphology }\label{line_ratio}

The smoothed, continuum-subtracted [O\,I]$_{63}$ maps for each target of our sample are shown in Figs.\,\ref{fig:hh111_svs13}, \ref{fig:hh26_hh34_hh30}, and \ref{fig:e_map_hh30_blue}. Green stars  in the maps indicate the position of the individual driving source (see Table\,\ref{table:objects}). Yellow boxes enclose the carefully selected regions along the expected protostellar outflow where flux measurements were taken (Tab.\,\ref{table:main_results}).

In the following paragraphs, we briefly describe the morphology of the [O\,I]$_{63,145}$ maps for each target individually.\\

\textbf{HH111:} Prominent [O\,I]$_{63}$ emission is detected on the continuum source HH111 IRS itself (box $B_1$). The  [O\,I]$_{63}$ emission at HH111 IRS is highly concentrated and slightly extended alongside the outflow  axis. The far-infrared counterpart of the optical jet is  firmly apparent only in the [O\,I]$_{63}$ map (inside $B_2$) and reveals a potential clumpy structure within its bright emitting region of $\sim$\,$ 45\arcsec$ projected length. At 420\,pc, this corresponds to $\sim$ 0.09\,pc or $\sim$ 18900\,AU.  For better orientation, the positions in right ascension of a few known optical knots are marked as blue lines \citep[notation from][]{reiputh_1989}. The bulk of the detected [O\,I]$_{63}$ emission of the jet body comes from the innermost part of the jet, meaning it is concentrated on the outflow axis and emerges between knots F to L. Between knots M and N, the [O\,I]$_{63}$ emission declines significantly and appears as a narrow emitting region that is connected to the clump-like structure between knots N and O. Some [O\,I]$_{63}$ emission is detected further downstream at the edge of the mapped outflow region and is located almost at knot P.  A gap of very little [O\,I]$_{63}$ emission between the source emission and the jet periphery can be recognised (between $B_1$ and $B_2$).  \\

\textbf{SVS13:} Predominant [O\,I]$_{63}$ and almost no [O\,I]$_{145}$ emission is detected at the driving source of SVS13 itself (box $B_1$). The source  [O\,I]$_{63}$ emission is extended and slightly aligned towards the jet. Blue crosses in the [O\,I]$_{63}$ map indicate the positions of HH7-11 \citep[coordinates taken from][]{bally_1996}. We detect strong emission centred in front of HH7 in both [O\,I]$_{63,145}$ maps (box $B_3$), whereas HH8-10 are faintly detected only in the [O\,I]$_{63}$ map (box $B_2$)  enclosing a cavity of no  [O\,I] emission. HH11 is not detected in [O\,I].  
 Significant [O\,I]$_{145}$ emission is detected about 20$\arcsec$ north-west of the driving source. This possible counter jet at P.A.$\sim$ $155^\text{o}$ lies outside the region mapped in [O\,I]$_{63}$. \citet{lefevre_2017} detected a wiggling H$_2$/CO jet at this P.A.\\

 \textbf{HH34:}  Three potentially interesting emission regions (here labelled $K_1$, $K_2$, $K_3$) are seen in the obtained [O\,I]$_{63}$ map. Serving as orientation, we mark the positions of knots E and L in between which the well-known optical jet prominently seen in [SII] towards HH34S is located (nomenclature adopted from \citet{heathcote_1992} with blue crosses). The position of the jet driving source HH34 IRS coincides with the brightest emission within $K_1$. The detected [O\,I]$_{63}$ line at HH34 IRS is slightly blue shifted (Fig.\,\ref{fig:all_minispectra}), whereas towards HH34N a red-shifted component within $K_1$ is apparent. Looking at the obtained spectra at the location of $K_2$ and $K_3$ (Fig.\,\ref{fig:all_minispectra}), we notice that the line fit suggests a red-shifted outflow towards HH34S. Physically, this is puzzling, since the jet towards HH34S is blue shifted. The morphology of this [O\,I]$_{63}$ emission at $K_2$ and $K_3$ would be difficult to explain, if this emission is connected to the outflow. The most obvious explanation then could be that this emission is part of a backflow along the cocoon of material surrounding the jet \citep{norman_1990, cabrit_1995}. Alternatively,  noise at 63\,$\upmu$m towards longer wavelengths mimics an emission line so that the line fit procedure falsely identifies this noise feature as a red-shifted [O\,I]$_{63}$ line.

\textbf{HH26:} We find almost no [O\,I]$_{63,145}$ emission on the driving source HH26 IRS itself (box $B_1$). However, three knots  (here labelled $K_1, K_2, K_3$) of significant emission are arranged along the outflow axis in the [O\,I]$_{63}$ map.  At 145\,$\upmu$m $K_1$ and $K_2$ seem to be one emitting region. The location of $K_1$ and $K_2$ coincides with knot C of the HH26A/B/C chain \citep[see nomenclature in][]{chrysostomou_2000}. Extended [O\,I]$_{63}$ emission at HH26A (here $K_1$ and $K_2$).   The very faint blue-shifted [O\,I]$_{63}$ emission at $K_3$ appears to be rather non-physical, since it lies in the red-shifted outflow lobe \citep{davis_1997, dunham_2014}. As in the case of HH34, we therefore interpret this emission seen at $K_3$ as a noise feature. \\ }

\textbf{HH30:} Since HH30 was the least luminous target within our sample, our obtained [O\,I]$_{63,145}$ maps are very noisy with low signal-to-noise ratios (SNR $\sim$ 0.5). Thus, structures appearing in the [O\,I]$_{63}$ map are not trustworthy.

Due to the  medium quality of our obtained [O\,I]$_{63}$ maps of HH111, SVS13, HH34, and HH26, we chose to put several aperture boxes of interest in their corresponding maps for flux measurements.  The position and size of the aperture boxes were chosen arbitrarily with three constraints. First, the  isolated region has to be larger than the blue channel spatial beam size to ensure negligible flux losses due to diffraction, secondly they encompass the jet region with relatively high signal-to-noise ratios, and thirdly a box  encloses a physically meaningful region, for example, the jet, the driving source, or a region with bright line emission. 

We determined the flux within each aperture box by averaging all enclosed spaxels to a representative spaxel on which the model function (Sect.\,\ref{data_reduction}) is fitted. The parameter $A$ (and its error) from this fit is then scaled with the box dimensions to get the reported flux values in Table\,\ref{table:main_results}.  

\subsection{Atomic mass-flux rates}\label{sec:mass_flux_rates}

 Mass-flux rates $\dot{M}_\text{jet}$ can be derived from direct jet observations via different methods \citep[see e.g.][]{cabrit_2002, dougados_2010, dionatos_2020}. It is tempting to derive mass-loss rates of our targets from the obtained [O\,I]$_{63}$ maps as it has become common practice in the field of star formation. Basically, two different approaches are worth considering here (Sects. \ref{sec:mass_flux_rates_HM89} and \ref{sec:mass_flux_rates_geometry}). Both methods have their specific limitations and caveats, which are discussed in  Sect.\,\ref{sec:caveats_HM89}.  
\subsubsection{$\dot{M}_\text{jet}$ from a shock model}\label{sec:mass_flux_rates_HM89}

 Assuming that the observed [O\,I]$_{63}$ line luminosity $L(\text{[O\,I]}_{63})$ is connected with a dissociative J-shock cooling region coming from one decelerated wind shock, we could utilise the results of the \citet{hollenbach_1985} and \citet{hollenbach_1989} papers, namely that the mass-loss rate is predicted to be proportional to the [O\,I]$_{63}$ luminosity. In this particular case, the [O\,I]$_{63}$ emission is (amongst  other cooling lines, e.g. [C\,II]$_{158}$, [Fe\,II]$_{35}$, [Fe\,II]$_{26}$, [Si\,II]$_{35}$   ) the dominant coolant in the post-shock gas, meaning that each particle entering the shock radiates about $E=3/2k_\text{B}T$ in the [O\,I]$_{63}$ line for temperatures in the range of $T\sim$\,$100-5000\,\text{K}$. Specifically, from their shock model \citet{hollenbach_1989}, hereafter HM89, predict the relation    
\begin{equation}\label{equ:shock_model}
\dot{M}_\text{jet}^\text{shock} = 10^{-4}\left(L\left([\text{OI]}_{63} \right)/L_\odot  \right) M_\odot \,\,\text{yr}^{-1}
\end{equation} 
to be valid over a wide range of shock parameters, provided that  $n_0\times v_\text{shock} \lesssim 10^{12}\,\text{cm}^{-2}\,\text{s}^{-1}$ ($n_0$ pre-shock density, $v_\text{shock}$ shock velocity). This method of measuring the mass-loss rate could potentially be  quite  powerful, because only one fairly easy-to-measure quantity enters Eq.\,\ref{equ:shock_model}. If the HM89 model is not applicable, the mass-loss rates  calculated  via Eq.\,\ref{equ:shock_model} are unusable and errors cannot be quantified. It would be interesting to test the validity of Eq.\,\ref{equ:shock_model} once again, since the \citet{hollenbach_1989} paper improved collisional strengths and element abundances are available, and new chemical networks could be included.

\subsubsection{$\dot{M}_\text{jet}$ from jet geometry and the [O\,I]$_{63}$ line luminosity}\label{sec:mass_flux_rates_geometry}
 
 Without any assumptions about the origin of the observed [O\,I]$_{63}$ emission, we could follow a similar analysis to the one performed by \citet{hartigan_1995} estimating the mass-loss rate from the [O\,I]$_{63}$ line luminosity and other jet parameters such as the flow velocity or the physical length of the jet. The forbidden [O\,I]$_{63}$ line tracing the warm jet component ($T\sim 300-5000$\,K) is mainly excited via atomic hydrogen collisions.
Since we cannot infer the gas density from the $j_{63}/j_{145}$ line ratio, we could follow \citet{nisini_2015} in the assumption that the collider density is close to the critical density. Detailed calculations found in Appendix B lead to the following estimate on the mass-loss rate:
 \begin{equation}\label{equ:line_luminosity}
  \left( \frac{\dot{M}_\text{jet}^\text{lum} }{M_\odot  \text{yr}^{-1} } \right) =  \left(3.3-6.7 \right)  \times 10^{-3}    \cdot   \left( \frac{v_\text{t} }{\text{km/s}} \right) \left( \frac{ '' }{ \theta  }\right)   \left( \frac{\text{pc}}{ D} \right)\left( \frac{L(\text{[O\,I]}_{63})}{L_\odot}\right).
  \end{equation}
Here, $v_\text{t}$ is the component of the jet velocity on the plane of sky and $\theta$ is the angular size of the jet.\\   We point out that Eq.\,\ref{equ:line_luminosity} comes with some uncertainties such as the unknown level population, often poorly constrained jet velocities and propagating errors that come from the distance measurements. These uncertainties may add up to a total uncertainty of one order of magnitude.  

\section{Discussion}\label{sec:discussion}

 In the context of protostellar outflows, the HM89 formula (Eq.\,\ref{equ:shock_model}) has been commonly used to derive mass-loss rates from the [O\,I]$_{63}$ line luminosity. This method is strikingly simple and seems to give reasonable results \citep[e.g.][]{nisini_2015, watson_2016, mottram_2017, dionatos_2018}. However, the derived mass-loss rates may be physically meaningless, if the underlying assumptions of the HM89 model are demonstrably violated in the region of interest \citep{hartigan_2019} or other observed quantities are inconsistent with the HM89 model predictions \citep[e.g.][]{green_2013}. In order to prevent a blind exploitation of this  model we briefly discuss the detected [O\,I] emission (Sect.\,\ref{sec:[OI]_schematics}) and compile some general limitations of the HM89 method (Sect.\,\ref{sec:caveats_HM89}). We point out that the discussion about the uncertainty of the observed [OI] emission concerns both methods presented  here determining mass-loss rates.

\subsection{Schematic views on the observed [O\,I] emission}\label{sec:[OI]_schematics}

\textbf{HH111:}  We interpret the extended on-source emission as shock-excited gas in the interaction region of a quasi-spherical wind/outflow coming from HH111 IRS (and the disc) with the ambient cloud (Fig.\,\ref{fig:hh111_svs13_figure}).  Presumably, several spatially unresolved, internal shocks within the jet body are the driving agents of this bright [O\,I]$_{63}$ emission. This argumentation is supported by optical and near infrared observations \citep[e.g.][]{reipurth_1997, davis_2001}, which reveal the presence of multiple knots with bow shock morphologies within the jet body. Adopting the knot notation introduced by \citet{reiputh_1989}, we deduce that the knots F$-$O appear as one emitting region in our obtained [O\,I]$_{63}$ maps due to their low spatial resolution. Several different physical mechanisms have been proposed explaining the origin of the knots \citep[see e.g.][]{raga_1990, micono_1998, reipurth_2001}. The gap of very little [O\,I]$_{63}$ emission between the source emission and the jet periphery can be attributed to very high obscuration at the jet base \citep{reipurth_1997} or due to a quiet interaction region, where the highly collimated jet expands almost freely into interstellar space. This has been proposed for the HH34 jet \citep{buehrke_1988}, which shows similar morphology at optical forbidden lines such as [SII] or H$\alpha$.
Interestingly, the observed [O\,I]$_{63}$ emission along the jet body (Fig.\,\ref{fig:cut_hh111}) follows the [Fe\,II]$1.64\upmu$m route rather than the H$_2$ ($2.12\upmu$m); see Fig.\,12 in \citet{nisini_2002} or Fig.\,2 in \citet{podio_2006} for comparison. At knot F,  [Fe\,II] and [O\,I] peak and show a gradually decreasing trend towards knot L. In H$_2$ however, three separate, almost triplet-like peaks are located roughly at knots F, H, and L. This suggests that the [Fe\,II] and [O\,I] emission is potentially connected to the same shock regions within the jet body. Some of these internal shocks may even be strong enough to dissociate H$_2$ molecules (from the flowing jet material or the ambient shocked medium) leading to the difference in morphology seen in [Fe\,II] and H$_2$.  \citet{nisini_2002} noted that H$_2$ emission might be connected to C-shocks in bow wings. However, the origin of the H$_2$ emission is still under debate. See, for example, the discussion in \citet{bally_2007}. \\
 In [Fe\,II], the jet body is brighter than the source, whereas in [O\,I] the opposite is the case. This is in line with the above-mentioned interpretation of different shock conditions present in these two regions. At the source, a less powerful quasi-spherical wind shocks the dense, cold ambient medium, and the \citet{hollenbach_1989} shock conditions prevail in good approximation. By contrast, various internal shocks (potentially J-type or C-type, dissociative or non-dissociative) in the highly collimated jet result in a very complex emission region with radiative cooling zones, internal working surfaces and bow shocks. As a reflection, the HH111 jet is detected in various emission lines from ions, atoms and  molecules indicating the presence of several gas components at different temperatures and densities. It remains puzzling that we see bright [O\,I] emission about $\sim54\arcsec$ away from the source at knot O, where only very faint [Fe\,II] and H$_2$ is detected.

\begin{figure} 
\resizebox{\hsize}{!}{\includegraphics[trim=0 0 0 0, clip, width=0.9\textwidth]{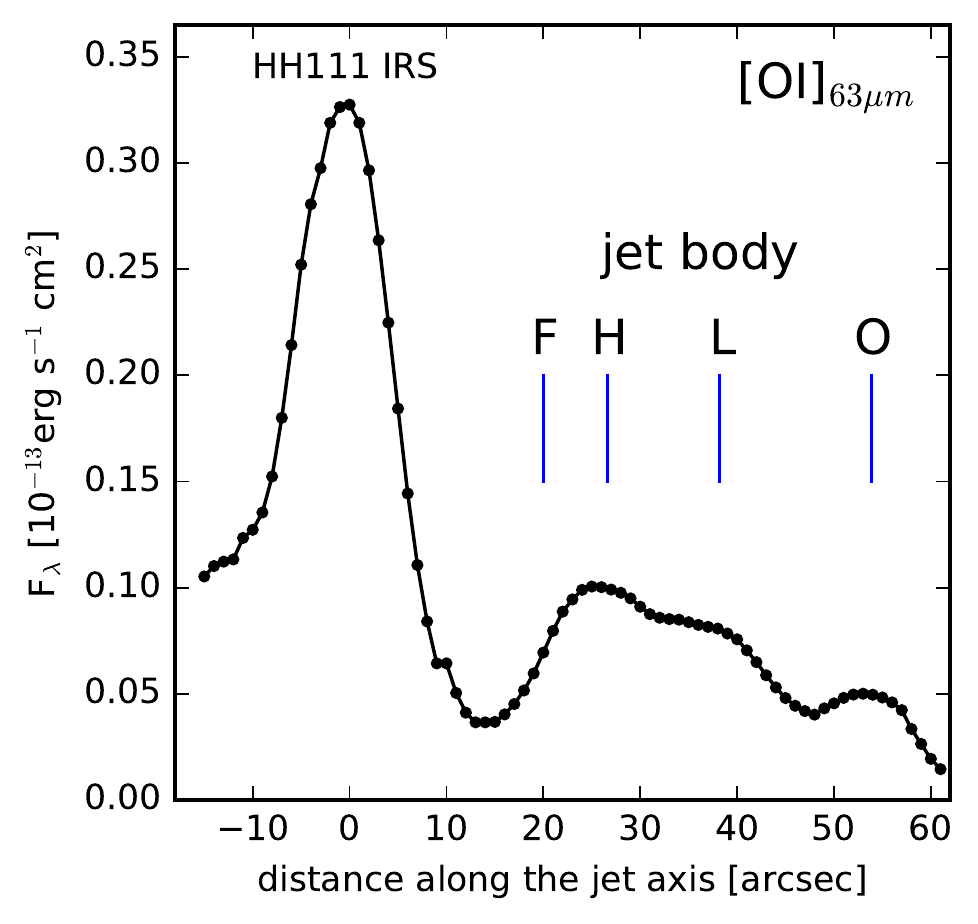}}
\caption{\small{Spatial distribution of observed [O\,I]$_{63}$ emission along the HH111 outflow axis with HH111 IRS as zero position. For comparison with NIR observations at [Fe\,II]$1.64\upmu$m and H$_2$ ($2.12\upmu$m) see  \citet{nisini_2002} and \citet{podio_2006}. }}\label{fig:cut_hh111} 
\end{figure}

\begin{figure*}   
\centering
\subfloat{\includegraphics[trim=60 25 32 70, clip,width=0.95\textwidth]{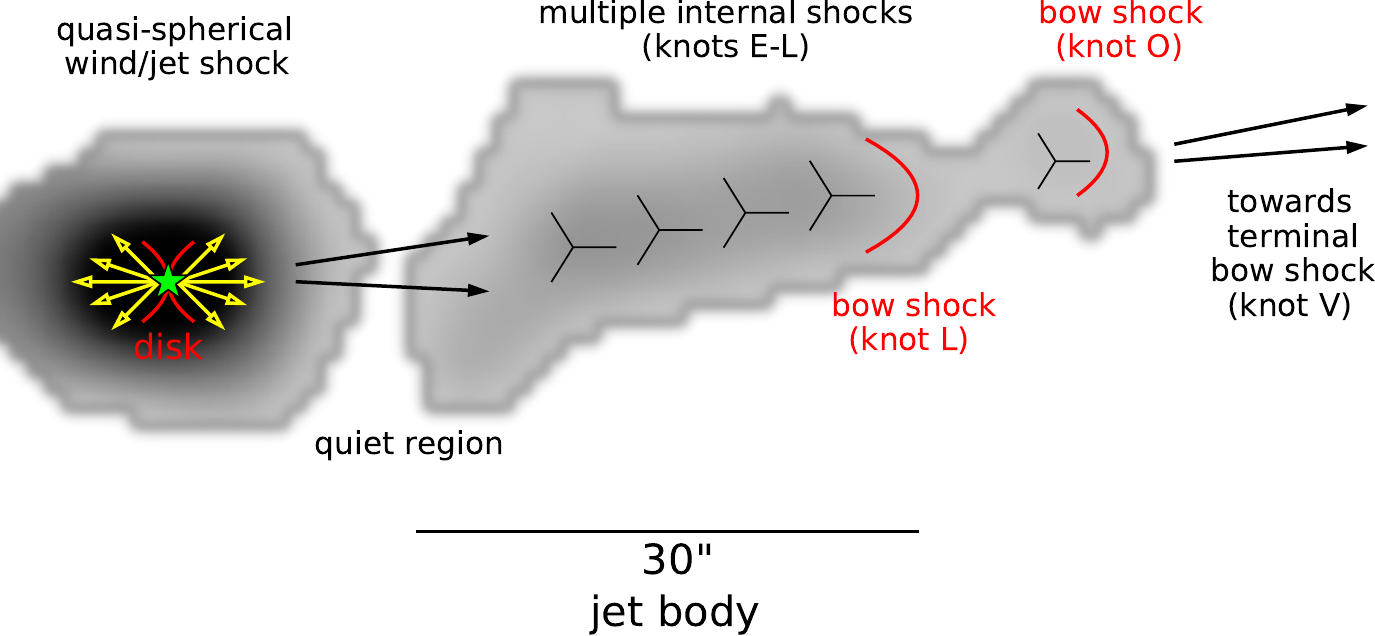}} 
\hfill
\subfloat{\includegraphics[trim=0 0 0 0, clip, width=0.95\textwidth]{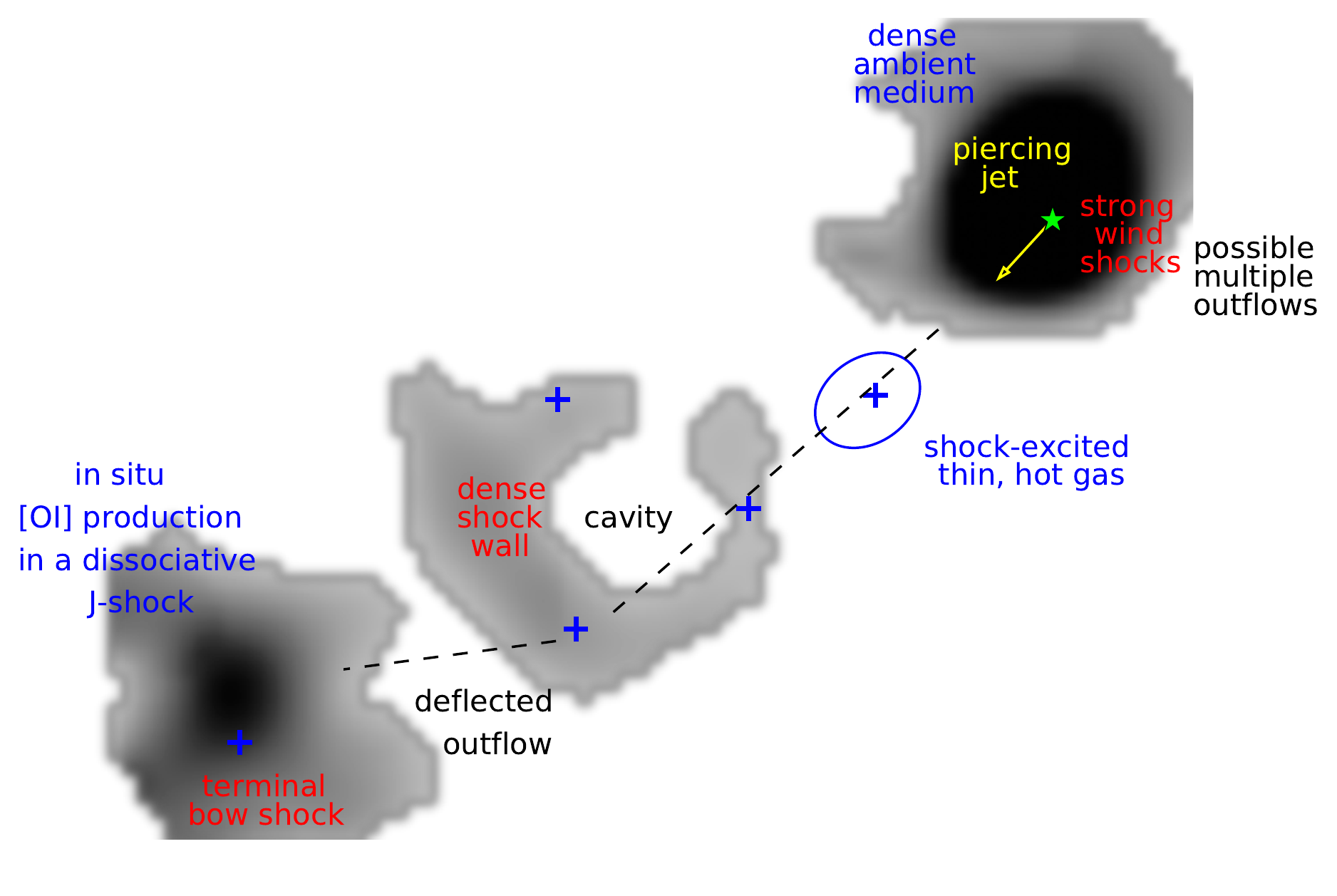}} 
\caption{\small{Schematic of the HH111 and SVS13 outflows.}}\label{fig:hh111_svs13_figure}
\end{figure*} 

\textbf{SVS13:}  \citet{dionatos_2017} recently mapped the SVS13 region in NGC 1333 in [O\,I]$_{63}$ and [CII]$_{158}$ with  Herschel/PACS.  In comparison, the [O\,I]$_{63}$ map presented in this study has a higher spatial resolution in a smaller field of view.     
 Since only faint [CII]$_{158}$ emission is detected in the HH7-11 region \citep{dionatos_2017}, we interpret the detected [O\,I]$_{63}$ emission as mostly coming from shock-excited gas connected to the outflow (Fig.\,\ref{fig:hh111_svs13_figure}). The overall [O\,I]$_{63}$ morphology matches the near-infrared H$_2$ map surprisingly well \citep{chrysostomou_2000,khanzadyan_2003}.   
    \\
 Fortunately, the region associated with HH7-11 has also been mapped in great detail, with HST revealing a consistent schematic view of the HH7-11 outflow (see Fig.\,12 in \citet{hartigan_2019}).   HH11 is not detected in [O\,I], which is consistent with it being a high-excitation region mostly seen in H$\alpha$.  In this context, it is noteworthy that for HH7, HH8, and HH10, [Fe\,II] and H$_2$ peak at roughly the same locations, whereas for HH11, [Fe\,II] peaks further downstream of the outflow (see Fig.\,6 in \citet{khanzadyan_2003}). This may suggest the presence of  shocked, thin, hot gas located at HH11 being able to dissociate molecular hydrogen at the apex of the bow shock. The detection of bright [O\,I]$_{63}$ emission at the location of HH7 strongly supports the notion of it being the terminal bow shock region of the HH7-11 outflow \citep[e.g.][]{smith_2003, hartigan_2019}. HH7 has a remarkable complex internal substructure (HH7A, 7B, 7C, spatially unresolved in our maps) from which the detected [O\,I]$_{63}$ emission in principle can arise. Furthermore, a potentially present Mach disc in HH7 was reported by \citet{noriega-crespo_2002}. Based on near-infrared H$_2$ observations, \citet{smith_2003} offered two shock model predictions for the imaging of the HH7 region at [O\,I]$_{63}$ \citep[see Figs.\,13 and 14 in][]{smith_2003}. The observed [O\,I]$_{63}$ emission, which is an intense compact knot together with a slightly blue-shifted line profile at HH7, is consistent with the dissociative J-type paraboloidal bow shock model. Following this line of reasoning, a significant amount of [O\,I]$_{63}$ emission may be produced in situ, that is, in the dissociative J-shock region, where CO or H$_2$O molecules are broken apart \citep{flower_2010}. However, recent spectroscopic observations of pure rotational H$_2$ lines at HH7 are more in agreement with a non-dissociative C-type molecular shock \citep{neufeld_2019}.   \citet{molinari_2000} reported signatures of both C-type and J-type shocks in the HH7-11 region, illustrating the complex shock structure of the HH7-11 outflow. \\
The innermost region of SVS13 is particularly interesting, since it  exhibits several astrophysical features, for example, H$_2$O maser emission \citep{haschick_1980}, multiple continuum sources forming a complex hierarchical system \citep[VLA3, VLA4, VLA4B,][]{rodriguez_1999, anglada_2000}, multiple detected outflows \citep[e.g.][]{noriega-crespo_2002, lefevre_2017}, and outburst events \citep{eisloeffel_1991}. \citet{hodapp_2014} revealed the presence of a micro-jet traced by shock-excited [Fe\,II] and a series of expanding bubble fragments seen in H$_2$. It has been speculated that an observed outburst event in 1990 \citep{eisloeffel_1991} may be the origin of these shell-like structures \citep[e.g.][]{hodapp_2014, gardner_2016}. Since we detect most of the [O\,I]$_{63}$ emission from that inner region, we interpret this originating from bow shock fronts of the bubble and the interaction zone where the micro-jet potentially pierces the bubble.  However, strong wind shocks from one or more continuum sources can also be responsible for the [O\,I]$_{63}$ emission at SVS13. \\
The diffuse and extended [O\,I]$_{63}$ emission seen in our obtained maps at HH8 and HH10 can be interpreted as jet deflection region \citep{bachiller_2000, hartigan_2019}, that is, a location where the outflow strikes the ambient medium leading to a substantial change in direction. In this scenario, HH7 appears to be off the HH11-HH10-HH8 chain due to that deflection. \\
The HH9 knot features no [Fe\,II] and only very faint H$_2$ emission \citep{khanzadyan_2003}. We detect some [O\,I] emission vaguely around HH9. Due to its location at the  cavity wall  around the HH7-11 outflow \citep{hartigan_2019}, we suspect that some entrained or deflected material turbulently shocks the ambient medium.

 \textbf{HH34:} We suspect that the detected [O\,I]$_{63}$ emission close to HH34IRS is linked to shock-excited gas in a jet/counter jet outflow region (Fig.\,\ref{fig:hh34_figure}). This conclusion is supported by near-infrared observations showing strong on-source emission in [Fe\,II] and H$_2$ \citep[e.g.][]{garcia_lopez_2010, davis_2011}. A potential disc around HH34 IRS \citep{rodriguez_2014} might contribute to the detected [O\,I]$_{63}$ emission.     \\ Compared with HH111, no far-infrared counterpart of the optical jet is seen between knots E and L. This is surprising, since there are several demonstrable similarities between the HH111 jet and the HH34 jet \citep[e.g.][]{reipurth_2002}. \\
 The HH34 jet is most prominently seen in [SII]$\lambda 6716$, [O\,I]$\lambda 6300$, less bright in H$\alpha$, and features numerous knots within the jet body \citep{bacciotti_1999, reipurth_2002, podio_2006}. The relatively strong optical [SII]$\lambda 6716$ emission in the HH34 jet indicates low shock velocities in the emitting gas \citep{hartigan_1994}. Since the [O\,I]$\lambda 6300$ line is prominently detected in the jet, we conclude that atomic oxygen is copiously present in the flow region and could in principle give rise to the far-infrared [O\,I]$_{63}$ line.   In the near-infrared, the jet is prominently seen in [Fe\,II] and H$_2$ \citep[e.g.][]{podio_2006, garcia_2008, antoniucci_2014}, and   [Fe\,II]   peaks where [SII]$\lambda 6716$ peaks.  So, both lines ([Fe\,II], [S\,II]$\lambda 6716$) are likely to be excited in J-shocks at the apices of the  internal bow shocks \citep{podio_2006, antoniucci_2014}. \citet{podio_2006} measured the  ionisation fraction ($x_e \sim 0.05-0.17$), electron density ($n_\text{e} \sim 10^3 \text{cm}^{-3}$), temperature ($T_e \sim 1.3\times 10^4\text{K}$), and total density ($n_\text{H} \sim 10^3-10^4\text{cm}^{-3}$) along the jet. These values are consistent with the assumed shock conditions of \citet{hollenbach_1989}.  So, the non-detection of a [O\,I]$_{63}$ jet can  either be a result of the too low shock velocities within the HH34 jet, or the [O\,I]$_{63}$ jet is indeed present, but too faint to be detected. We suspect that the HH34 jet is detectable in [O\,I]$_{63,}$ provided there are deeper exposures.\\

\textbf{HH26:} The non-detection of [O\,I]$_{63}$ at HH26IRS is consistent with the interpretation given by \citet{antoniucci_2008} that the jet driven by HH26IRS is mainly molecular, meaning that it is mostly seen in H$_2$ \citep[e.g.][]{davis_2002, chrysostomou_2002} or CO \citep{dunham_2014}. Interestingly, no [Fe\,II] but strong H$_{2}$ was detected at HH26IRS in the near-infrared \citet{antoniucci_2008}, hinting to a low jet density \citep{davis_2011}.  Following this line of reasoning, the extended [O\,I]$_{63}$ emission at HH26A (here $K_1$ and $K_2$) supports the conclusion that this region represents a shock-excited region (Fig.\,\ref{fig:hh26_figure}), where the jet has struck the ambient medium and thus is indeed a deflection region as proposed by \citet{chrysostomou_2002}. In this scenario, HH26C is interpreted as the deflected, terminal bow shock, which was not mapped here. Spectroscopic observations at different locations within the HH26 region \citep{benedettini_2000, giannini_2004} support this assumption that the observed [O\,I]$_{63}$ emission in the blue lobe towards HH26A is mainly due to shock excitation, that is, not due to the presence of a strong FUV field \citep{benedettini_2000}.

\begin{figure}[htb!] 
\resizebox{\hsize}{!}{\includegraphics[trim=14 41 22 43, clip, width=0.9\textwidth]{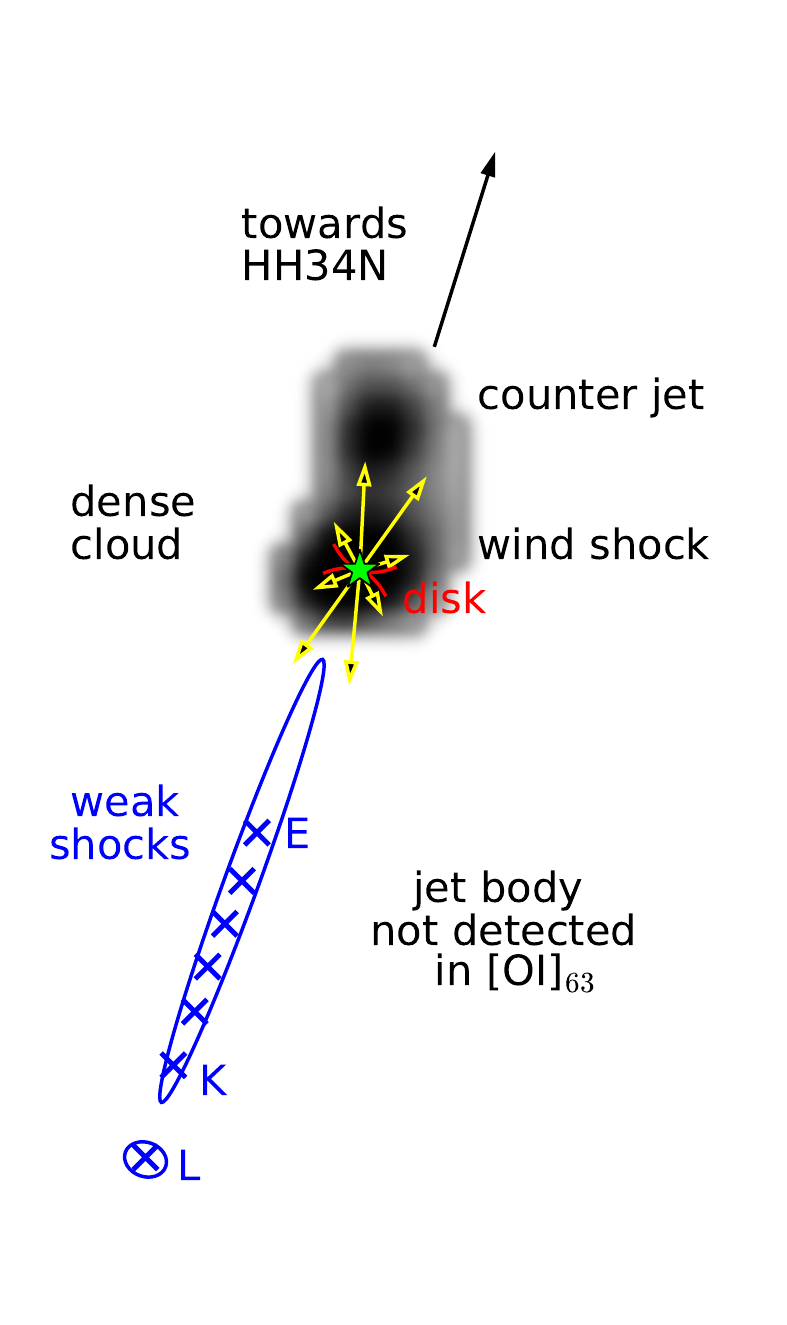}}
\caption{Schematic of the HH34 outflow.}\label{fig:hh34_figure} 
\end{figure}

\begin{figure}[htb!] 
\resizebox{\hsize}{!}{\includegraphics[trim=35 75 65 60, clip, width=1.0\textwidth]{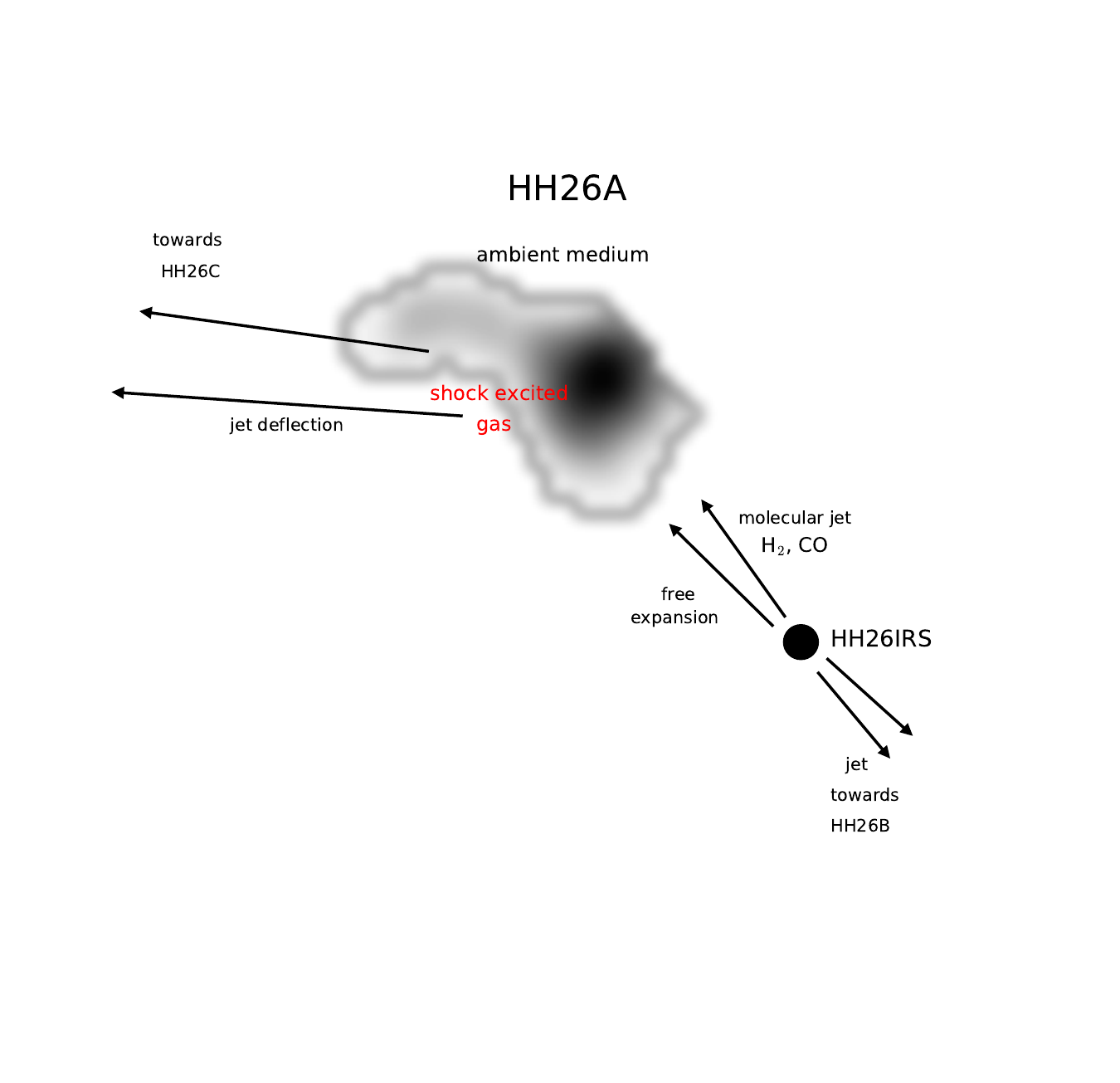}}
\caption{Schematic of the HH26 outflow.}\label{fig:hh26_figure} 
\end{figure}


\subsection{Caveats on the derived mass-loss rates}\label{sec:caveats_HM89}
 
 The crucial assumption of the HM89 model is that all the observed [O\,I]$_{63}$ emission comes from one decelerated wind shock. This explicitly excludes emission from possible multiple shocks, slow shocks, bow shocks, or deflection regions. Observations at high spatial resolution together with comparable observations at other shock tracers (e.g. H$\alpha$, H$_2$, [Fe\,II]) may provide deeper insights into the number of shocks, the presence of a Mach disc (the deceleration shock), or the overall geometry of the shock region. Without these potentially valuable observations, the HM89 formula has to be applied with strong reservations, since the global shock structure is undetermined. In the case of HH26 and HH8/HH10, in the SVS13 region  there are indeed indications that the detected [O\,I]$_{63}$ emission comes from a deflection region (see Sect.\,\ref{line_ratio}) and not from a wind shock. Thus, as already  emphasised by \citet{hartigan_2019}, the HM89 formula will probably give inaccurate mass-loss rates in these cases.  \\ 
 On the other hand, if the jet material passes through several, spatially unresolved shocks, an unknown amount of [O\,I]$_{63}$ luminosity emerges from these multiple shocks, and thus the HM89 formula potentially overestimates the mass-loss rate. This is certainly  true in the case of the HH111 jet observed in this study as discussed in Sect.\,\ref{sec:[OI]_schematics}.  Theoretically, the HM89 formula could be adjusted, taking into account the unknown number of shocks \citep{dougados_2010, nisini_2015}, and this correction factor may be on the order of unity \citep{nisini_2015}. \\
If, on the other hand, parts of the wind flow without any interactions within the ambient medium, HM89 underestimates the mass-loss rate \citep{cohen_1988}. This interaction-free component of the jet will inevitably  be untraceable.  The best case scenario would be that both effects cancel each other out by chance.   \\ 
 Furthermore, [O\,I]$_{63}$ emission can have various origins, for example a present disc or a PDR region. In order to disentangle and quantify all possible contributions, specific line ratios in the mid- and far-infrared, for example  [O\,I]$_{63}$/[O\,I]$_{145}$, [C\,II]$_{158}$/[O\,I]$_{63}$ \citep{nisini_1996, kaufman_1999, flower_2010}, [O\,I]$_{63}$/[S\,I]$_{25}$ \citep{benedettini_2012}, [Si\,II]$_{35}$/[Fe\,II]$_{26}$, or [Fe\,III]$_{23}$/[Fe\,II]$_{26}$ \citep{watson_2016} can be exploited and compared with model predictions.   Unfortunately, the valuable [O\,I]$_{145}$ detection is at noise level in our target sample, and other emission lines were not simultaneously observed, so we must speculate about the physical origin of the detected [O\,I]$_{63}$ emission. Luckily, in the case of SVS13, \citet{molinari_2000} measured [OI]$_{63}$ and [OI]$_{145}$ line fluxes at the driving source and at HH7.  According to their measurements, the [OI] line ratio is quantified by $j_{63}/j_{145}\sim 22.5$ at the driving source SVS13,   and it is $j_{63}/j_{145}\sim 28.5$ at HH7. These values are perfectly consistent with predictions from shock models, meaning that depending on pre-shock densities and shock velocities, values between 10 and 35 are expected \citep{hollenbach_1989}. \\ 
 The assumption of the shock origin of the [O\,I]$_{63}$ line is also strongly supported by several lines of observational evidence.  Herschel observations of similar protostellar outflow sources  demonstrated that most of the [O\,I]$_{63}$ emission emerges from dissociative J-shock regions located at the apex of bow shocks \citep{vanKempen2010, benedettini_2012, podio_2012, karska_2013}.   Additionally,  there are several observational studies that estimate the total [O\,I]$_{63}$ emission coming from the disc in such sources to be only a few percent \citep{podio_2012, watson_2016}. \\
In conclusion, detecting a substantial amount of [O\,I]$_{63}$ at the jet driving sources HH111 IRS, SVS13, and HH34 IRS as extended blob-like emission, we suggest that the bulk of this [O\,I]$_{63}$ emission is connected to a wind shock supporting the applicability of the HM89 formula only in these three cases (see Table\,\ref{table:main_results_II}). For those three cases, both methods lead to very similar mass-loss rates, that is, they only differ by a factor of 2 at most and are to the order of $\dot{M}  \sim 2\times 10^{-6} M_\odot\text{yr}^{-1}$. In comparison, the dubious regions show broader differences by a factor of $2-3$ in the derived mass-loss rates. In the case of HH34, \citet{heathcote_1992} and \citet{hartigan_1994} used optical tracers to derive mass-loss rates to the order of $\dot{M} \sim 10^{-7} M_\odot\text{yr}^{-1}$. Similar mass-loss rates are found in HH111 jet \citep{hartigan_1994, lefloch_2007}.  We derive significantly higher mass-loss rates in both cases. This points to the conclusion that the bulk of the jet material resides in the warm, neutral component of the jet. \citet{molinari_2000} used far-infrared spectra of the SVS13 region to derive mass-loss rates with the HM89 method. Their mass-loss rates are substantially higher at the source than our measurements.  However, \citet{molinari_2000} took spectra from a region that includes HH10 and HH11. Since we detected some [OI]$_{63}$ emission at HH10, we conclude that they have overestimated the [OI]$_{63}$ emission at the driving source leading to too high a mass-loss rate.


{\renewcommand{\arraystretch}{1.2}
\begin{table*}
\caption{\small{[O\,I] mass-flux rates and comparison with $\dot{M}_\text{acc}$. $\dot{M}_\text{jet}^\text{lum}(\text{OI})$ and  $\dot{M}_\text{jet}^\text{shock}(\text{OI})$ are calculated via Eq.\,\ref{equ:line_luminosity} and Eq.\,\ref{equ:shock_model},  respectively. Green cell colours (light) indicate the regions where the HM89 model is likely to be applicable, and red cells (dark) mark the doubtful cases (see Sect.\,\ref{sec:caveats_HM89}).}}\label{table:main_results_II}
\centering
\begin{tabular}{c c c c c c c c}
\hline\hline
Obj. & Region & Box & $\theta$ & $v_\text{t}$ &  $\dot{M}_\text{jet}^\text{lum}(\text{OI})$  & $\dot{M}_\text{jet}^\text{shock}(\text{OI})$ &  $\dot{M}_\text{loss}^\text{literature}$ \\[1.5pt]
     &        &     & ('')  & (km/s) & $(10^{-7}M_\odot\text{yr}^{-1})$  & $(10^{-7}M_\odot\text{yr}^{-1})$   & $(10^{-7}M_\odot\text{yr}^{-1})$  \\[1.5pt]
\hline  
\small{HH111} & \tiny{HH111IRS}        & $B_1$ & 20 & 270\tablefootmark{a} & $26-53$ & \cellcolor{green}$22.9-26.4$ &  $-$  \\ [1.5pt]
              & \tiny{jet (knots F-O)} & $B_2$ & 45 & 260\tablefootmark{a} & $6-12$ & \cellcolor{red}$10.0-15.9$  &   $1.8-4$\tablefootmark{e}     \\[3pt]
\hline
\small{SVS13} & \tiny{SVS13}           & $B_1$ & 22 & 270\tablefootmark{b} & $25-51$ & \cellcolor{green}$13.1-16.0$  &  $38-66$\tablefootmark{f} \\[1.5pt]
              & \tiny{HH8-11}          & $B_2$ & 40 & 340\tablefootmark{b} & $8.5-17.2$ & \cellcolor{red}$6.1-8.0$  &   $\lesssim 1-10$\tablefootmark{g}         \\[1.5pt]
              & \tiny{HH7}             & $B_3$ & 21 & 400\tablefootmark{b} & $18-37$ & \cellcolor{red}$6.5-7.2$     &  48\tablefootmark{h}       \\[1.5pt]
\hline
\small{HH26}  & \tiny{HH26A}           & $B_2$ & 28 & 100\tablefootmark{b} & $5-9$ & \cellcolor{red}$13.6-18.5$     & $18-21$\tablefootmark{i}    \\[1.5pt]
\hline
\small{HH34}  &\tiny{counter jet, HH34IRS} & $B_1$ & 26 & 160\tablefootmark{c} & $11-23$ & \cellcolor{green}$20.7-27.5 $ &   $0.5-5$\tablefootmark{j}   \\[1.5pt]
\hline
\small{HH30}  & \tiny{HH30IRS}             & $B_1$ & 20 & 190\tablefootmark{d} & $<0.1-0.2$ & \cellcolor{red}$<0.45 $ & $0.002-0.02$\tablefootmark{k}    \\[1.5pt]
\hline\hline
\end{tabular}
\tablefoot{
\tablefoottext{a}{\citet{hartigan_2001}}
\tablefoottext{b}{\citet{chrysostomou_2000}}
\tablefoottext{c}{\citet{eisloeffel_1992}}
\tablefoottext{d}{\citet{mundt_1990}}
\tablefoottext{e}{\citet{hartigan_1994, lefloch_2007}}
\tablefoottext{f}{\citet{molinari_2000, dionatos_2020}}
\tablefoottext{g}{estimate based on \citet{snell_1981, lizano_1988, molinari_2000, dionatos_2020} }
\tablefoottext{h}{\citet{molinari_2000} }
\tablefoottext{i}{\citet{benedettini_2000} }
\tablefoottext{j}{\citet{hartigan_1994, heathcote_1992, bacciotti_1999, garcia_2008, antoniucci_2008, nisini_2016}}
\tablefoottext{k}{\citet{mundt_1990, bacciotti_1999} }
}
\end{table*}

 

\section{Conclusions}

We have presented SOFIA FIFI-LS observations of five protostellar Class I objects and their outflows (HH111, SVS13, HH26, HH34, HH30) in the [O\,I]$_{63,145}$ transitions. Our maps were used to detect shock-excited regions that are connected to protostellar outflows (e.g. a low-excitation atomic jet component, bow shocks, or wind shocks).   
Our main findings can be summarised by the following points.

\smallskip
\noindent
Strong [O\,I]$_{63}$ emission was detected at the driving sources HH111IRS, SVS13, and HH34IRS, and almost none was detected at HH26IRS. Bright on-source detection of [O\,I]$_{63}$ in these cases may arise from the protostellar outflow, interacting with the ambient medium and leading to shock excitation, meaning a wind shock. Thus, in these three cases (HH111IRS, SVS13, HH34IRS) the \citet{hollenbach_1989} shock model assumptions most likely prevail, justifying utilisation of the HM89 relation to derive mass-loss rates.

\smallskip
\noindent
The  optical jet at about 15$\arcsec$ west of  HH111IRS (e.g.\,\citet{reipurth_1989_discovery}) is strikingly apparent at [O\,I]$_{63}$ with an extent of about 45$\arcsec$. On the contrary, almost no [O\,I]$_{63}$ was detected at the optical HH34 jet south-east   between HH34IRS and knot L in \citet{eisloeffel_1992}.  

\smallskip
\noindent
Prominent [O\,I]$_{63}$ emission is noticeable at the terminal bow shock region HH7.  The detected [O\,I]$_{63}$ line at HH26A ($K_1, K_2$) and HH8/HH10 likely emerges from a jet deflection region and not a wind shock.

\smallskip
\noindent
A potential [O\,I] counter jet about 22$\arcsec$ north-west of SVS13 at P.A.$\sim 155^\text{o}$ is detected in [O\,I]$_{145}$. Unfortunately, this region was not fully mapped in [O\,I]$_{63}$.

\smallskip
\noindent
No [O\,I] emission was detected at HH30.  Since it was the faintest target in our sample, we suspect that a deeper exposure may lead to a successful [O\,I] detection in this case. 

\smallskip
\noindent
We determined continuum fluxes of HH111IRS, SVS13, and HH34IRS at $63\,\upmu\text{m}$ and $145\,\upmu\text{m}$, whereas for HH26IRS this was only possible at $145\,\upmu\text{m}$ (values are reported in Table\,\ref{table:continuum_fluxes}).

\smallskip
\noindent
The observed outflow rates of our low-mass Class I sample are to the order of $\dot{M}_\text{jet}\sim 10^{-6} M_\odot\text{yr}^{-1}$, which is considerably higher than typical outflow rates found in jets from low-mass classical T\,Tauri stars \citep[$\dot{M}_\text{jet}$ $\sim$ $10^{-7}-10^{-9}$ $M_\odot\text{yr}^{-1}$,][]{frank_2014}. This finding is consistent with \citet{caratti_2012}, who found lower mass-loss rates for more evolved sources. 

\smallskip
\noindent
We find that both methods applied to determine mass-loss rates (Sects.\,\ref{sec:mass_flux_rates_HM89} and \ref{sec:mass_flux_rates_geometry}) lead to similar values in most of the cases, that is to the same order of magnitude, even though both methods have dissimilar deficiencies. However, considering the discussed caveats (Sect.\,\ref{sec:caveats_HM89}), this result might be fortuitous and only in the cases of HH34IRS, HH111IRS, and SVS13 physically reliable.

\smallskip
\noindent
Comparing the obtained mass-loss rates with estimates from the literature (Table\,\ref{table:main_results_II}), we find that all the listed values scatter within one order of magnitude. No clear tendencies towards a systematic overestimate or underestimate of our mass-loss rates are discernable.

\smallskip
\noindent
Mass-loss rates can in principle be calculated from the observed [O\,I]$_{63}$ line luminosity utilising the \citet{hollenbach_1989} shock model, only if the intrinsic assumptions of this shock model are fulfilled in the region of interest. Specifically, the [O\,I]$_{63}$ must actually come from one decelerated wind shock. Simultaneous observations at other emission lines (e.g. [O\,I]$_{145}$, [C\,II]$_{158}$, [S\,II]$_{35}$, [Fe\,II]$_{26}$) may support the interpretation of the shock origin of the [O\,I]$_{63}$.   
 

\begin{acknowledgements}  
We thank the referee for his/her substantial comments and helpful criticism, which have changed this paper quite a bit. We thank Bill Vacca for providing us with the ATRAN-models.
This research is based on observations made with the NASA/DLR Stratospheric Observatory for Infrared Astronomy (SOFIA). SOFIA is jointly operated by the Universities Space Research Association, Inc. (USRA), under NASA contract NNA17BF53C, and the Deutsches SOFIA Institut (DSI) under DLR contract 50 OK 0901 to the University of Stuttgart.
This work has been supported by the German Verbundforschung grant 50OR1717 to JE.

\end{acknowledgements}

\bibliographystyle{aa} 
\bibliography{papers}  

\begin{appendix} 

\section{Brief outflow descriptions of the observed targets}\label{appendix:target_information}

\textbf{HH111:} This iconic Herbig-Haro object is located in the L1617 dark cloud in the Orion B cloud complex at a distance of approximately 420\,pc \citep{zucker_2019}. A remarkable one-sided optical jet appears at about $15\arcsec$ west of its driving source and extends to a distance of roughly $40\arcsec$ \citep{reipurth_1992}. The total extent of the HH111 highly collimated jet complex is about $370\arcsec$ \citep{reipurth_1999}. Observations at 3.6\,cm indicate the optically obscured driving source of the jet to be IRAS 05491+0247  \citep[VLA-1 in][]{reipurth_1999}. Since its discovery \citep{reipurth_1989_discovery}, the HH111 system  has been extensively studied across  the  entire spectral wavelength range to investigate the physical properties of the jet \citep[e.g. see introduction in][and references therein]{noriega-Crespo_2011}. NIR observations in the K-Band at various lines of [Fe\,II]  and  H$_2$ suggest the existence of a second outflow (HH 121) almost perpendicular to HH 111  \citep{gredel_1993}. The driving source of HH111 might be a binary \citep{reipurth_1999}. A well-collimated   molecular outflow in CO is associated with the HH 111 jet  \citep[e.g.][]{reipurth_und_olberg_1991, cernicharo_1996, lee_2000}.\\\\
\textbf{HH26:} HH26 is located in the L1630 dark cloud  of the Orion B molecular cloud \citep{herbig_1974}, at a distance of $D\sim$\,$420\,$pc \citep{zucker_2019}. It is part of a complex region HH 24-26 that shows ongoing star formation: for example, two embedded   protostars HH24MMS  \citep{chini_1993}, HH25MMS  \citep{bontemps_1995}, and four bright NIR sources  SSV59, SSV60, SSV61, and  SSV63 \citep{strom_1976}. HH26IR is an embedded Class I protostar  driving the HH26A/C/D chain (also brightly seen in SDSS9 images) and the associated molecular outflow \citep{davis_1997}. A small-scale H$_2$ jet  driven by  HH26IR was described in  \citet{davis_2002} and \citet{chrysostomou_2008}. \citet{antoniucci_2008} investigated the accretion and ejection properties of HH26IRS via NIR spectra ($H$ and $K$ bands). \citet{benedettini_2000} observed the HH26IR source as well as the HH26IR outflow with the ISO spectrometers (covered wavelength range: 45--197$\upmu$m), from which a diffuse ($n_{\text{H}_2} \sim$\,$10^4\text{cm}^{-3}$) and  warm ($T\sim 1800$\,$\text{K}$) gas component was detected. \citet{caratti_2006} derived several jet  parameters (e.g. $T\sim 2350-3500\,\text{K}$) of HH26 from the $\text{H}_2$ and [Fe\,II] line analysis.\\\\
\textbf{HH34:} The well-known and likewise spectacular HH34 parsec scale outflow powered by HH34IRS \citep{eisloeffel_1997} is located in L1642 in the Orion A cloud  \citep[$D\sim$\,$430\,\text{pc}$,][]{zucker_2019}. The associated optical jet consists of two symmetrically placed large bow-shocks, namely HH34 N \citep[discovered by][]{buehrke_1988}  and HH34 S, between which a chain of several knots is identified \citep{reipurth_1986}. Proper motion studies of these knots  suggest an inclination angle of $20^\text{o}-30^\text{o}$ \citep{eisloeffel_1992, heathcote_1992}. Close to the driving source HH34IRS, a highly collimated jet of about $30\arcsec$ in projected size pointing to HH34 S is detected in the optical as well as in the IR \citep{reipurth_2002, antoniucci_2014}. \citet{antoniucci_2008} investigated the accretion and ejection properties of HH34IRS via NIR spectra ($H$ and $K$ bands). \citet{nisini_2016} also derived mass-accretion rates from X-shooter line observations (350--2300\,nm). 
\\\\
\textbf{SVS13:} Located at a distance of about 235\,pc in the L1450 Perseus cloud \citep{hirota_2008, hirota_2011}, SVS13 (IRAS 03259+3105) is the driving source of the well-studied HH7-11 outflow \citep[e.g.][]{bachiller_2000}.  HH7-11 is a famous chain of similarly shaped optically bright nebulosities located south-east of SVS13, with HH7 at the end of this string \citep{herbig_1974}. A nearby Class 0 protostar SVS 13B is placed south-west of SVS13 with a separation of $14\farcs 5$ \citep{bachiller_1998}. Radio observations revealed two other nearby radio sources \citep[VLA 3 and VLA 20, e.g.][]{rodriguez_1997, rodriguez_1999}. \citet{anglada_2000} found evidence via radio observations that SVS13 is a close binary system with a separation of $0\farcs 3$.  \citet{molinari_2000} obtained FIR spectroscopic observations of the HH7-11 flow with ISO and detected several FIR lines ([O\,I]$_{63\upmu\text{m},145\upmu\text{m}}$, [S\,II]$_{34.8\upmu\text{m}}$, [C\,II]$_{158\upmu\text{m}}$ and molecular H$_2$, CO, H$_2$O), indicating that dissociative (J-type) and non-dissociative (C-type) shocks are simultaneously present along the flow.
\\\\
\textbf{HH30:} The prototypical jet/disc system associated with HH30 is situated in the dark molecular cloud L1551 at a distance of 140\,pc in Taurus \citep{mundt_1983}. \citet{vrba_1985} detected the exciting infrared source that features a flared accretion disc seen edge-on \citep{burrows_1996, stapelfeldt_1999}. Observations in the optical reveal a prominent jet/counter-jet structure  \citep[e.g.][]{mundt_1988, graham_1990, estalella_2012}. Observations in [S\,II] close to the driving source show a knotty structure within the jet \citep{mundt_1990}. \citet{anglada_2007} models the HH30 jet structure seen in [S\,II] images as a wiggling ballistic jet, suggesting that either the orbital motion of the jet source around a primary or the precession of the jet axis is responsible. Morphological and photometric variability of HH30 have been reported \citep[e.g.][]{burrows_1996, stapelfeldt_1999, watson_2007, watson_2008, duran_2009}, in particular in the brightness ratio of the upper (north-northeast) and lower (south-southwest) nebulae.  

 \newpage
\section{Mass-loss rates from the [O\,I]$_{63\upmu\text{m}}$ line}

In this paragraph, we derive an alternative relation to the one found by \citet{hollenbach_1985}  that connects mass-loss rates with the  [O\,I]$_{63}$ line luminosity, (Eq. 8 in this paper).\\
The derivation is logically divided into three parts. In Sect. \ref{sec:analytical_solution}, we solve the rate equations for the atomic OI system. In the subsequent Sect. \ref{sec:critical_density}, we utilise this solution and constrain the relevant level population from which the [O\,I]$_{63}$ line emerges. Following \citet{hartigan_1995}, we then exploit our findings of Sect. \ref{sec:critical_density} to express the [O\,I]$_{63}$  line luminosity in terms of observable jet parameters such as its projected length and velocity.
 
 \subsection{Analytical solution of the three-level system}\label{sec:analytical_solution} 
 
\begin{figure}[h!]
\resizebox{\hsize}{!}{\includegraphics[trim=0 0 42 0, clip, width=0.9\textwidth]{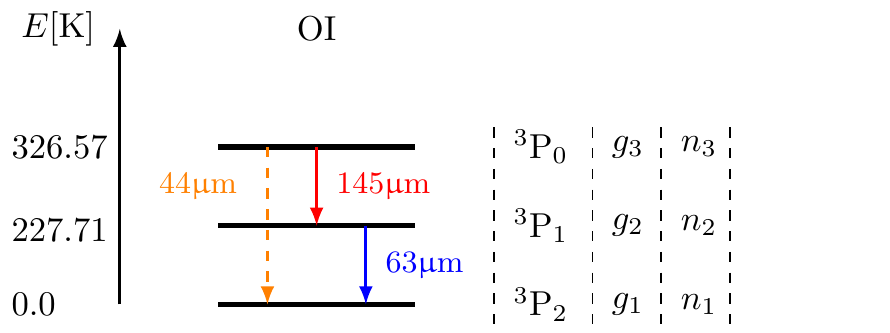}}
\caption{\small{The lowest three energy levels $^3\text{P}_{2,1,0}$ in the OI system. Notation: The $g_i$ are  the statistical weights and $n_i$ stand  for number densities of oxygen atoms in the levels $i=1,2,3$.}}\label{fig:level} 
\end{figure}

Due to LS-coupling, neutral oxygen  can be approximated energetically as a five-level system associated with atomic terms in ascending order of energy: $^3\text{P}_2,\,\, ^3\text{P}_1,\,\,^3\text{P}_0,\,\,^1\text{D}_2,\,\,^1\text{S}_0$.  
However, non-LTE calculations performed in \citet{nisini_2015} show that for temperatures below $T\sim$\,$5000\,$K it is sufficient to take into account only the three lowest lying levels (Fig.\,\ref{fig:level}) since the  higher levels ($^1\text{D}_2$, $^1\text{S}_0$) are barely populated in this case.  The level population $n_i$  with $i = 1,2,3$  stands for the number density of oxygen atoms in the $i^{\text{th}}$ state ($[n_i] = 1\,\text{cm}^{-3}$). The total number density of oxygen atoms is then $n(\text{O})\approx n_1 + n_2 + n_3,$ and it follows
 \begin{align}
 \left( \frac{n_2}{n(\text{O})} \right) &= \left[ 1+ \left( \frac{n_1}{n_2} \right) + \left( \frac{n_3}{n_2} \right)\right]^{-1}.
  \end{align}
 Assuming statistical equilibrium and neglecting radiation fields, we can solve the three rate equations \citep[e.g.][]{liseau_2006}:
 \begin{align}
 \left(  \frac{n_2}{n_3} \right) & = \frac{C_{13}  \left(C_{32} + A_{32} \right) +  C_{12}  \left( C_{31} + C_{32} + A_{31} + A_{32} \right) }{C_{12}  C_{23} + C_{13}   \left(C_{21} + C_{23} + A_{21}  \right) }, \\
 \left(  \frac{n_1}{n_2} \right) & = \frac{\left( C_{31} + A_{31} \right) \left( C_{21} + C_{23} + A_{21}  \right) + \left( C_{32} + A_{32} \right) \left( A_{21}+C_{21}\right)   }{\left( C_{32} + A_{32} \right) \left( C_{12} + C_{13} \right) + C_{12}  \left( A_{31} + C_{31} \right)   }.
   \end{align}
 
Here, we introduce the Einstein coefficients $A_{ul}$ for the spontaneous emission ($A_{21}=  8.91\times 10^{-5}\, \text{s}^{-1}$, $A_{32}= 1.75\times 10^{-5}\, \text{s}^{-1}$, $A_{31}  = 1.34 \times 10^{-10}\, \text{s}^{-1}$, NIST Atomic Spectra Database\footnote{\url{https://physics.nist.gov/PhysRefData/ASD/levels_form.html}}) and the total collisional rates $C_{ul}$ ($C_{lu}$), which record all collisional contributions from the upper (lower) to the lower (upper) level ($[C_{ul}] = [C_{lu}]= 1\,\text{s}^{-1}$). We expect atomic hydrogen to be the decisive collisional partner, since warm  interstellar gas at $T<5000\,\text{K}$ is mostly composed of atomic hydrogen and almost neutral, that is, collisions with electrons are negligible.  
 From the Boltzmann statistics, the collisional excitation rate coefficients  can be determined via  
\begin{equation}
q_{lu}^\text{\tiny{H}}(T)      = q_{ul}^\text{\tiny{H}}(T) \,\left( \frac{g_u}{g_l}\right) \,\text{exp}\left(-\frac{E_{lu}}{k_\text{B} T}\right), \quad  E_{lu} = E_u -E_l > 0.
\end{equation}
Here, $k_\text{B}$ refers to Boltzmann’s constant.  
Based on quantum scattering calculations, \citet{lique_2018}\footnote{Taken from \url{https://home.strw.leidenuniv.nl/~moldata/datafiles/oatom@lique.dat}.} provided  collisional rate coefficients $q_{ul}^\text{H}(T)$ for atomic oxygen.

  \subsection{Critical density}\label{sec:critical_density}
  
The critical density $n_\text{crit,p}$ for a given energy level $k$ may be defined as the density of the collision partner  p 
at which collisional transitions (excitations and deexcitations) are equal to the effective deexcitation by spontaneous decay from that level $k$ \citep{osterbrock_book}, that is, 
 \begin{equation} 
  n_\text{crit,p}(k; T) := \frac{\sum_{i<k} A_{ki} }{\sum_{i\neq k} q_{ki}^\text{p}(T)} .
   \end{equation}
 
Hence, for $T$ $\sim$ $300-8000\,\text{K,}$ critical densities are at the range of $n_\text{crit,H}\sim 2.6\times 10^{5}\,\text{cm}^{-3}$ -- $5.5\times 10^4\,\text{cm}^{-3}$. It is reasonable to assume that the gas density in the warm, dense jet component traced by the [O\,I]$_{63}$ line is close to the critical density. Indeed, if the gas density were much lower than the critical density, then the $^3\text{P}_1$ level would not be sufficiently populated and consequently the [O\,I]$_{63}$ line would not be detected. On the other hand, if the density were much higher than the critical density, the level population would not depend on the density anymore; instead, high-density limit conditions would prevail.  

So, provided that the gas density is close to the critical density, we can constrain the population ratio $n_2/n(\text{O})$. Figure\,\ref{fig:level_population} shows the 
analytical solution of $n_2/n(\text{O})$ for five selected temperatures in the range of $T\sim 300-8000\,\text{K}$ together with their critical densities (straight lines). Since all corresponding ratios of $n_2/n(\text{O})$ lie in a narrow population stripe (highlighted in grey), we estimate the ratio $n_2/n(\text{O})$ to be in the range of 0.1--0.2.

\begin{figure} 
\resizebox{\hsize}{!}{\includegraphics[trim=0 0 0 0, clip, width=0.9\textwidth]{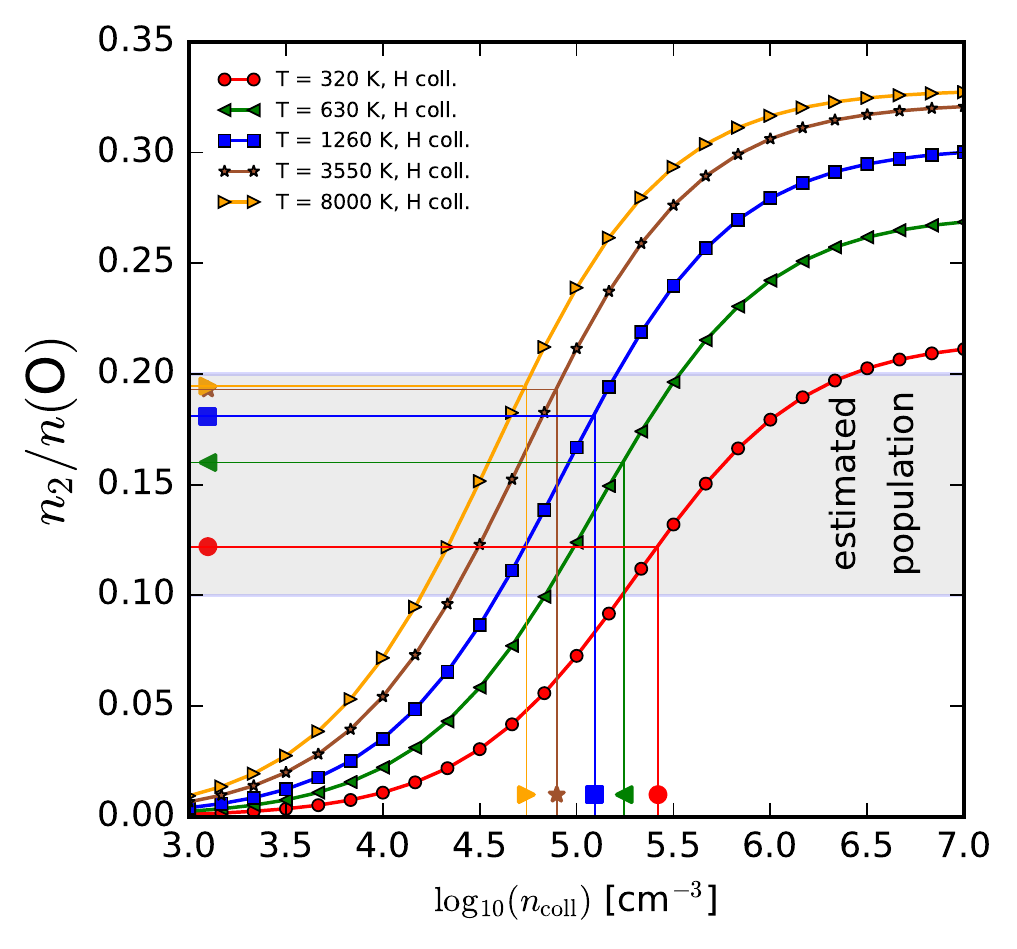}}
\caption{From the analytical solution for the three-level system, we estimate $n_2/n(\text{O}) $ assuming that the gas density of the warm jet component traced by the [O\,I]$_{63}$ line is close to the critical density.}\label{fig:level_population} 
\end{figure}


\subsection{The [O\,I]$_{63}$ line luminosity}

The [O\,I]$_{63}$ line luminosity is given by
\begin{equation}
  L(\text{[O\,I]}_{63})/L_\odot    =   \left( h\nu_{63} \, A_{21}/L_\odot  \right)   \cdot N_2  ,
\end{equation}
with $L_\odot =3.828 \times 10^{33}\,\text{erg}\,\text{s}^{-1}$ as  solar  luminosity, $h\nu_{63}   = 3.1438\times 10^{-14}\,\text{erg}$ as the energy of a photon with wavelength $\lambda = 63.1852\,\upmu\text{m}$, and $N_2$ as total number of oxygen atoms in level $^3\text{P}_1$. We now express $N_2$ in terms of element abundances and proper astronomical units in the following way:

\begin{equation}
N_2   = \left( \frac{N_2}{N(\text{O})}\right)\left( \frac{N(\text{O})}{N(\text{H})}\right)\left( \frac{N(\text{H})}{ N_\text{tot} }\right)\left(  \frac{M_\odot }{ \mu \, m_\text{H} } \right)\left(  \frac{M_\text{tot}  }{ M_\odot } \right) .  
 \end{equation}
Here, $N(\text{O})$ is the total number of oxygen atoms, $N(\text{H})$ is the total number of hydrogen atoms, $N_\text{tot}$ is the total number of atoms, $M_\text{tot}$ is the total mass, $\mu$ is the molecular weight ($\mu \approx 1.24$ for neutral atomic gas, \citet{hartigan_1995}), $m_\text{H} = 1.6735\times 10^{-24}\,\text{g}$ is the mass of a hydrogen atom, and $M_\odot= 1.99\times 10^{33}\,\text{g}$ as solar mass. Combining Eqs. B.6 and B.7 and taking the literature values $N(\text{O})/N(\text{H})   = 4.90\times 10^{-4}$ from \citet{asplund_2009}, $N(\text{H})/N_\text{tot} = 0.921$ from \citet{hartigan_1995},  and exploiting that $N_2/N(\text{O})=n_2/n(\text{O}),$ we get 
\begin{equation}
\left(  \frac{M_\text{tot}  }{ M_\odot }\right)    = \frac{3.16\times 10^{-3} }{   \left( \frac{n_2}{n(\text{O})}\right) }  \,  \left(\frac{L(\text{[O\,I]}_{63})}{L_\odot} \right) . 
 \end{equation}
Following \citet{hartigan_1995}, we make use of the equation  
 \begin{equation}
 \dot{M} =  M_\text{tot}  v_\text{t}/l_\text{t},
 \end{equation}
 where $l_\text{t}\approx \theta\cdot D$ (with $\theta$ as angular distance) is the projected size of the jet on the plane of sky and $v_\text{t}$ is the component of the velocity in the plane of sky.    
The four quantities $v_\text{t}$, $\theta$, $D,$  and $L(\text{[O\,I]}_{63})$  can be obtained from observational data, whereas the the population density ratio  $n_2/n(\text{O})$ is an imprecise function of the temperature $T$ and the hydrogen density $n_\text{H}$. However, assuming  that the gas density is close to the critical density, we can employ the result of Sect.\,\ref{sec:critical_density}  leading  to the useful relation
 \begin{equation}
  \left( \frac{\dot{M}}{M_\odot  \text{yr}^{-1} } \right) =  \left(3.3-6.7 \right)  \times 10^{-3}    \cdot   \left( \frac{v_\text{t} }{\text{km/s}} \right) \left( \frac{ '' }{ \theta  }\right)   \left( \frac{\text{pc}}{ D} \right)\left( \frac{L(\text{[O\,I]}_{63})}{L_\odot}\right).
  \end{equation}
 The derived equation B.10 represents an alternative to the \citet{hollenbach_1985} relation to estimate the mass-loss rate from the [O\,I]$_{63}$ line luminosity.


  \section{Continuum maps}\label{appendix:continuum_maps}

\begin{figure*} 
  \centering
  \subfloat{\includegraphics[trim=0 33 0 0, clip, width=0.9\textwidth]{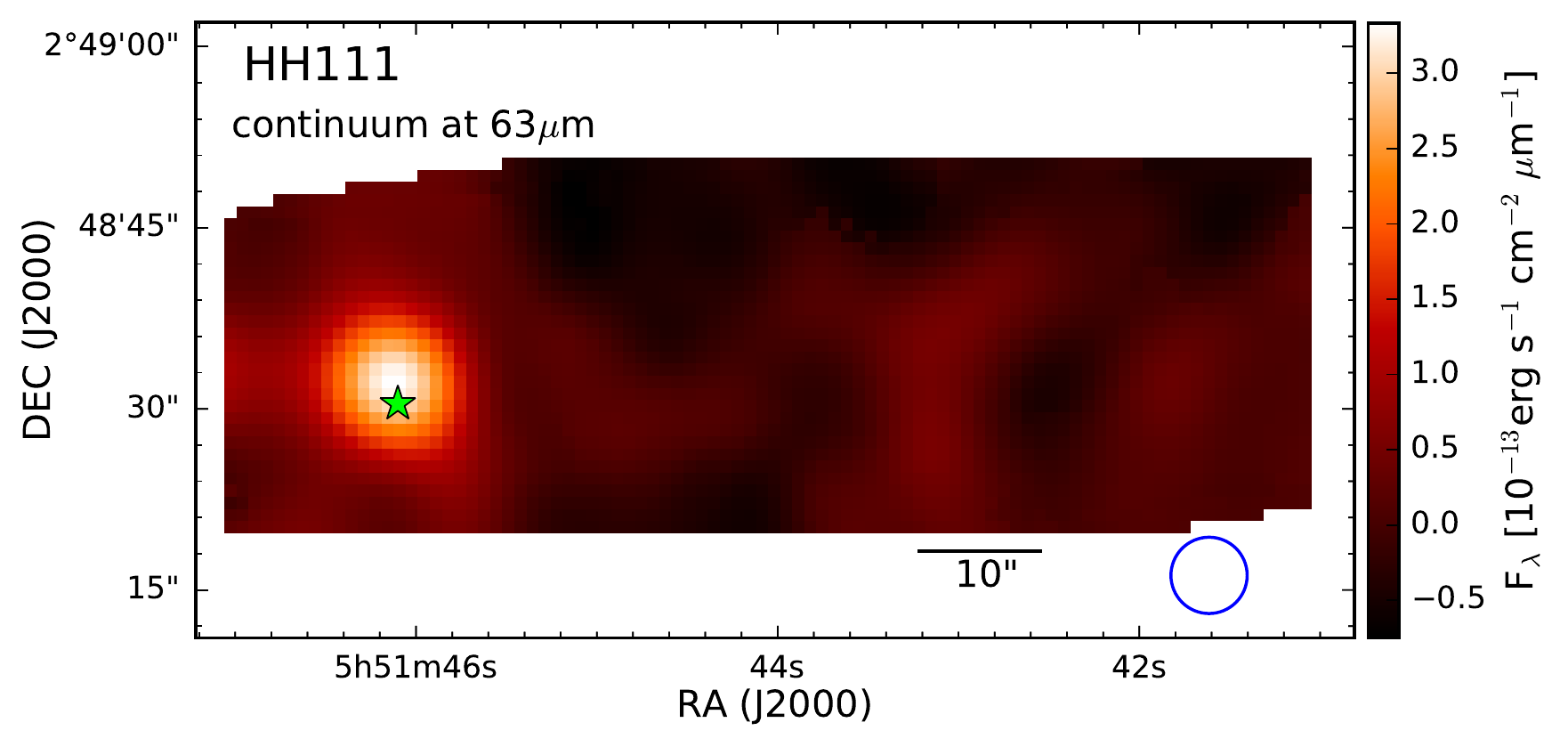}\label{fig:f1_continuum}}
  \hfill
  \subfloat{\includegraphics[trim=0 0 0 5, clip, width=0.9\textwidth]{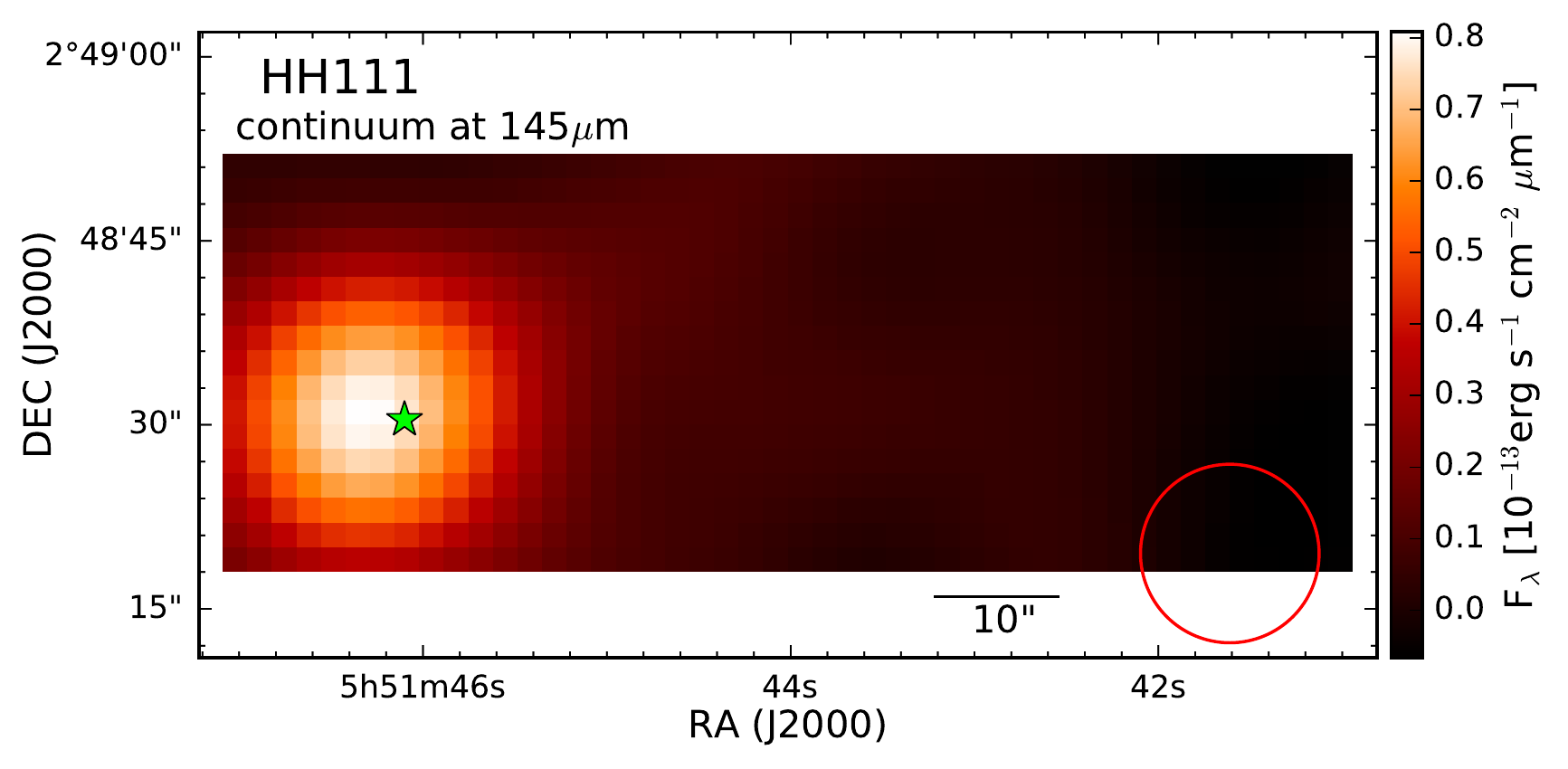}\label{fig:f2_continuum}}
  \caption{Continuum maps of HH111.}
\end{figure*}

\begin{figure*} 
  \centering
  \subfloat{\includegraphics[trim=0 33 0 0, clip, width=1.0\textwidth]{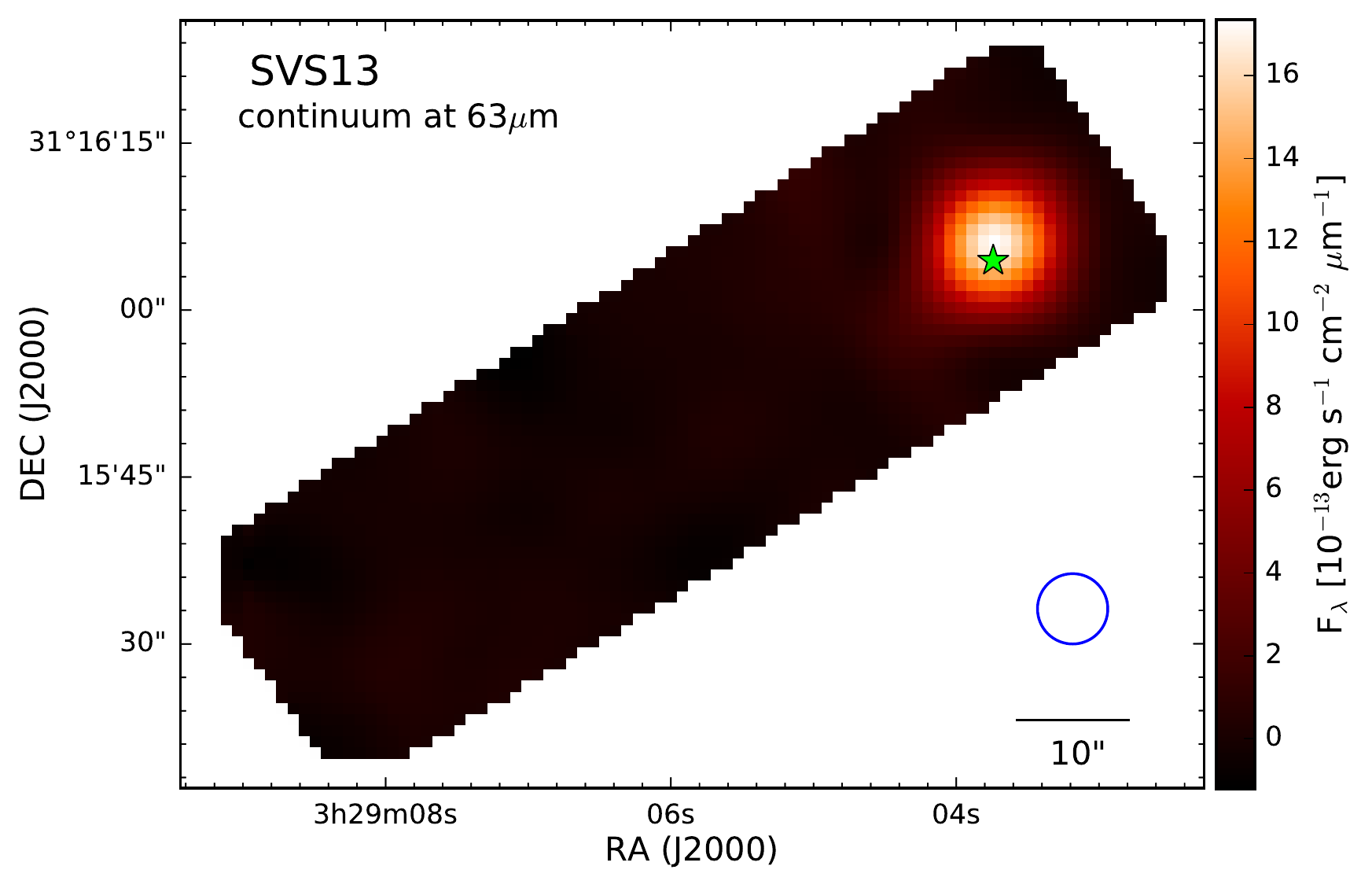}\label{fig:svs13_continuum_blue}}
  \hfill
  \subfloat{\includegraphics[trim=0 0 3 0, clip, width=1.0\textwidth]{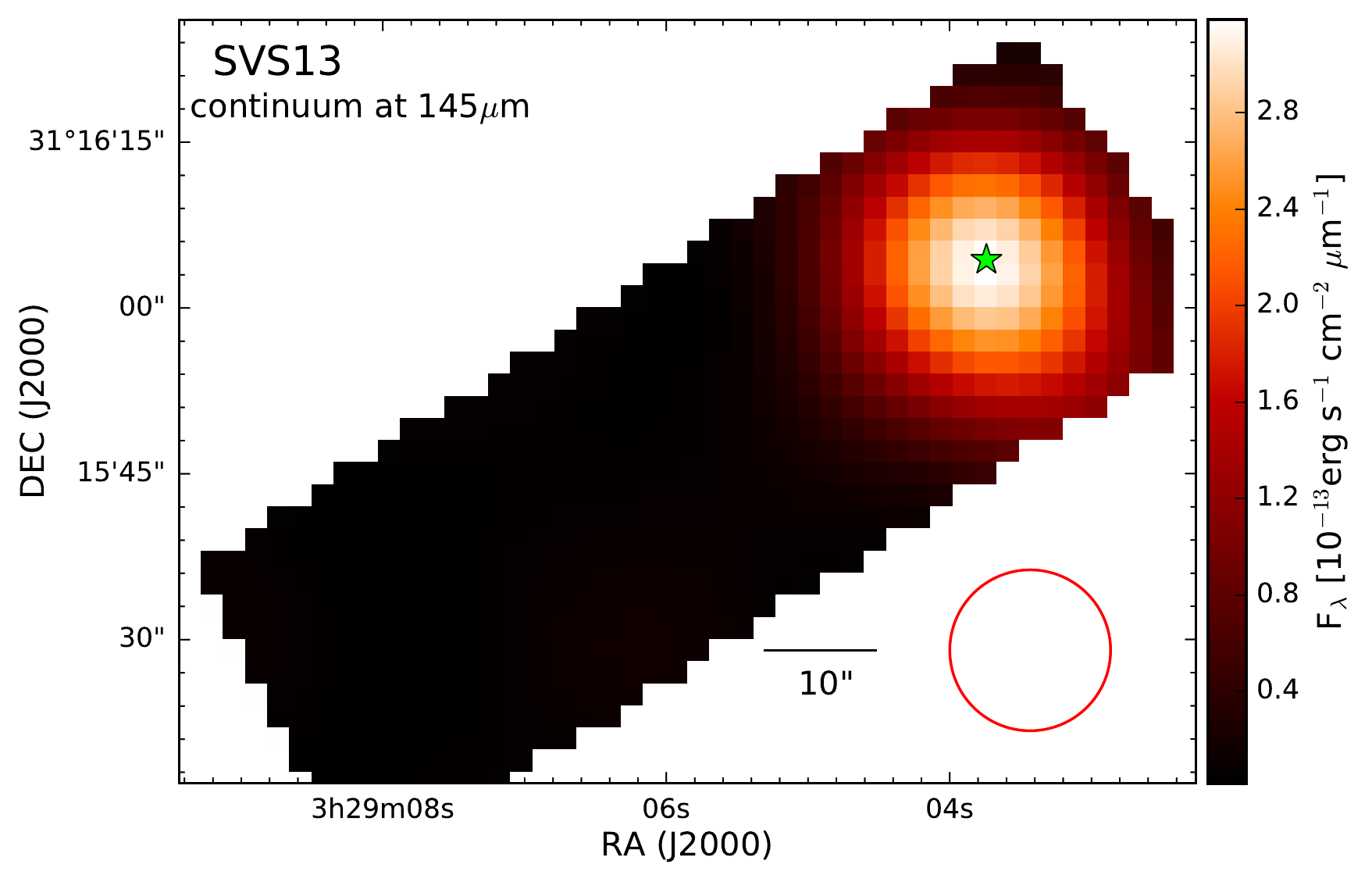}\label{fig:svs13_continuum_red}}
  \caption{Continuum maps of SVS13.}
\end{figure*}

\begin{figure*} 
  \centering
  \subfloat{\includegraphics[trim=0 0 0 0, clip, width=0.6\textwidth]{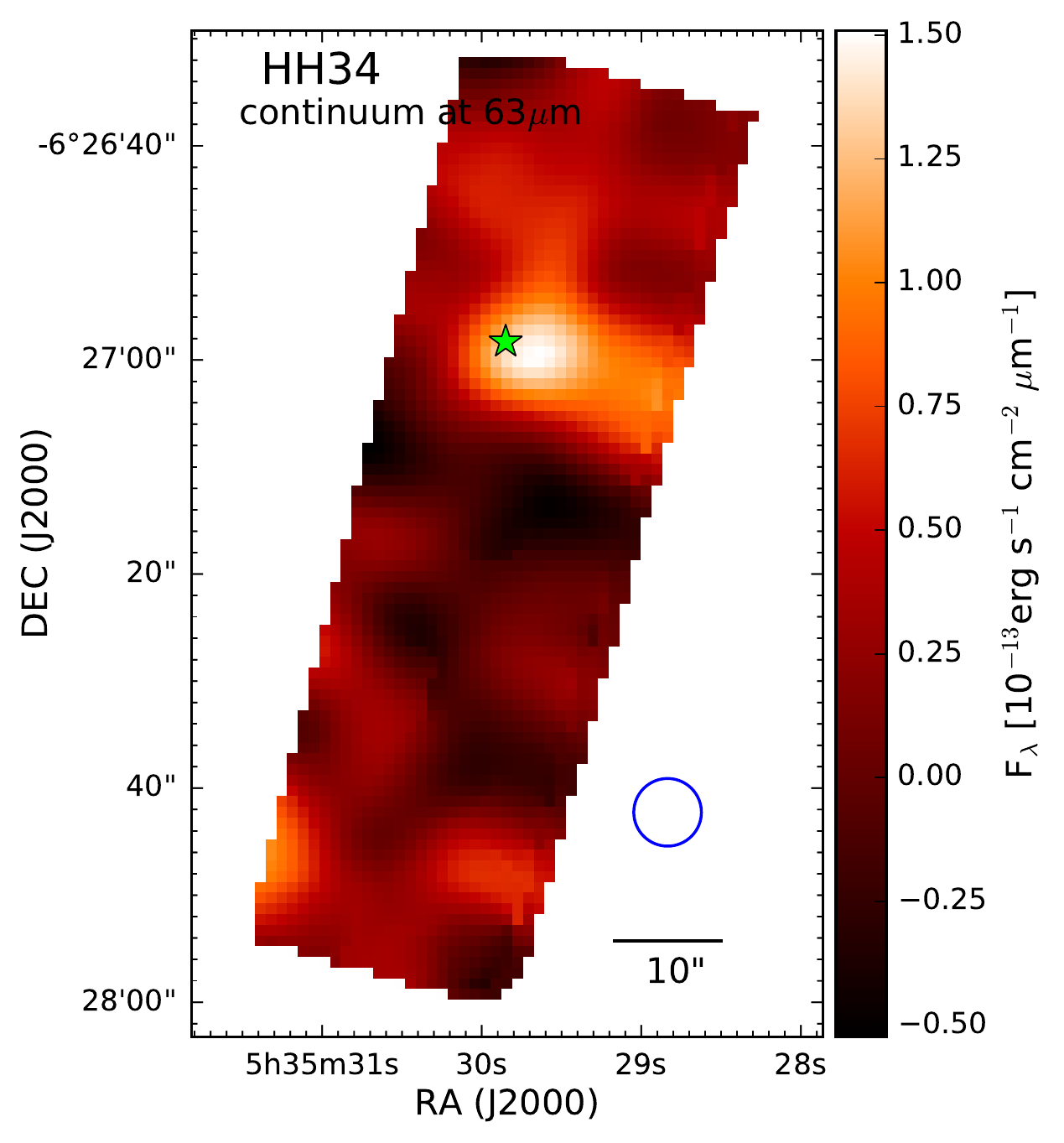}\label{fig:hh34_continuum_blue}}
  \hfill
  \subfloat{\includegraphics[trim=0 0 0 0, clip, width=0.6\textwidth]{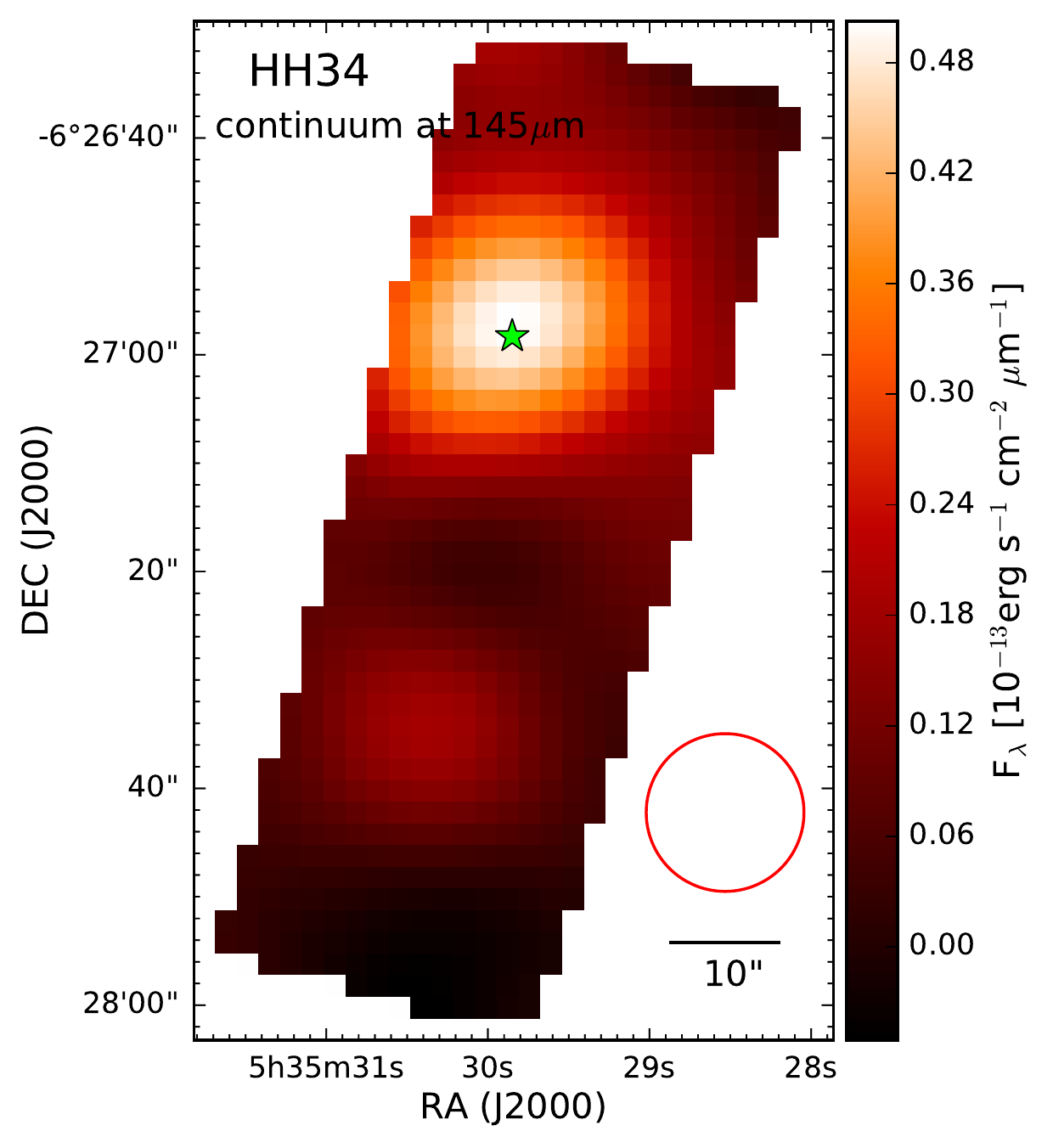}\label{fig:hh34_continuum_red}}
  \caption{Continuum maps of HH34.}
\end{figure*}

\begin{figure*} 
  \centering
  \subfloat{\includegraphics[trim=0 33 0 0, clip, width=0.78\textwidth]{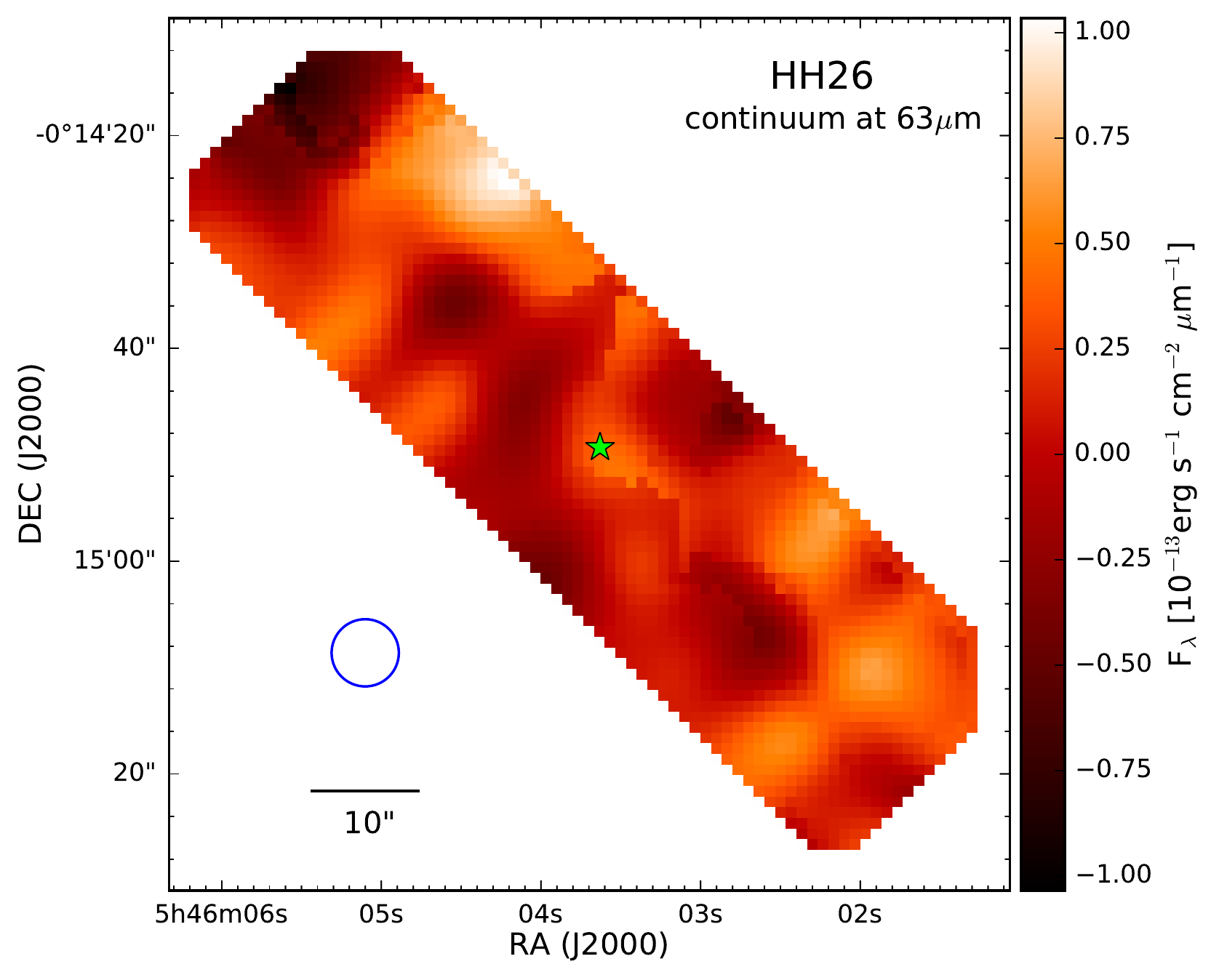}\label{fig:hh26_continuum_blue}}
  \hfill
  \subfloat{\includegraphics[trim=0 0 5 0, clip,  width=0.78\textwidth]{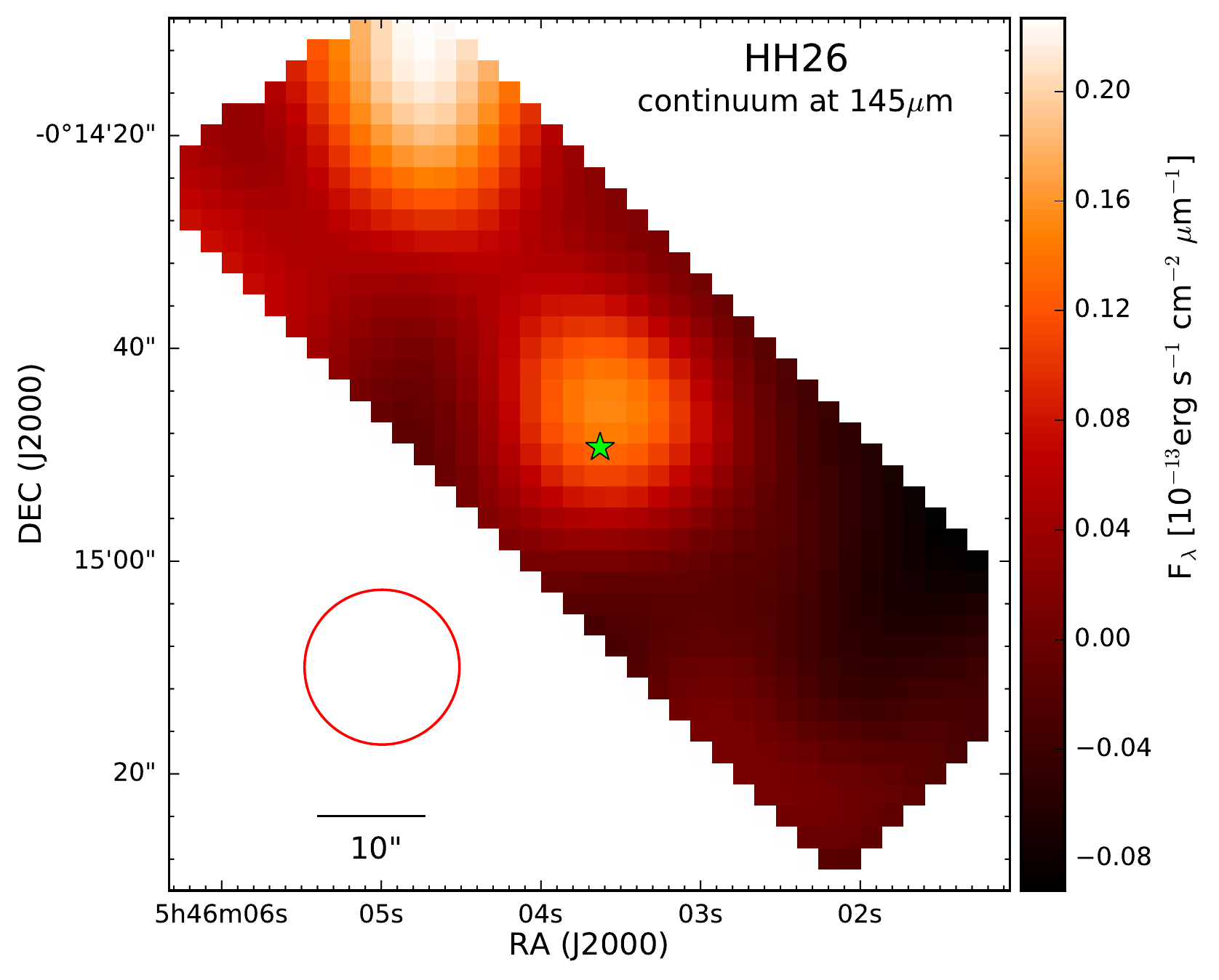}\label{fig:hh26_continuum_red}}
  \caption{Continuum maps of HH26.}
\end{figure*}

\begin{figure*} 
  \centering
  \subfloat{\includegraphics[width=0.7\textwidth]{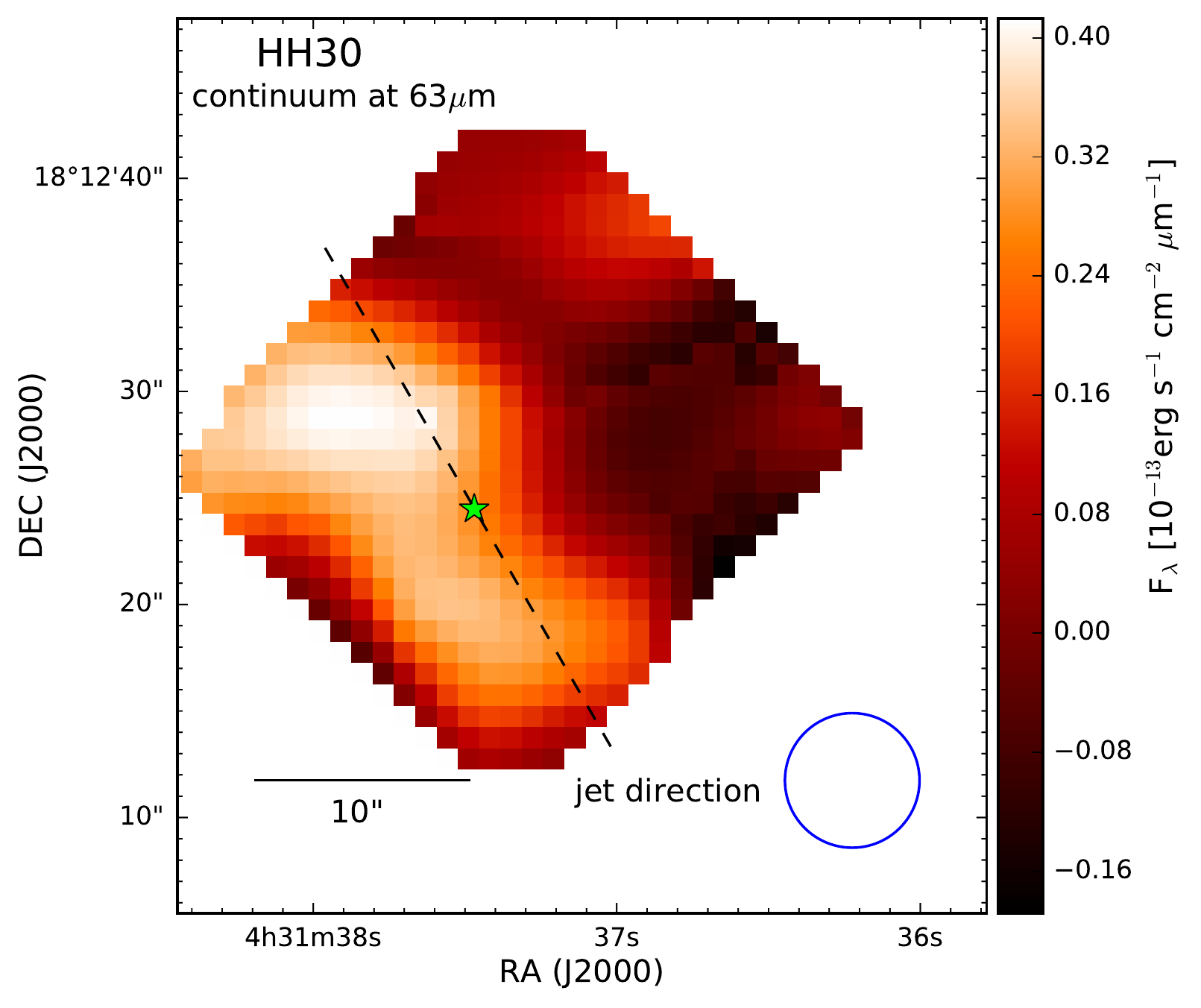}\label{fig:hh30_continuum_blue}}
  \hfill
  \subfloat{\includegraphics[trim=0 0 0 0, clip, width=0.7 \textwidth]{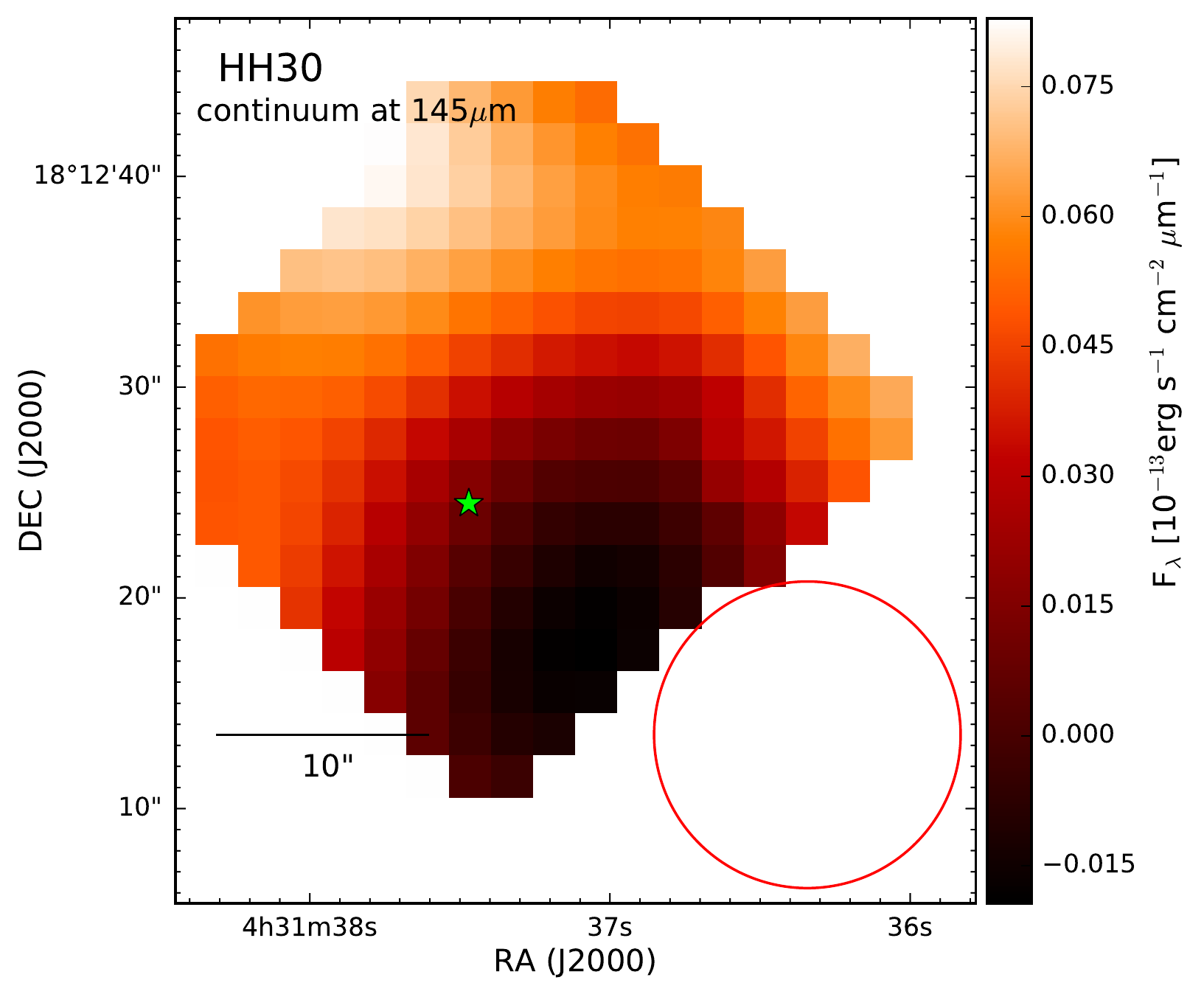}\label{fig:hh30_continuum_red}}
  \caption{Continuum maps of HH30. The dashed black line indicates the jet direction at P.A. $30^\text{o}$ \citep{pety_2006}.}
\end{figure*} 

 \end{appendix}

\end{document}